\documentclass[12pt]{article}
 \usepackage[dvips]{graphicx}
 \def \bpi{\mbox{\boldmath $\pi$}}
 \def \bsigma{\mbox{\boldmath $\sigma$}}
 \def \btau{\mbox{\boldmath $\tau$}}
 \def \blambda{\mbox{\boldmath $\lambda$}}
 \def \bM{\mathbf{M}}
 \def \bG{\mathbf{G}}
 \def \b\r{\mathbf{r}}
 \def \bJ{\mathbf{J}}
 \def \bL{\mathbf{L}}
 \def \bn{\mathbf{n}}
 \def \bp{\mathbf{p}}
 \def \bP{\mathbf{P}}
 \def \bR{\mathbf{R}}
 \def \bV{\mathbf{V}}
 \def \bgamma{\mbox{\boldmath $\gamma$}}
 \def \bepsilon{\mbox{\boldmath $\epsilon$}}
 
\begin{document}
 \begin{flushright} GEF - TH  11/09  \\ October 2009
\end{flushright}
 \begin{center}
 \noindent {\bf{The general QCD parametrization and the hierarchy of its parameters.\\ (Why 
some simple
 models of hadrons work so well)}}
  \vskip 15 pt
 G.Dillon and G.Morpurgo
 \end{center}
 Universit\`a di Genova and Istituto Nazionale di Fisica Nucleare, Sezione
 di Genova.
 \footnote{e-mail: dillon@ge.infn.it \hskip0.5cm ;\hskip0.5cm
 morpurgo@ge.infn.it}
 \vskip 15 pt
 \baselineskip 12pt
 \vskip 20 pt
 \noindent 1.Introduction.\\
 2.The derivation of the general QCD parametrization.\\
 3.The GP baryon mass formulas.\\
 4.The magnetic moments of the octet baryons to 1st order in flavor breaking.\\
 5.The $\Delta \rightarrow p+\gamma$ decay, the magnetic moments of the $\Delta$'s and 
related results.\\
 6.Double counting on inserting explicit pion fields in the QCD Lagrangian.\\
 7.The hierarchy of the parameters.\\
 8.The parametrization of the masses of the lowest Ps and V meson nonets.\\
 9.The radiative $V\rightarrow P + \gamma$ meson decays.\\
 10.The baryons electromagnetic mass differences and the Coleman-Glashow equation.\\
 11.Two relations: a) Between the charge radii of $p,n$,$\Delta^{+}$; b)Between the radii 
of\\ \indent $\pi^{+},K^{+}$,$K^{0}$.\\
 12.Parametrization of the $\rho\gamma, \,\omega\gamma$ and $\phi\gamma$ couplings:
    Why $f_{\rho\gamma}$:$f_{\omega\gamma}$ differs from 3:1 only\\ \indent by $\sim 10\%$ in 
spite of flavor breaking.\\
 13.The GP and chiral theories: A few remarks.\\
 14.Comparing the GP with the $1/N_{c}$ method; some comments.\\
 15.Appendix I-A field theoretical derivation of the GP.\\
 16.Appendix II.A summary of the main steps in the derivation of the spin-flavor\\
 \indent dependence of the GP terms.\\
 \\ \\ \\ \\ \\ \\ \\ \\
 We dedicate this article to the memory of Giuseppe Franco Bassani, past president of the 
SIF.

 \newpage

 \vskip 20 pt \noindent {\bf 1. Introduction.}\\

    This survey summarizes
 the ``General QCD Parametrization" (GP), a procedure derived exactly from QCD, to understand 
and predict many hadron properties; it applies to the
 low energy region where, obviously, the perturbative method is out of question.\\
 \indent The GP was formulated \cite{mo89} to explain why a naive description -such as the 
simple non relativistic quark model (NRQM)\cite{mo65}- can reproduce semi-quantitatively many 
properties of hadrons.
 A classical example is that of fig.1, where a two parameter NRQM
 description of the magnetic moments of the lowest octet baryons is compared with the 
experimental data.\\
 \indent Indeed the reason why the NRQM was so successful
 remained an open question till the development of the GP.{\footnote{Some years later the 
same problem was studied by L.Durand et al.
 via chiral QCD \cite{du02}(compare also \cite{du01, du03, du05}). See also 
Ref.\cite{der75}(De Rujula et al.), including a discussion
 of the relation between the NRQM and a Fermi-Breit approximation of the QCD Hamiltonian.}
  A similar problem (why does it work?) existed for other simple models
 (e.g. the MIT bag model \cite{chodos} -more limited, but still of interest), also providing 
a simple description of several facts.
 The GP solves these problems. Short summaries of the GP were given
 previously \cite{dm97}. This report amplifies them.\\
 \indent The starting point of the GP is this: Many hadron properties (e.g. magnetic
 moments, masses, electromagnetic (e.m.) mass differences, semileptonic decays, strange quark 
contribution to the proton
 e.m. form factor etc.) can be parametrized exactly in the spin-flavor space exploiting only 
a few general properties of QCD.
  These are:\\
   \indent a) Flavor breaking is due only to the mass term in the Lagrangian,\\
   \indent b) Only quarks carry electric charge,\\
   \indent c) Exact QCD eigenstates can be put in correspondence (for baryons) to a set of 
three quark-no gluon states
   and (for mesons) to a set of quark-antiquark-no gluon states,\\
   \indent d) The flavor matrices in the electromagnetic (e.m.) and in the flavor breaking 
term of the QCD Lagrangian
   commute.\\
  \indent \textit{However, although the above properties are essential in what follows, they 
would not lead too far
    without exploiting the  "hierarchy" of the parameters \cite{mo92}}, mentioned below (see 
point 3) and illustrated
   later (Sect.7).\\
  \indent As we will see, the``General QCD Parametrization" has three important features}:\\
  \indent 1) The spin-flavor structure of the hadron properties, expressed in terms
  of 2x2 Pauli matrices, is similar to that of the NRQM; yet the GP is fully relativistic, 
although the procedure to
  derive it from QCD is non covariant.\\
  \indent 2) In spite of the complexity of the QCD interaction terms,
   the number of \textit{important} additive terms appearing in the parametrization of 
several hadronic
   properties is often rather small, much smaller than one might have expected a priori.
   One reason for this will be stated in the next point (3).\\
   \indent 3) When all the values of the coefficients of each term in the parametrization 
(the parameters)
   can be extracted from the experimental data, it turns out that the terms of increasing 
complexity
   (for the meaning of ``complexity" compare Sect.7) are multiplied by decreasing parameters; 
we call this \textit{the
   hierarchy of the parameters}. The hierarchy reduces the number of significant terms needed 
to reproduce the data,
   thus explaining the success of simple models like the NRQM.

   \begin{figure}
    \begin {center}
    \includegraphics[width=9cm,height=12cm]{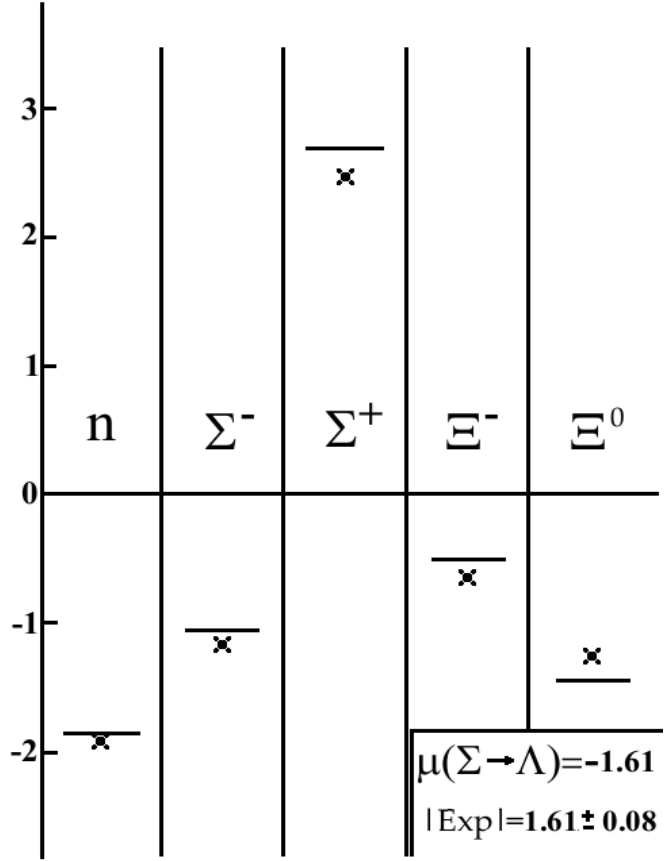}
     \end{center}
   \caption{\footnotesize{The measured magnetic moments of the baryons compared with the 
values (solid lines) calculated with the simple two-parameter formula (Eq.1 in 
Ref.\cite{mo89}) of the NRQM with $\mu=2.793$ and $A=0.96$ (that is $a=0.65$) having used as 
input only the proton ($+2,793$)  and $\Lambda$ ($=-0.613\pm 0.004$) magnetic moments. 
[Calculated and (measured) values are neutron $=-1.86$ ($-1.913$); $\Sigma^-=-1.04$ 
($-1.160\pm 0.025$); $\Sigma^+=+2.68$ ($+2.458\pm 0.040$); $\Xi^-=-0.05$ ($-0.6507\pm 
0.0025$); $\Xi^0=-1.43$ ($-1.250\pm 0.014$); $\mu(\Sigma\rightarrow \Lambda=-1.61$) 
($|\mathrm{expt}|=1.61\pm 0.08$).]}}
   \end{figure}

     4) The hierarchy allows also to define properly the notion of constituent quarks (first 
introduced in \cite{mo65} as dressed quarks
   with their cloud of $\overline{q}q$ and gluons).\\
   \indent The fact that the general QCD parametrization is derived
   exactly from QCD using only the above properties ($a$ to $d$)
   has the following implication: Any specific approximation (good or bad)
   to the exact QCD Lagrangian \textit{sharing the simple properties a) to d), listed above
   (think e.g. to some forms of chiral QCD Lagrangian)} also may lead to results with the 
same
   general structure as those of the GP (though the values of the parameters so obtained are 
not necessarily correct).\\
   \indent By this we mean that if one performs a complete calculation with a
   specific approximate Lagrangian (say a chiral one) and/or with lattice QCD, one may obtain 
specific
   numerical values for the parameters mentioned
   above. But it is not evident that the values so obtained are better than those resulting 
from the GP,
    by fitting the largest parameters empirically and using the hierarchy
   mentioned above to estimate the order of magnitude of the remaining ones. Whether
   this is so or not depends on the chiral theory used, on the approximations
   performed in the chiral calculation etc.\\
   \indent To anticipate an example of application of the GP we write below a formula (to be 
derived in Sect.3) obtained with the GP \cite{mopl92}
   for the lowest octet + decuplet baryon masses. This formula -Eq.(1)- improves the
   Gell Mann-Okubo formula; the particle symbols stay for their masses and $T$ is a 
definition of the quantity appearing
    in the equation on the left. We will derive the Eq.(1) at the end of Sect.3. The standard 
Gell Mann-Okubo formula
     is the same as Eq.(1), without the $T$ term.
   \begin{equation}
   \frac{1}{2}(p+\Xi^{0})+ T = \frac{1}{4}(3\Lambda
   +2\Sigma^{+}-\Sigma^{0})\hspace{2cm}
    T\equiv\Xi^{*-} - \frac{1}{2}(\Omega +\Sigma^{*-})
   \end{equation}
   The fit to the data depends on whether we use the conventional or the pole values for the 
masses of the resonances appearing in $T$:
   \begin{eqnarray}&&
    l.h.s.= 1132.36\pm 0.7\, MeV, \quad \quad  r.h.s.= 1133.93\pm\, 0.04 MeV \,\quad\quad 
(conventional)\nonumber\\
    &&l.h.s.= 1133.86\pm 1.25\, MeV,\quad \quad  r.h.s.= 1133.93\pm\, 0.04 MeV \quad\quad 
(pole)
   \end{eqnarray}
    (the above numbers need no corrections from the e.m. contributions to the masses). A 
similar formula (except for the e.m.
    corrections) was re-obtained only later - by a series of chiral calculations (compare: 
Ref.\cite{du02}, its Erratum Ref.\cite{duE02},
     and Ref.\cite{dimo03}).\\
    \indent Two other points should be mentioned:\\
    \indent a) The GP does not need, for its derivation, the use of unproven, though widely 
adopted, assumptions. For instance it is often
    implied that chiral dynamics is needed to relate QCD to constituent quark models. This is 
incorrect: The relationship
    between QCD and constituent quark models depends on the hierarchy mentioned above and is 
much more general (See Sect.7).\\
    \indent Neither chiral dynamics nor the notion of pions as quasi-Goldstone bosons
    play any role in deriving the general parametrization that relates the quark descriptions 
(both current and constituents) to QCD.\\
    \indent b) We will consider only problems of hadrons with light quarks $(u,d,s)$. The 
fact that the NRQM works quantitatively has always been
    considered miraculous for the light quarks (although to understand its working raises 
problems also for the hadrons
     with non zero charm or beauty). It appears from the GP that the reason why the NRQM 
works has nothing to do with the low
     velocity of quarks inside the hadrons, that was implied in \cite{mo65}; $u,d,s$ quarks 
can move as fast as one likes.\\
    \indent Finally, besides clarifying why some models work,
    the general parametrization (with its hierarchy of parameters) leads to a variety of new 
results, as we shall see.

    \vskip 30 pt
    \noindent {\bf 2. The derivation of the general QCD parametrization.}\\

    In the past section we exemplified in fig.1 (magnetic moments) how the NRQM describes 
semi-quantitatively the data.
    Here we show how the general parametrization is derived from QCD. It is simpler to 
discuss the masses of the
    lowest $\bf{8}$ + $\bf{10}$ baryons, instead of the magnetic moments. Thus we illustrate 
the GP first with
   the masses and consider later the magnetic moments (and other problems). However most of 
this section, even if referring
   to the masses, is intended to be a general introduction; it is closely related to 
Ref.\cite{mo89} and, in part, to Ref.\cite{dimo96}.\\
   \indent The mass $M_{B}$ of a baryon $B$ is the expectation value of the exact QCD 
Hamiltonian $H_{QCD}$ in
   its lowest exact QCD eigenstate $\mid\Psi_{B}\rangle$ in the rest system:
   \begin{equation}
                M(B)= \langle\Psi_{B}\mid H_{QCD}\mid\Psi_{B}\rangle
   \end{equation}
   Of course $\mid\Psi_{B}\rangle$, being the exact state of a strongly interacting system of 
quarks and
   gluons, is a superposition of an infinite number of Fock states,
   starting with three quarks. The superposition includes four quarks and one antiquark, 
three quarks plus one gluon and so on. Schematically:
   \begin{equation}
   \mid\Psi_{B}\rangle = \mid qqq\rangle + \mid qqqq\bar{q}\rangle + \mid qqq,\,Gluons\rangle 
+......
   \end{equation}
   where the ellipsis stands for the sum of an infinite number of additional states; the 
amplitudes that multiply each state
   (depending on the momenta, spins, flavors, colors of the intervening quarks, antiquarks 
and gluons) have been left understood.\\
   \indent We now introduce an auxiliary Hamiltonian $\mathcal{H}$, non-relativistic, 
operating
    $only$ in the $3q$ sector; the operator $\mathcal{H}$ (we call it the ``model 
Hamiltonian") has the only purpose
    of providing a set of baryon states - to be called the $model$ (or $auxiliary$) states 
$\mid \Phi_{B}\rangle$.
    $\vert\,\Phi_{B}\rangle$ is constructed (as in the NRQM) with just three quarks and no 
gluon.\\
   \indent We now write the exact state $\vert\,\Psi_{B}\rangle$ in (4) as:
   \begin{equation}
   \mid\Psi_{B}\rangle = V\mid\Phi_{B}\rangle
   \end{equation}
   where $V$ is some (very complicated) unitary transformation. In principle $V$ can be 
expressed in terms of
   $H$ and $\mathcal{H}$ using the adiabatic construction of the bound states. (See
   the Appendix I, where the construction of $V$ is related to the Gell Mann-Low adiabatic 
procedure and to the
   $U(0,-\infty)\,$ Dyson operator). Using $V$, the Equation (3) giving the mass $M_{B}$ of 
$B$, can be rewritten:
   \begin{equation}
   M_{B}= \langle \Phi_{B}\mid V^{\dagger}H_{QCD}V\mid \Phi_{B}\rangle
   \end{equation}
   \indent The difference between Eq.(6) and Eq.(3) is the following: In Eq.(6)
   the states are simple; the complexity of the states (4) is transferred to $V$. This has an 
advantage which is
   basic in the procedure: $V^{\dagger}H_{QCD}V$ is indeed a complicated operator, but since 
it
   has to act only on the coordinates (space, spin, flavor, color) of the $three$ quarks 
present in the state $\vert\Phi_{B}\rangle$,
    it must be (after contraction of all the field operators) necessarily a function of these 
coordinates only. In what follows the three
   quarks in $\vert\Phi_{B}\rangle$ will be numbered $1,2,3$. Thus, after the elimination of 
all the creation and destruction operators,
   $V^{\dagger}H_{QCD}V$ behaves as a color singlet three body operator acting on $1,2,3$.\\
   \indent In Eq.(3) $H_{QCD}$ transforms -under space rotations- as a scalar and the same is 
true for
   $V^{\dagger}H_{QCD}V$ because $V$ is invariant under rotations (it is expressed in terms 
of $H_{QCD}$ and $\mathcal{H}$, both
   rotationally invariant). As to $\Phi_{B}$ in Eq.(6), it depends on how we select the model 
Hamiltonian $\mathcal{H}$.
   By choosing $\mathcal{H}$ as the \textit{simplest, most naive, most unrefined} N.R. quark 
model Hamiltonian,
   the parametrization of $\langle \Phi_{B}\mid V^{\dagger}H_{QCD}V\mid \Phi_{B} \rangle$ is 
considerably simplified. We
   select $\mathcal{H}$ so that, for the lowest octet and decuplet baryons, the wave 
functions $\Phi_{B}$ have a non relativistic
   space-spin structure with the following properties: (1) $\Phi_{B}$ is the product of a 
space-spin structure
   $X(\b\r_{1},\b\r_{2},\b\r_{3})$ symmetrical in $\b\r_{1},\b\r_{2},\b\r_{3}$, times a 
spin-flavor part $W_{B}(1,2,3)$,
   times a color singlet factor $C(1,2,3)$; (2) The space part has orbital angular momentum 
$\bL=0$; thus
   $X \equiv X_{L=0}(\b\r_{1},\b\r_{2},\b\r_{3})$.\\
   \indent Altogether we have for the baryon $B$:
   \begin{equation}
   \Phi_{B}= X_{L=0}(\b\r_{1},\b\r_{2},\b\r_{3})\cdot W_{B}(1,2,3)\cdot C(1,2,3)
   \end{equation}
   where $C(1,2,3)$ is the color factor.\\
   \indent The assumption $\bL=0$ and the symmetry of the space wave function imply 
automatically that the spin factors
   $W_{B}(1,2,3)$ have the $SU_{6}$ structure. Omitting the color factor $C(1,2,3)$ one has, 
for instance ($S$ in Eq.(8)
    below means symmetric in $1,2,3$)\footnote{Of course $\Delta$ has an
   appreciable width and, therefore, is not an exact eigenstate of $H_{QCD}$ but this is 
irrelevant here and will be considered later.}:
   \begin{equation}
W_{p}^{\uparrow}=(18)^{-1/2}S[\alpha_{1}(\alpha_{2}\beta_{3}-\alpha_{3}\beta_{2})u_{1}u_{2}
d_{3}]; \quad\quad
    W_{\Delta^{++}}^{\uparrow}=\alpha_{1}\alpha_{2}\alpha_{3}u_{1}u_{2}u_{3}
    \end{equation}
    The $W_{B}$'s of the other $\mathbf{8}$ and $\mathbf{10}$ states are constructed 
similarly.\\
    \indent Two remarks must be added:\\
    \indent The first is on the spin functions:  Because the operator $V$
     must be written in terms of creation and destruction operators of Dirac particles, the 
spinors appearing in
    $W_{B}(1,2,3)$ must be four component spinors; otherwise the operation $V$ would not be 
defined.
    This is achieved by completing the Pauli spinors of the model wave function with two 
zeros in the lower components.
    This is compatible with our non-relativistic choice of the model Hamiltonian 
$\mathcal{H}$. In fact  $\mathcal{H}$ operates
    on two-component Pauli spinors, but it is formally possible to extend the space of such 
spinors to that of
    four component spinors, provided that $\mathcal{H}$ is extended without connecting the 
space of the upper
    and lower components and giving zero when operating on the latter (compare 
\cite{mo89}).\\
    \indent The second remark is related to the $X_{L=0}(\b\r_{1},\b\r_{2},\b\r_{3})$ in the 
wave function.
    Why $X_{L=0}$ does not carry a baryon index $B$, specifying the baryon we are 
considering? The answer is that we selected a flavor
      independent $\mathcal{H}$, with the masses of the quarks $u,d,s$ in $\mathcal{H}$ 
chosen equal,
     at some intermediate value between those of the $u,d$ and $s$ masses. The whole flavor 
breaking is assigned to the
     $V$ operator; recall that $V$ depends on the \textit{exact} QCD Hamiltonian $H_{QCD}$; 
this contains
     the flavor breaking due to the mass differences between $u,d,s$. (We might have made a 
different choice,
      constructing the model Hamiltonian $\mathcal{H}$ with a flavor dependence, but we 
preferred to assign also this task to $V$).\\
     \indent Consider now briefly the flavor dependence of $H_{QCD}$. The mass term in the 
QCD Hamiltonian is for the $u,d,s$ quarks:
     \begin{equation}
     \int d^{3}\b\r\,\, m\lbrack \bar{u}_{R}(x)u_{R}(x) + \bar{d}_{R}(x)d_{R}(x)\rbrack 
+(m+\Delta m) \bar{s}_{R}(x)s_{R}(x)
     \end{equation}
     where the index R refers to the mass renormalization point.\\
     \indent In \cite{mo89} we never used explicitly the values of the quark masses $m$ (the 
quarks were then indicated as $\mathcal{P},
     \mathcal{N}, \lambda$) nor a value of $\Delta m$, but we had in mind $m$ values 
(corresponding to a renormalization point
      around $\Lambda_{QCD}$) of the order $300-400\,MeV$ for $\mathcal{P},\mathcal{N}$ and 
$450-550\,MeV$ for $\lambda$,
     such that $\Delta m/m \cong 1/3$. Subsequently \cite{dimo96}
     we re-analyzed the results of \cite{mo89} renormalizing the quark masses at the 
conventional $q=1\,GeV$ (we used then
     $u,d,s$ symbols for quarks, with $m_{u},m_{d}$=a few MeV). As it appeared (and as it 
will emerge also here) the GP procedure
      is independent of the choice of the mass renormalization point. In fact in 
\cite{dimo96} we confirmed and
      extended significantly the results of Ref.(\cite{mo89}). \footnote{In Ref.\cite{mo89} 
the ratio  $\Delta m/m$ (associated to
     the magnitude of flavor breaking) determined from the baryon masses and the baryon octet 
magnetic moments - was found
     in both cases $\cong 0.3$. In Ref.\cite{dimo96} (compare its Sect.II) $\Delta m/m \cong 
m_{s}/m$ is, of course, totally different
     (say, in the range $8-25$), but the flavor breaking parameter turned out to be $\approx 
m_{s}/(3\Lambda_{QCD})\approx 0.3$, the same value
     obtained in Ref.\cite{mo89}. In the present paper, for simplicity, the quark symbols 
will be always $u,d,s$
     (and, only if necessary, we will specify the mass renormalization point in the QCD 
Lagrangian).} \\
   \indent Coming back to the description of the model state of a baryon in momentum space, 
we obtain (performing in Eq.(7)
    a Fourier transformation of $X_{L=0}$):
   \begin{equation}
    \mid\Phi_{B}\rangle = \sum_{\bp,w}C^{B}_{w_{1},w_{2},w_{3}}(\bp_{1},\bp_{2},\bp_{3})
    a^{\dagger}_{\bp_{1},w_{1}}a^{\dagger}_{\bp_{2},w_{2}}a^{\dagger}_{\bp_{3},w_{3}}
    \mid 0\rangle
   \end{equation}
    where $\sum_{\bp,w}$ stays, as indicated, for the sum over all the \textbf{p}'s and 
$w$'s. In Eq.(10) the
     $a^{\dagger}_{\bp,w}$ are creation operators of quarks of
    momentum \textbf{p} and spin-flavor-color index $w$, $\mid 0\rangle$ is the vacuum state 
in a Fock space of quarks and
     gluons. $\mid\Phi_{B}\rangle$ belongs to a given representation of $SU_{3}$ (flavor); in 
it
    the masses of the three quarks $u,d,s$ are taken equal and
    their values identified with their QCD values (in the unbroken flavor limit) at any 
convenient mass renormalization point.\\
    \indent Because we are in the baryon rest frame, 
$C^{B}_{w_{1},w_{2},w_{3}}(\mathbf{p}_{1},\mathbf{p}_{2},\mathbf{p}_{3})$ in Eq.(10)
    contains a factor $\delta(\bp_{1}+\bp_{2}+\bp_{3})$.\\
    \indent We now come back to the unitary correspondence $V$ between the model
    state $\mid\Phi_{B}\rangle$ and the exact QCD state $\mid\Psi_{B}\rangle$.
     Recall that, for the lowest octet and decuplet baryons, $\vert \Phi_{B}\rangle$ is a 
simple 3 quark S state, while, of
    course, $\vert \Psi_{B}\rangle$ (4) contains in addition all kinds of
    $q\overline{q}$ and gluon states and, moreover, it certainly has terms with a non 
vanishing
    orbital angular momentum $L$. For convenience we rewrite the Eqs.(4),(5) as:
    \begin{equation}
    V\vert\Phi_{B}\rangle\, =\, \mid qqq\rangle + \mid qqqq\overline{q}\rangle + \mid 
qqq,Gluons\rangle +......
    \end{equation}
    Obviously each state in the sum (11) has the same conserved quantum numbers as those of
    $\vert \Psi_{B}\rangle$. The transformation $V$ has the tasks:\\
    \indent a) Of dressing the simple three quarks state $\mid\Phi_{B}\rangle$, that is of 
transforming $\mid \Phi_{B}\rangle$
     into an infinite sum of states of $q, \bar{q}$ and gluons,
    the only restriction being that such states have the correct quantum numbers;\\
    \indent b) Of introducing configuration mixing in
    $L$; this is not present in $\mid\Phi_{B}\rangle$ which, as stressed, is a pure $L=0$ 
state;\\
    \indent c) Of transforming the Pauli spinor quark states in $\mid\Phi_{B}\rangle$ into 
Dirac 4-spinors.\\
    \indent As already mentioned, in the Appendix I we will show how the existence of $V$ can 
be
    established in field theory. Now we proceed directly to the parametrized baryon mass 
formula (and to the other
    properties of hadrons). Calling $M_{B}$ the mass of a baryon of the lowest $\mathbf{8}$ 
or $\mathbf{10}$, we start from Eqs.(3),(6)
    that we recall below:
    \begin{equation}
    M_{B}= \langle\Psi_{B}\mid H_{QCD}\mid \Psi_{B}\rangle = \langle\Phi_{B} \mid
    V^{\dagger}H_{QCD}V\mid \Phi_{B}\rangle
    \end{equation}
    \indent The general form of the parametrization of any \textit{physical property} (e.g. 
the baryon \textbf{8+10} masses to be discussed
    now, or the magnetic moments or any other property) is independent of the selection of a 
specific
    $\vert\Phi_{B}\rangle$. That is, independently of this selection,
     the calculated $M_{B}$ is, in all cases:
    \begin{equation}
    M_{B}=\langle\Phi_{B}\mid V^{\dagger}H_{QCD}V\mid\Phi_{B}\rangle \equiv \langle W_{B}\mid 
parametrized \ mass \mid W_{B}\rangle
    \end{equation}
    In Eq.(13) the last term \textit{defines} what we call the ``parametrized mass" of $B$; 
that is, it displays the general parametrization of
    the mass of $B$.\\
    \indent We now outline the calculation of the expectation value of 
$\,V^{\dagger}H_{QCD}V$ in the state $\mid \Phi_{B}\rangle $;
    this amounts to construct explicitly the last term of Eq.(13), thus expressing the masses 
as the expectation values of a spin-flavor
    operator in the $3q$ spin-flavor functions, $W_{B}$. Because $\Phi_{B}$ in (13) is a 3 
quark state, the only part of $V^{\dagger}H_{QCD}V$ that
    contributes in Eq.(13) is its projection $\tilde{H}$ in the $\mid 3q\rangle$ Fock sector:
    \begin{equation}
    \tilde{H} = \sum_{3q}\sum_{3q'}\mid 3q\rangle \langle 3q\mid V^{\dagger}H_{QCD}V\mid 
3q'\rangle \langle 3q'\mid,
    \end{equation}
    where the sums in (14) are now on all possible 3-quark, no gluon Fock states. After 
normal ordering of all the creation and
    destruction operators in $\tilde{H}$ and their contraction with those arising from 
$\mid\Phi_{B}\rangle$ and $\langle\Phi_{B}\mid$
    (see Eq.(13)), the operator $\tilde{H}$ becomes a function of only the spin-flavor-space 
variables of the three quarks in
    $\vert \Phi_{B}\rangle$; thus parametrizing $\tilde{H}$ means to construct the most 
general scalar operator in the
    space ${\b\r}_{i}$, spin $\bsigma_{i}$'s, flavor $f_{i}$'s and color operators of the 
three quarks ($i=1,2,3$), keeping obviously
    only the terms with a non-vanishing expectation value in $\mid \Phi_{B}\rangle$.
    (We identified already the quarks in the model states $\mid \Phi_{B}\rangle$ as those 
appearing in the QCD Lagrangian at the
    selected renormalization point in the no flavor breaking limit). Thus it is 
straightforward
    to contract the creation and destruction operators in $V^{\dagger}H_{QCD}V$ with
    those in the auxiliary states $\mid \Phi_{B}\rangle$. This leads to the last term of 
Eq.(13). After this
    contraction the projection $\tilde{H}$ of the field operator $V^{\dagger}H_{QCD}V$
    in the 3-body sector becomes a scalar (i.e. a rotation invariant) function of the space
    ${\b\r}_{i}$, spin ${\bsigma}_{i}$, flavor $f_{i}$ and color operators of the three 
quarks.
    One must write the most general expression of that operator. We call it $\tilde{H'}$ (a 
different symbol must of course be used
    because $\tilde{H}$ (Eq.(14)) operates in Fock space, whereas $\tilde{H'}$, obtained 
after contraction of the field operators,
    is just a 3-body quantum mechanical operator).\\
    \indent The number of independent scalar operators in the spin-flavor space of $3$ quarks 
is of course finite.
    We use for them the symbol $Y_{\mu}(\bsigma,f)$, where the index $\mu$ specifies the 
operator to which we refer.
    Thus \textit{the most general operator of the space and spin-flavor variables is 
necessarily}
    \begin{equation}
              \tilde{H'} = \sum_{\mu}R_{\mu}(\mathbf{r},\mathbf{r'})Y_{\mu}(\bsigma,f)
    \end{equation}
    where $R_{\mu}(\b\r,\b\r')$ are operators in the coordinate space of the three quarks 
and: $\b\r\equiv(\b\r_{1},\b\r_{2},\b\r_{3}),
    \hspace{0.1cm}
    \b\r^{'}\equiv(\b\r^{'}_{1},\b\r^{'}_{2},\b\r^{'}_{3})$. In the baryon rest system the 
combination $\sum_{i}\b\r_{i}$
    does not intervene; only ($\b\r_{i} - \b\r_{k} $) and ($\b\r_{j}- 
(1/2)(\b\r_{i}+\b\r_{k})$) appear [$i\neq j\neq k$].\\
    \indent To calculate the masses we must calculate the expectation value of
    $\tilde{H'}$, Eq.(15), on $\mid\Phi_{B}\rangle$, given by Eq.(10)\footnote{One might ask 
why we do not calculate the
    $(masses^{2})$ operating similarly with the $(Hamiltonian^{2})$; one might do so, except
    that the $(Hamiltonian^{2})$ is most probably non renormalizable. (Think, in ordinary 
quantum mechanics, of using the square
    of the Hamiltonian of the hydrogen atoms to calculate its levels!).} (Recall that 
$\Phi_{B}$ is the product of a space part
    $X_{L=0}(\b\r_{1},\b\r_{2},\b\r_{3})$ with orbital angular momentum zero
    and a spin-flavor-(color) factor $W_{B}(1,2,3)$ carrying the whole \textbf{J}).
     Then the mass $M_{B}$ results:
    \begin{equation}
     M_{B}= \sum_{\mu}\langle X_{L=0}(\b\r)\mid R_{\mu}(\b\r,\b\r^{'})\mid 
X_{L=0}(\b\r^{'})\rangle \langle W_{B}\mid Y_{\mu}(\bsigma,f)\mid
            W_{B}\rangle
    \end{equation}
    that is
    \begin{equation}
      M_{B} = \sum_\mu k_{\mu} \langle W_{B}\mid Y_{\mu}(\bsigma,f)\mid W_{B}\rangle \equiv 
\langle W_{B}\mid``parametrized\ mass"\mid W_{B}\rangle
    \end{equation}
     where the coefficients $k_{\mu}$ are:
    \begin{equation}
                     k_{\mu}= \langle{X_{L=0}(\mathbf{r}) \mid 
R_{\mu}(\mathbf{r},{\mathbf{r}}')\mid X_{L=0}({\mathbf{r}'})\rangle }
    \end{equation}
    Because the $X$'s in Eq.(18) have $L=0$, the operators $R_{\mu}(\b\r,\b\r^{'})$ must be 
rotation invariant; also, because
    $\tilde{H'}$ is a scalar, the $Y_{\mu}(\bsigma,f)$'s in Eq.(17) must be flavor-dependent 
scalar operators constructed with the
    spins $\bsigma_{i}$ 's of the three quarks. Eqs.(17)(18) give the most general form of 
the \textit{``parametrized masses"} of the
    lowest baryons \textbf{8} and \textbf{10}.\\
    \indent To conclude, we summarize the contents of this Section: The two main  QCD steps 
leading to the Eq.(19) of next Sect.3, that
     will give explicitly the``\textit{parametrized mass}" of the \textbf{8+10} baryons are: 
1) The masses are calculated as the expectation
    values of the exact $H_{QCD}$ Hamiltonian in the exact states $\vert \, \Psi_{i} 
\rangle$; 2) Such states $\vert \, \Psi_{i} \rangle$ are
    related by $\mid\Psi_{B}\rangle = V\mid\Phi_{B}\rangle$, [Eq.(5)], to a set of auxiliary 
(model) states $\vert \,\Phi_{i}\rangle$.

    \vskip 30pt
    \noindent {\bf 3. The GP baryon mass formulas.} \\

      We now write the general expression for the ``parametrized mass" of the octet and 
decuplet baryons. After integrating
    on the space coordinates, the ``parametrized mass"
    is a function only of the spin and flavor operators of the three quarks in the model 
state. Indicating by  $P^{s}$ the projector in the flavor
    space of the $s$ quark ($P_{i}^{s}s_{i}=s_{i},\, P^{s}_{i}u_{i}=P^{s}_{i}d_{i}=0$), that 
is $P^{s}_{i}\, \equiv diag(0,0,1)$, one has:
    \begin{eqnarray}&&
       ``parametrized\,\,mass"= M_{0} +B\sum_{i}P^{s}_{i} 
+C\sum_{i>k}(\bsigma_{i}\cdot\bsigma_{k})+                                                            
D\sum_{i>k}(\bsigma_{i}\cdot\bsigma_{k})(P^{s}_{i}+P^{s}_{k}) \nonumber\\
  &&+E\sum^{i>k}_{i\neq k\neq j}(\bsigma_{i}\cdot\bsigma_{k})P^{s}_{j} + a\sum_{i>k} 
P^{s}_{i}P^{s}_{k}+b\sum_{i>k}(\bsigma_{i}\cdot \bsigma_{k}) P^{s}_{i}P^{s}_{k} \nonumber\\
  &&+ c\sum^{i>k}_{i\neq k\neq j}(\bsigma_{i}\cdot\bsigma_{k})P^{s}_{j}(P^{s}_{i}+P^{s}_{k})+ 
d P^{s}_{1}P^{s}_{2}P^{s}_{3}
    \end{eqnarray}
    where $M_{0}, B, C,D,E,a,b,c,d$ are coefficients (or, as we call them, 
\textit{parameters}).\\
    \indent A few comments: Barring -at the moment- the e.m. and isospin corrections, the 
number of masses of the lowest octet
    and decuplet baryons is 8 and the parameters ($M_{0}, B, C, D, E, a, b, c, d$) in Eq.(19) 
are $9$; thus Eq.(19) can certainly be satisfied
    [in the expressions of the masses only $(a+b)$ intervenes].Also, spin scalars 
($\bsigma_{1}\times\bsigma_{2})\cdot \bsigma_{3}$
    cannot contribute in Eq.(19) as shown in Eq.(26) of Ref.\cite{mo89}; in fact they
    are absent in Eq.(19). Note also that $(P^{s}_{i})^{n}$=$(P^{s}_{i})$ for any $n$ (this 
implies that the GP
    records automatically, to all orders in $n$, the flavor breaking contributions additive 
in the quarks).\footnote{Explicit formulas relating
     the \textbf{8} and \textbf{10} baryon masses to the coefficients in Eq.(19) are given in 
Sect.XII of \cite{mo89} and
     in the Appendix B of \cite{dimo96}. We will transcribe them in part in the Appendix II 
here.}\\
    \indent The parametrization (19) holds for any magnitude of the flavor breaking term 
$\Delta \,m$ (which is included to all orders),
     and is independent of the choice of the quark mass renormalization point. Finally we 
stress the interest of the following fact.
      We will see by fitting the masses that in the Eq.(19)
      the magnitude of the parameters decreases strongly moving to terms with increasing 
number of indices; this is what we call the hierarchy of the parameters. \\
     \indent The parameters, in MeV, of Eq.(19) fitting the baryon masses, using
     the pole -1st line- (for discussions of the pole parameters see 
\cite{lichtdi74,di92,dimo96}) or the conventional -2nd line- values
     for the resonances, are:
    \begin{eqnarray}&&\cr
    M_{0}&&\,\, B\quad \,\quad C\quad\qquad\qquad D  \quad\quad\qquad (a+b)\qquad \quad E 
\qquad  \quad \qquad   c\qquad \qquad d \\
    1076&&   192,\,\, 45.6\pm 0.3,\,\, -13.8\pm 0.3,\,\,\,-16\pm 1.4,\,\,5.1\pm 0.3,\quad 
-1.1 \pm 0.7,\,\,\,\, 4\pm 3 \,\,\nonumber \\
    1086&&   184,\,\, 49.2\pm 0.3,\,\, -16.4\pm 0.2,\,\,\, -7.5\pm 0.8,\,2.5\pm 0.2,\quad 
\,\,  3.1\pm 0.4,\,\,\,-5.7\pm 2\nonumber
    \end{eqnarray}
    \indent The hierarchy of the parameters is evident; the values (20) decrease rather 
strongly  with increasing complexity
    of the accompanying spin-flavor structure, so that, neglecting $c$ and $d$ in Eq.(19), 
one obtains -see below- the mass formula Eq.(1)
    of Sect.1. Its agreement with the data (to 1/1000, Eq.(2)) confirms the smallness of 
$c,d$.\\
    \indent Although the subject of the hierarchy (and the question of the correct choice of 
normalization for the coefficients
    in an expression like Eq.(19)) will be discussed in Sect.7, a few remarks may be useful 
also here.\\
    \indent Both $C$ and $D$ multiply terms of type $\bsigma_{i}\cdot \bsigma_{k}$ but the 
value of $D$ should be depressed with respect
     to that of $C$
    because the $D$ terms includes a flavor breaking factor while the $C$ term does not. In 
fact it is $\mid D/C \mid\,\, \approx 1/3$.
    The $E$ term ($E\sum^{i>k}_{i\neq k\neq j}(\bsigma_{i}\cdot\bsigma_{k})P^{s}_{j}$) has a 
spin structure similar to that of the $D$ term
    but its flavor factor has a different index, so that three quarks are involved; this can 
be interpreted as the exchange of an
    additional gluon and produces a further reduction. Depending on the use of the 
conventional or pole values of the masses for the resonances,
    the additional gluon reduction factor for $E$ is $\approx 0.22$ (conventional) or 
$\approx 0.37$ (pole).\\
    \indent Another remark: \textit{Limiting to first order flavor breaking},
    the Eq.(19) simplifies and contains only 5 parameters instead of 8; thus it reduces to:
    \begin{eqnarray}&
    ``parametrized \, mass" \equiv M(1) = 
M_{0}+B\sum_{i}P^{s}_{i}+C\sum_{i>k}(\bsigma_{i}\cdot\bsigma_{k}) + \\
     &D\sum_{i>k}(\bsigma_{i}\cdot\bsigma_{k})(P^{s}_{i}+P^{s}_{k}) +E\sum^{i>k}_{i\neq k\neq 
j}(\bsigma_{i}\cdot \bsigma_{k})P^{s}_{j} \nonumber
    \end{eqnarray}
    With five parameters and eight masses, we get in this approximation, three mass 
relations: 1) The Gell Mann-Okubo formula for the octet ;
    2) The two equal spacing formulas of Gell Mann for the decuplet: $\Omega 
-\Sigma^{*}=\Xi^{*}-\Sigma^{*}=\Sigma^{*} -\Delta$. \\
    \indent To obtain the generalized Gell Mann-Okubo formula (1) (Sect.1)\cite{mopl92} we 
consider the 2nd order flavor breaking terms present in the
    general parametrization [Eq.(19)]; we neglect the $3$-quark term with coefficient $c$, 
which is very small (see Eq.(20)); this smallness is
     due to the hierarchy; in other words we keep only the 2nd order term $(a+b)$. Because it 
is easy to check that the combination of
     masses $T$ appearing
    in Eq.(1) is $T=-(a+b)/2$, we obtain the Eq.(1) and establish that it is correct to 2nd 
order in flavor breaking (except for the neglect of $c$).
    For seeing that the result is free from electromagnetic corrections compare the footnote 
9 of Ref.\cite{mopl92}.

    \vskip 30pt
     \noindent{\bf 4. The magnetic moments of the octet baryons to 1st order in flavor 
breaking.}\\

    To display other results of the GP we consider now the magnetic moments of the lowest 
octet baryons. In fact the first
    motivation of this work was to understand why the NRQM two parameters formula ($\mu$ is 
the proton magnetic moment= $2.793\,\mu_{N}$):
    \begin{equation}
    \mathbf{\mathcal{M}}=\mu \sum(\frac{2}{3}\bsigma^{\mathcal{P}} - 
\frac{1}{3}\bsigma^{\mathcal{N}} - \frac{1}{3}\bsigma^{\lambda})
     + (A/3)\sum\bsigma^{\lambda}
    \end{equation}
    worked so well, as shown in fig.1 (Sect.1). It is convenient, as done originally 
\cite{mo89}, to perform
    this calculation of the magnetic moment $\mathcal{M}$ in the rest frame of the baryon; we 
will use the standard formula
    (non covariant, but correct in the rest frame!):
    \begin{equation}
       \mathbf{\mathcal{M}} = (1/2) \int d^{3}\mathbf{r}\,\big (\mathbf{r} \times 
\mathbf{j}(\mathbf{r})\big)
    \end{equation}
    where $\mathbf{j}(\b\r)$ is the space part of the electromagnetic current $j_{\mu}(x)$ at 
time $t=0$:
    \begin{equation}
    j_{\mu}(x)=(ie)[\frac{2}{3}\bar{u}(x)\gamma_{\mu}u(x)-\\
    \frac{1}{3}\bar{d}(x)\gamma_{\mu}d(x)-\frac{1}{3}\bar{s}(x)\gamma_{\mu}s(x)]\\
    \equiv(ie)[\bar{\Psi}(x)(\lambda_{3} + \frac{1}{3}\\
    \lambda_{8})\gamma_{\mu}\Psi(x)]
    \end{equation}
    \indent Of course $e(1/2)(\lambda_{3} + \frac{1}{3}\lambda_{8})$ is the charge $eQ \equiv 
e[\frac{2}{3}P^{u}-\frac{1}{3}P^{d}-\frac{1}{3}P^{s}]$;
     the $P^{q}$'s are projectors on $u,d,s$. [Frequently $eQ_{i}$ will replace $eP^{q}_{i}$, 
the notation used in \cite{mo89}].\\
    \indent The magnetic moments of the octet baryons $B$ are \footnote{We use now for the 
magnetic moments the notation $M_{z}$
     instead of $\mathcal{M}$ - there should be no confusion with the masses.}:\\
    \begin{equation}
     M_{z}(B)= \mu\sum_{\nu}\langle X_{L=0}(\b\r)\mid R_{\mu}(\b\r,\b\r')\mid 
X_{L=0}(\b\r')\rangle \langle W_{B}\mid \bG_{\nu}(\bsigma,f)\mid W_{B}\rangle
    \end{equation}
    From Eq.(25) we obtain the ``parametrized magnetic moments" calling:
    \begin{equation}
    g_{\nu}={\mu}\langle X_{L=0}(\b\r)\mid R_{\nu}(\b\r,\b\r')\mid X_{L=0}(\b\r')\rangle
    \end{equation}
    and writing:
    \begin{equation}
    M_{z}(B)=\sum_{\nu}g_{\nu}\langle W_{B}\mid\bG_{\nu}(\bsigma,f)\mid W_{B}\rangle \equiv 
\langle W_{B}\mid``parametr.\, magn.\, mom."\mid W_{B}\rangle
    \end{equation}
    Here we write the GP of the magnetic moments only to first order in flavor breaking.
    (For the baryon masses the parametrization was exact to all orders in $P^{s}$).
    Keeping only terms linear in $P^{s}$ the GP of the magnetic moments of the baryon octet, 
is for each baryon $B$,
     a linear combination of $7$ terms. It is:\\
    \begin{equation}
    M_{z}(B)=\langle W_{B} \vert \sum_{\nu =0}^7 g_\nu ({\bf G}_\nu)_z \vert
    W_{B}\rangle
    \equiv \langle W_{B}\vert \sum_{\nu =1}^7 \tilde g_\nu ({\bf G}_\nu)_z
    \vert W_{B}\rangle
    \label{moct}
    \end{equation}
    We will explain in a moment why the two expressions in Eq.(28) - the first containing 
eight terms and
    the second seven - are identical. In Eq.(28) the $W_{B}$'s ($\equiv W_{B}(1,2,3)$) are, 
of course,
    the same defined in Sect.3. As to the ${\bf G}_\nu$'s, they are:
    \begin{equation}
    \begin{array}{lcl}
    \label{Ginu}
{\bf G}_0=Tr[QP^s]\sum_i\bsigma_i \quad{\bf G}_1=\sum_i Q_i\bsigma_i \quad{\bf G}_2=\sum_i 
Q_i P_i^s\bsigma_i \quad{\bf G}_3=\sum_{i\ne k} Q_i\bsigma_k\ \\
{\bf G}_4=\sum_{i\ne k} Q_i P_i^s \bsigma_k \quad \,{\bf G}_5=\sum_{i\ne k} Q_k P_i^s 
\bsigma_i \quad \,{\bf G}_6=\sum_{i\ne k} Q_i P_k^s \bsigma_i \\
{\bf G}_7 =\sum_{i\ne j\ne k} Q_i P_j^s \bsigma_k
    \end{array}
    \end{equation}
    \indent As we will see, the coefficient $g_{0}$ of $\bG_{0}$ is expected to be 
significantly smaller than those of the other terms;
    therefore here the effect of the term $\bG_{0}$ is unimportant.\\
    \indent Forgetting $g_{0}$ one might then obtain the seven coefficients ($g_{1}$ to 
$g_{7}$) from the seven magnetic moments;
    but it is unnecessary
    to discuss now how small is $g_{0}$, because the relation between the eight terms, 
displayed below in Eq.(30), allows to express
    all the magnetic moments of the octet
    baryons in terms of only 7 parameters $\tilde{g}_{i}$ , with i=1 to 7.
    The sums over $i,j,k$ in Eq.(29) extend from $1$ to $3$ and
    $Q_i$ is the quark charge. Although
    we wrote $M_{z}(B)$ (\ref{moct}) in two forms, the two coincide, due to the relation 
(\ref{rel}), easily verifiable (compare the proof
    at the end of this section).\\
    \begin{equation}
    {\bG}_0= -\frac{1}{3}{\bG}_1+ \frac{2}{3}{\bG}_2 - \frac{5}{6}{\bG}_3+
    \frac{5}{3} {\bG}_4+\frac{1}{6}{\bG}_5+\frac{1}{6}{\bG}_6+\frac{2}{3}{\bG}_7
    \label{rel}
    \end{equation}
    The Eq.(30) leads to the following relations between the $\tilde g_\nu$'s in
   the right hand side of Eq.(28) and the $g_\nu$'s in the middle expression:
   \begin{equation}
    \begin{array}{lclcl}
   \label{relg}
   \tilde g_1=g_1-(1/3)g_0 &;&\tilde g_2=g_2+(2/3)g_0&;&\tilde
    g_3=g_3-(5/6)g_0 \\
   \tilde g_4=g_4+(5/3)g_0 &;&\tilde g_5=g_5+(1/6)g_0 &;&\tilde
    g_6=g_6+(1/6)g_0 \\
   \tilde g_7=g_7+(2/3)g_0
   \end{array}
   \end{equation}
 \indent The magnetic moments of the octet baryons expressed via
Eqs.(28,29) in terms of the $\tilde g_\nu$ 's are given by the following expressions (now the
baryon symbol indicates its magnetic moment).
\begin{equation}
\begin{array}{c}
\label{mfg}
p=\tilde g_1\\
n=-(2/3)(\tilde g_1-\tilde g_3)\\
\Lambda =-(1/3)(\tilde g_1-\tilde g_3+\tilde g_2-\tilde g_5) \\
\Sigma^+=\tilde g_1+(1/9)(\tilde g_2-4\tilde g_4-4\tilde g_5+8\tilde
g_6+8\tilde g_7)\\
\Sigma^-=-(1/3)(\tilde g_1+2\tilde g_3)+(1/9)(\tilde g_2-4\tilde
g_4+2\tilde g_5-4\tilde g_6-
4\tilde g_7)\\
\Xi^0=-(2/3)(\tilde g_1-\tilde g_3)+(1/9)(-4\tilde g_2-2\tilde
g_4+4\tilde g_5-8\tilde g_6+10\tilde
g_7)\\
\Xi^-=-(1/3)(\tilde g_1+2\tilde g_3)+(1/9)(-4\tilde g_2-2\tilde
g_4-8\tilde g_5-2\tilde g_6-
2\tilde g_7)\\
(\Sigma\Lambda) = -(1/\sqrt{3})(\tilde g_1 -\tilde g_3 + \tilde g_6 -\tilde g_7)
\end{array}
\end{equation}
From the first seven Eqs. one obtains
\begin{equation}
\begin{array}{lclclcl}
\label{g1} \tilde g_1=2.793 &;&\tilde g_2=-0.934 &;& \tilde
g_3=-0.076 &;&
\tilde g_4=0.438 \\
\tilde g_5=0.097 &;& \tilde g_6=-0.147 &;& \tilde g_7=0.154 &&
\end{array}
\end{equation}
If written with the $g_\nu$'s rather than with the $\tilde
g_\nu$'s, using (31), the formulas (32) are
changed all in the same way, by the addition of $(-1/3)g_{0}$ to
the r.h.s. of all expressions (e.g.
$n=-(1/3)g_0-(2/3)(g_1-g_3)$, etc.). The last Eq. $(\Sigma\Lambda)$ stays
unchanged. From it one gets $\mu(\Sigma\Lambda) = -1.48\pm 0.04$; the experimental
value is $-1.61\pm 0.08$; errors are still large.\\ \indent We underline that the structure 
of the ${\bf G}_\nu$'s
is similar to that of a NRQM, but the expression (\ref{moct}) for $M_z(B)$ is an exact
consequence of full QCD (to first order in flavor breaking).
We emphasized this often \cite{mo89},\cite{dimo96}, but we repeat it to avoid misinterpreting 
the
Eqs.(28,29) as a sort of generalized NRQM. In fact the results of the GP are
exact consequences of QCD; they include all effects of virtual $q\bar q$
pairs and gluons, as well as those of configuration mixing and the relativistic ones. \\
\indent Note the relative dominance of
$\tilde g_1$ and $\tilde g_2$ (related to the $p$ and $\Lambda$) in the sum in 
Eq.(\ref{moct})
(see the numbers in the Eq.(30)); this \textit{explains} why the fit of the naive NRQM:
\begin{equation}
M_z(B)=\langle W_B \vert \tilde g_1({\bf G_1})_z + \tilde g_2({\bf
G_2})_z \vert W_B \rangle
\end{equation}
is fairly good.\footnote{Interrupting briefly this comparison, we recall
that, as well known, \cite{oku} the $SU_{3}$ expression of the e.m current, to 1st order
in flavor breaking, has just seven parameters. How does this compare with the results here?
The point is that, as the derivation of the GP shows, each term in it
has a definite dynamical meaning (corresponding to a certain class of Feynman diagrams) that 
produces
a hierarchy of the parameters, with some parameter expected to be larger than others.
Instead the dynamics plays no role in the $SU_{3}$ treatment, that only displays (group 
theoretically)
a list of the possible terms.}
From Eq.(33) we can now explore the details of the hierarchy (the main parameters, related to 
$p$ and $\Lambda$ have been already mentioned):
\textit{The average value of the reduction factor resulting from the presence of two 
different indices in a sum (we call it sometimes also the
      one gluon exchange reduction factor) derived from the values of
      $\vert\tilde g_{6}\vert, \vert\tilde g_{5}\vert, \vert\tilde g_{4} \vert$} [the 
parameters multiplying terms
      with two different indices, except ${\bf G}_3$ and ${\bf G}_7$ -see below]
      is $0.25$, having adopted $0.33$ for the flavor reduction factor derived from 
$\vert\tilde g_{2}\vert$ and $\vert\tilde g_{1}\vert$.
      We will proceed using $0.33$ for both the flavor and the ``one gluon exchange" 
reduction factors.
     The maximum discrepancy between the estimated and empirical values is 2.5 for the 
$\vert\tilde g_{\nu}\vert$'s with
     $\nu=4,5,6$. Above we did not consider $\tilde g_{7}$ for the reasons explained in 
Ref.\cite{mo92}.\\
     \indent  The above values seem consistent, but a serious exception is $\vert\tilde g_{3} 
\vert \simeq 0.08$.
     This is much too small: One expects from the hierarchy $2.79 \cdot 0.33 \simeq 0.92$, an 
order of magnitude larger.
     In Sect.5 we will indicate a possible solution to this question.\\
     \indent In extracting the hierarchy from the orders of
     magnitude of the $\tilde{g}_{\nu}$ 's, we neglected, in the equations above, the 
presence of $g_{0}$; the reason, we repeat, is that
     its effect on the order of magnitude of the parameters discussed above is expected to be 
rather small. We will come back on this in Sect.7
      but here we give the order of magnitude of $g_{0}$.\\
    \indent A first estimate of $g_{0}$ is obtained from the recent measurements of the 
$\overline{s}s$ contribution to the
     magnetic moment of the proton. Essentially the question is: How much do the 
$\overline{s}s$ pairs
     inside the proton contribute to its magnetic moment? One can answer to this question 
with the GP, calculating
      \cite{dimo07}  the magnetic moment of the proton to \textit{all orders in flavor 
breaking}. One can thus show that
       the contribution of the $s\overline{s}$ to the proton magnetic moment is related to 
the value of $g_{0}$.
       One obtains $\vert g_{0}Tr[QP^{s}]\vert$ in the interval $1/10 \div 1/30$. (Another 
determination of a Trace term -similar order of magnitude-
        will be obtained, using data from the ratio between the $\rho-\gamma$ and 
$\omega-\gamma$ couplings in Sect.12).
       Incidentally, these values are one order of magnitude larger than our previous 
estimates; but they are still small enough
        to confirm the fact that the \textit{Trace terms} were indeed negligible when we 
neglected them.
        We add that the  $g_{0}$ given above leads to a very small (non measurable) 
contribution of the $s\overline{s}$ pairs
      to the proton magnetic moment at small values of $q^{2}$. This agrees with the 
experiments
       \cite{acha} and also with a lattice plus chiral evaluation \cite{australia}.\\
     \indent \textbf{Proof of Eq.(30)}. Define:\\
    $\sum_{q}\equiv 
Q_{i}\bsigma_{i,z},\qquad\qquad\,\sum_{s}\equiv(1/3)\sum_{i}P^{s}_{i}\bsigma_{i,z}$\\
    and consider the expectation values of the above $z$ components
    for all octet baryons. One can check that the equation:\\
    $(1+S)\sum_{q} - 3Q\sum_{s}=(-2/3) + (5/3)Q -(2/3)S +(4/3)QS$\\
   (where $Q$ is the
    baryon charge and $S$ the baryon strangeness) holds for all the octet baryons (compare 
the table I in Ref.\cite{mo89},p.3004). Observe that the $z$
    component of $\bG_{7}$ given in Eq.(29) can be written:
    $\,G_{7,z} = -S(Q+(1/3))-[3Q+2]\sum_{s}+S\sum_{q}$. Simplifying in a similar way all the 
other $\bG_{\nu}$'s (with $\nu$ from
     $0$ to $6$) one obtains the Eq.(30). To check this we list the expressions of the 
$G_{i,z}$ with i=1 to 6:     $G_{1,z}=\sum_{q};\quad
     G_{2,z}= -\sum_{s};\\ G_{3,z}= Q-\sum_{q};\quad G_{4,z}= -S+3\sum_{s};\quad G_{5,z}= 
-(3Q+1)\sum_{s};\quad G_{6,z}= -S\sum_{q}+\sum_{s}$.

    \vskip 30pt

    \noindent {\bf 5. The $\Delta \rightarrow p+\gamma$ decay, the magnetic moments of the 
$\Delta$'s and related results.}\\

    We now apply the GP to the subjects indicated in the title. As a byproduct we will obtain 
a possible explanation of why the ratio between the
    proton and neutron magnetic moments is so near to $-3/2$ (that is $\tilde g_{3}$ is so 
small), as noted in Sect.4. The contents
    of this section is again related to the hierarchy. Possibly it would have been better to 
discuss the hierarchy in general at this stage.
    We preferred to continue here with problems related to the magnetic moments; but one can 
switch now
     to Sect.7 on the hierarchy and come back.\\
    \indent For the $\Delta \rightarrow p+\gamma$ decay $M1$ transition see Ref.\cite{ds66} 
to which we refer also for details on the
    transition form factor. (For the E2 contribution to the transition see 
Ref.\cite{bemo65}). The GP treatment of the $M1$ transition is
     developed in Ref.\cite{mo89}; the formula of \cite{mo89}
     to calculate both the diagonal magnetic moments of the $\Delta$'s and the transition 
matrix elements $\Delta\rightarrow Nucleon +\gamma$,
     is Eq.(62), in Sect.9.\footnote{To avoid confusion we will call that Equation: 
($62_{1}$). Also we correct here again
     some misprints in the Eqs.$(63, 64, 66)_{1}$. In Eq.($63_{1}$)-third line- replace, in 
the square brackets,
      $\delta-\beta -2\gamma$ with $\delta -\beta$; in Eq.($64_{1}$) $\delta -\beta+2\gamma$ 
should be $\delta -\beta -4\gamma$;
     in Eq.($66_{1}$) write $F=\delta-\beta -4\gamma$.}
     Note that Eq.($62_{1}$) is obtained omitting the $\eta$ term in Eq.($61_{1}$); this is 
possible because the $\eta$ term
     just takes into account (Fermi-Watson theorem-compare, for this application, 
Ref.\cite{dich94}) the effect of the final state interaction
     in the matrix element $\Delta\rightarrow n\pi$. The omission of $\eta$ amounts to 
analyze the data after the extraction of such effects.
     Moreover Eq.($62_{1}$) does not include a term $\approx \bJ\cdot Tr[QP^{s}]$, 
negligible and not contributing to the
     $\Delta \rightarrow p+\gamma$ decay; thus we continue to ignore below the difference 
between the $g$'s and the $\tilde{g}$'s.\\
    \indent Using our present notation, the Eq.($62_{1}$) of \cite{mo89} can be rewritten:
    \begin{equation}
    \bM = \sum_{perm}\big[\alpha Q_{1} +\delta(Q_{2}+Q_{3})\bsigma_{1}\big] + \big[\beta 
Q_{1} +\gamma (Q_{2}+Q_{3})\big]\bsigma_{1}
    (\bsigma_{2}\cdot\bsigma_{3})
    \end{equation}
    \indent The sum over the perm(utations) in Eq.(35) means that to the term (123) displayed 
one has to add the terms (321) and
    (231); in Eq.(35) $\alpha,\beta,\delta,\gamma$ are four real parameters (the same as 
those of the Eq.($62_{1}$));
    an estimate of their magnitude can now be obtained from the hierarchy:
    \begin{equation}
     \vert \delta/\alpha \vert \simeq 0.33;\qquad\qquad  \vert\beta/\alpha\vert \simeq 
\vert\gamma/\alpha\vert \simeq (0.33)^{2} \approx 0.11
    \end{equation}
    As shown in \cite{mo89} the Eq.($62_{1}$) can be rewritten (recall that 
$\bG_{1}=\sum_{i}Q_{i}\bsigma_{i}$) :
    \begin{eqnarray}&&
    \bM=(\alpha-\delta)\bG_{1} +(\beta -\gamma)\Big[(1/4)(4\vert \bJ\vert^{2}-7)\bG_{1}+ 
(1/4)\bG_{1}(4\vert \bJ\vert^{2}-7)\Big]\nonumber\\&&+
    \Big[\delta -\beta -2\gamma +(1/2)\gamma(4\vert \bJ\vert^{2}-7)\Big]Q_{B}\cdot(2\bJ)
    \end{eqnarray}
    The Eq.(37) can also be used to calculate to higher orders in the hierarchy \cite{dimo03}
     the magnetic moments $p,n$ of proton and neutron; it gives (taking the expectation 
value):
     \begin{equation}
     p=\alpha -3\beta -2\gamma, \qquad\qquad\qquad n= -(2/3)(\alpha -\delta -2\beta +2\gamma)
     \end{equation}
     that is:
     \begin{equation}
     g_{1}=\alpha -3\beta- 2\gamma,  \qquad\qquad\qquad g_{3}= \delta - \beta -4\gamma
     \end{equation}
     \indent One can obtain $p/n\simeq-3/2$ if $\delta \simeq \beta +4\gamma$; it is 
interesting to note that the hierarchy
     (does not -of course- prescribe, but) allows
     this. Indeed $\delta$ multiplies a term with two indices while $\beta$ and $\gamma$ both 
multiply terms with three indices.
     If, indeed, $\delta \simeq \beta +4\gamma$ (it appears the only possible explanation), 
the fact that $p/n$ deviates so little from -(3/2)
     would be due to sheer chance. This is a case of historical interest because in
     1965 the successful prediction \cite{beg64},\cite{mo65}
     $\mu(p)/\mu(n)=-3/2$ was \textit{most important for the acceptance of the quark 
description}. That prediction could
      not have been made if this chance cancellation (due to $\delta \approx \beta + 
4\gamma$) had not occurred. \footnote{Also
      Leinweber et al. \cite{lein01} attribute to chance
      the smallness of the deviation of $\mu(p)/\mu(n)$ from $=-3/2$, as we had (previously) 
noted \cite{mo92}. However -though this is not relevant
      in our treatment- we do not share
      their claim that this conclusion is not possible in a constituent quark description. 
See Ref.\cite{moarXiv01}.}\\
      \indent We add now a few remarks on the $\Delta\rightarrow p+\gamma$ transition and on 
the magnetic moments of the $\Delta$'s.
     From Eq.(37) the matrix element $\mu(\Delta\rightarrow p\gamma)$ is:
     \begin{equation}
     \mu(\Delta\rightarrow p\gamma)=(2/3)\sqrt{2}(\alpha -\delta+ \beta -\gamma)
     \end{equation}
     The approximate equation just discussed $\delta \simeq \beta +4\gamma$ gives:
      \begin{equation}
     \mu(\Delta\rightarrow p\gamma)= (2/3)\sqrt{2}\mu(p)[1 +3(\beta-\gamma)/\mu(p)]
     \end{equation}
     If $\beta$ and $\gamma$ have equal values and opposite signs (with $\gamma<0$ as 
suggested by the previous relation
     $\delta \simeq \beta + 4\gamma$ and by the hierarchy),
     using $\vert\gamma/\alpha \vert \approx 0.11$ one gets:
     \begin{equation}
     \mu(\Delta\rightarrow p\gamma)\approx 1.7\cdot 2/3\sqrt{2}\mu(p)[1 
+3(\beta-\gamma)/\mu(p)]
     \end{equation}
     One more point. From the Eq.(37) one can also express in terms of 
$\alpha,\delta,\beta,\gamma$ the magnetic moments
     $\mu(\Delta)$ of the $\Delta$'s. It is:
     \begin{equation}
     \mu(\Delta^{Q})=(\alpha+2\delta +\beta +2\gamma)Q_{\Delta}= [\mu(p)+2\delta +4\beta 
+4\gamma]Q_{\Delta}
     \end{equation}
     The Eq.(43) (with the estimate of $\delta,\beta,\gamma$ given in Ref.\cite{dimo03}, 
Sect.VI) shows that the magnetic moment
      $\mu_{\Delta^{+}}$ of the singly charged $\Delta^{+}$
     is expected to be appreciably smaller than $\mu_{p}$, but the error in the estimate is 
still large.
     Note that it is quite generally $\mu(\Delta^{Q})=kQ +\xi$  with $\xi$ very small 
(Ref.\cite{dimoZSI})
     because $\xi$ is the coefficient of a Trace term (ignored in this section), depressed 
due to the need of exchanging several gluons.
     Although the estimate of $\xi$ in Ref.\cite{dimoZSI} was \textit{much} too small,
      the $\xi$ term remains negligible with respect to $kQ$ (for $Q\neq 0$!).
      In practice the magnetic moments of the $\Delta$'s are expected to be 
quasi-proportional to their charges
     (thus $\mu_{\Delta^{0}}\simeq 0$). It is most difficult (perhaps impossible), but it 
would be most
     interesting to have a measurement of $\mu_{\Delta^{0}}$; its deviation from zero would 
give directly the order of magnitude (for this case) of the coefficient of $Tr(QP^{s})$.

     \vskip 30 pt
      \noindent {\bf 6. Double counting on inserting explicit pion fields in the QCD 
Lagrangian.}\\

     We now digress to a problem, marginal with respect to our main line, but important to 
clarify
     some treatments appearing frequently in the literature. In fact any calculation based on 
a quark-gluon Lagrangian
     \textit{to which pion fields are added as explicit degrees of freedom} leads to terms 
duplicating those coming from the standard QCD
     Lagrangian. This can be shown by the GP; the doubly counted terms can be displayed 
explicitly.\\
     \indent We do this for the baryon magnetic moments but the argument holds generally. It 
follows that, if some pion exchange contribution is assumed,
      it is impossible in any way to extract univocally its amount. \\
     \indent In fact, one often introduces, in addition to the QCD Lagrangian, an explicit 
$\bar{q}q\pi$ coupling, say
     $\pi_{i}(\bar{q}(x)\gamma_{5}\lambda_{i}q(x))$, where $q(x)$ is the
     quark field. That is, in such treatments (e.g. Manohar and Georgi \cite{mange84}, 
Krivoruchenko \cite{krivo87})
     the pions appear in the Hamiltonian in
     addition to the quark and gluon fields of QCD; the same occurs phenomenologically in
     various bag models [Ref.\cite{brv80}]. In such treatments $\lambda_{1},\lambda_{2}$ (or, 
in $SU(2)$, $\tau_{x},\tau_{y}$) should enter
     in the calculation of some hadronic properties, in particular of the magnetic moments. 
For instance, in \cite{krivo87}
     the magnetic moments of proton and neutron contain, due to pion exchange, the spin 
flavor term:
     \begin{equation}
     \sum_{i\neq k}(\bsigma_{i}\times\bsigma_{k})(\btau_{i}\times\btau_{k})
     \end{equation}
    But the QCD Lagrangian (on which, of course, the GP is based) contains
    - also including electromagnetism and the flavor breaking mass
    term - only the flavor matrices $\lambda_{3}$ and $\lambda_{8}$.
    They commute and have a closed algebra. Thus from a QCD
    calculation, where pions intervene indirectly as $q\bar{q}$ aggregates, one
    should not get in the final result any $\lambda_{1}$ or $\lambda_{2}$; this \textit{seems 
to be} in
     contrast with Eq.(44).\\
     \indent Imagine, as an example, to calculate in pure QCD the
    magnetic moments, using Feynman diagrams. No matter how
    complicated is the calculation, no flavor $\lambda_{1},\lambda_{2}$
    should appear in the final result, because no $\lambda_{1}$ or
    $\lambda_{2}$ can arise from $\lambda_{3}$ and $\lambda_{8}$. Then how
    can the result (this is the \textit{apparent} ``paradox") contain $\tau_{x}, \tau_{y}$'s, that is
   $\lambda_{1},\lambda_{2}$?\\
  \indent Now the answer: As we will show, \textit{in contradiction
  with the above argument}, the term (44) is not logically in contrast with a pure QCD 
calculation [the detailed treatment is
  presented in Ref.\cite{dimoZS97}]. We will show that the term (44) {\em can} indeed be 
present because it can be
  identically rewritten as a sum of terms \textit{not containing} at all
  $\tau_{x}, \tau_{y}$. In short, as we will see, adding a term like (44) to a QCD
  calculation, is not forbidden. But it gives rise to a disturbing case of \textit{double 
counting}.
   We will show in what follows that this double counting
  is unavoidable in Lagrangians containing explicitly pions in addition to
  quarks and gluons, unless a rule (that, however, would amount to have
  solved QCD) is given to subtract from QCD some definite quark-gluon diagrams.\\
  \indent Let us first examine how the fact that Eq.(44) contains
  $\lambda_{1}$ and $\lambda_{2}$ (while only  $\lambda_{3}$ and $\lambda_{8}$ appear in QCD) 
is not, in principle,
  in contrast with QCD.\\
  \indent This is fairly simple (compare, for more details, the Sect.3 of \cite{dimoZS97}). 
Call $P_{x}^{ik}$ the Majorana exchange operator
  of the \textit{space coordinates of quarks $i,k$}. Let us, for example, consider the 
magnetic moments of $p,n$ (compare Sect.4) and refer
   to their calculation, described by the Eq.(25)(Sect.4).\\
  \indent In a QCD calculation it is possible that the Majorana $P_{x}^{ik}$  appears in the 
$R_{\nu}(\b\r,\b\r')$ factor
  of Eq.(25), producing terms of the type
  $\sum_{i\neq k}Q_{i}\bsigma_{i}P_{x}^{ik}$. We may calculate these terms in Eq.(25). First, 
let $P_{x}^{ik}$
  act on $X_{L=0}$. Because $X_{L=0}(\b\r_{1},\b\r_{2},\b\r_{3})$ is symmetric in all pairs 
$i,k$, the operation $P_{x}^{ik}$
  does not alter the result of the calculation of the parameters $g_{\mu}$.\\
  \indent Second, always in $SU(2)$ [we will extend to $SU(3)$ in a moment], consider the 
operation of $P_{x}^{ik}$ on the $W_{B}(1,2,3)$ factor.
   Write $P_{x}^{ik} = (1+\bsigma_{i}\cdot \bsigma_{k})(1+\btau_{i}\cdot \btau_{k})/4$,
  and, in Eq.(25), let $P_{x}^{ik}$ operate on the $W_{B}$ factor using the relation:
  \begin{equation}
  \sum_{i\neq k}Q_{i}\bsigma_{i}P_{x}^{ik}= \sum_{i\neq k}Q_{i}\bsigma_{i}(1+\bsigma_{i}\cdot 
\bsigma_{k})(1+\btau_{i}\cdot \btau_{k})/4
  \end{equation}
  with $Q_{i}=(1/2)\tau_{zi}+1/6$. Using the identities $\tau_{zi}(\btau_{i}\cdot\btau_{k})= 
\tau_{zk}-i(\btau_{i}\times\btau_{k})_{z}$
  and 
$\sigma_{zi}(\bsigma_{i}\cdot\bsigma_{k})=\sigma_{zk}-i(\bsigma_{i}\times\bsigma_{k})_{z}$, 
Eq.(45) becomes:
  \begin{equation}
  4\sum_{i\neq k}Q_{i}\sigma_{zi}P_{x}^{ik}=\sum_{i\neq k}\Big[(Q_{i}+Q_{k}) 
+(1/6)\big[(\btau_{i}\cdot\btau_{k})- 1]\Big]
  (\bsigma_{i}+\bsigma_{k})_{z}-(1/2)(\btau_{i}\times\btau_{k})_{z}(\bsigma_{i}\times
\bsigma_{k})_{z}
  \end{equation}
  In the steps leading to (46) we omitted the terms not contributing to the expectation 
values on real function $W_{i}$ (see Ref.\cite{mo89},sect.V).\\
  \indent Because it is $\sum_{i\neq 
k}[(\btau_{i}\cdot\btau_{k})-1](\bsigma_{i}+\bsigma_{k})_{z}=0$ for $p,n$ states,
  using, on the left hand side $P_{x}^{ik}X_{L=0}=X_{L=0}$, we obtain the identity, for 
$p,n$: \\
  \begin{equation}
   \sum_{i\neq k}(\btau_{i}\times\btau_{k})_{z}(\bsigma_{i}\times\bsigma_{k}) = 
-8\sum_{i}Q_{i}\bsigma_{i}
  +4\sum_{i\neq k}Q_{i}\bsigma_{k} = -8\bG_{1}+4\bG_{3}
  \end{equation}
  \indent  The above equation (47) shows that the expression $\sum_{i\neq 
k}(\bsigma_{i}\times\bsigma_{k})(\btau_{i}\times\btau_{k})$
  is already contained in the
  terms with coefficients $g_{1}$ and $g_{3}$ in the GP of the magnetic moments; that is the 
contribution from the pion
  term is indistinguishable and cannot be separated from the contributions of type $\bG_{1}$ 
and $\bG_{3}$ that already come from the quark gluon dynamics of QCD
  independently of the pion degrees of freedom. Adding (44) to $\bG_{1}$ and $\bG_{3}$ 
amounts to write twice the same terms (double counting).
  This argument is easily extended from $SU(2)$
  to $SU(3)$, that is from terms of type $\sum_{i\neq 
k}(\bsigma_{i}\times\bsigma_{k})(\btau_{i}\times\btau_{k})$
  to those of type $\sum_{i\neq k}(\bsigma_{i}\times\bsigma_{k})(\blambda_{i}\times 
\blambda_{k})$; obviously, if one keeps pion exchange,
  but excludes kaon exchange (as implied in Ref.\cite{mange84}), one should insert the 
appropriate projectors, that is one
  should consider the term $\sum_{i\neq k}(\bsigma_{i}\times\bsigma_{k})(\blambda_{i}\times 
\blambda_{k})(1-P^{s}_{i})(1-P^{s}_{k})$.
  For the detailed calculation consult the Ref.\cite{dimoZS97}. The result is:
  \begin{equation}
  \sum_{i\neq k}(\bsigma_{i}\times\bsigma_{k})(\blambda_{i}\times 
\blambda_{k})_{3}(1-P^{s}_{i})(1-P^{s}_{k}) = -8\bG_{1}+8\bG_{2}+4\bG_{6}
  +4\bG_{3}-4\bG_{5}-4\bG_{4}
  \end{equation}
  Once more we see that pion exchange is already included in the GP of QCD; in other words
  it is conceptually impossible to disentangle the pion term from the quark-gluon 
contribution. Theories that obtain
  agreement with experiment using this trick are useless. That is, if we do not know, a 
priori, the
  contribution of the pion exchange, we cannot separate it from terms due only to gluons. If 
we somehow
   assume to know the contribution of pion exchange,
   then we should be able to solve QCD so as to subtract from gluon exchange that part 
already accounted
   by pion exchange. But, then the introduction of pion degrees of freedom,
    even if quasi Goldstone \cite{mange84}, looks unnecessary.\\
   \indent Note finally the following: The whole argument is based on the equivalence of the 
Majorana
   exchange operator, that can occur in a pure quark-gluon QCD, to terms typical of pion 
exchange.
   This it is not limited to the magnetic moments; we used them here only as a convenient 
example.\\
   \indent Although the above calculations, using explicitly the Majorana exchange operator, 
show directly
   that it is meaningless to insert in a QCD Hamiltonian a pion-quark interaction,
   this result is, in fact, already contained in the simple argument at the beginning of this 
Section:
    The fact that only the flavor matrices
   $\lambda_{8}$ and $\lambda_{3}$ appear in the strong plus electromagnetic QCD Hamiltonian 
implies, since
   $\lambda_{8}$,$\lambda_{3}$ form a closed algebra, that in the exact expression of any 
quantity derived from the QCD
   Hamiltonian, no other flavor operator can be present. That is any effective Lagrangian 
depending explicitly
   on the pion field and thus on the $\lambda_{1},\lambda_{2}$ flavor matrices cannot 
reproduce the results of the original
   Lagrangian. This remark applies also to a statement
   on the Skyrmion model (considered by Gross -see \cite{mopt87}- as a step that contributed 
to clarify the success
   of the NRQM) and applies also, more importantly, to the effective Lagrangian written by 
Manohar and Georgi \cite{mange84} to explain the NRQM
   with chiral quarks, treating the pseudoscalar lowest nonet as quasi-Goldstone bosons and 
the other mesons as
   $\bar{q}q$ aggregates. A few words should be added on Ref.\cite{mange84}, the title of 
which is
   ``Chiral quarks and the non relativistic quark model". In writing a quasi-chiral 
Lagrangian of quarks and gluons plus pions, the latter
   seen as quasi-Goldstone bosons, the authors of Ref.\cite{mange84} implicitly assumed, of 
course, that their Lagrangian was equivalent, for
   the description of hadrons, to that of QCD. In fact their conclusion is that their 
effective Lagrangian explains the good results of the NRQM.
   We find this conclusion unwarranted for more than one reason. First the Authors should 
have been be able to prove the mathematical
   equivalence of their Lagrangian to that of QCD, at least for a set of low energy problems. 
This equivalence is often taken for granted but this
   is unjustified. The simultaneous presence of quarks and pions in their Lagrangian, leading 
to an unavoidable double counting
   in their theory, as shown above, duplicates in an uncontrolled way contributions that are 
obtained with quarks and gluons alone,
    creating serious consistency problems. We mentioned specifically this point because one 
often finds, as already stated,
    that \textit{the way} to understand the success of the NRQM is to start from a quasi 
chiral effective Lagrangian. Of course by leaving
    free in the quasi chiral effective Lagrangian a sufficient number of appropriately chosen 
parameters (as done in Ref.\cite{krivo87}) it is possible
    that some combination of them can be related to the parameters appearing in the GP. But, 
\textit{quite generally}, our conclusion \cite{dimoZS97}
    is that a Lagrangian with explicit pions cannot reproduce QCD.

   \vskip 30pt
   \noindent{\bf 7. The hierarchy of the parameters.}\\

   In Sections 3,4,5 we exemplified (for the baryon masses and magnetic moments) how the data 
indicate a ``hierarchy" of the GP parameters.
   At this stage, before examining other cases, we should give explicitly the rules for 
writing
   the expressions of the various quantities (masses, magnetic moments, etc.)  so that the 
parameters are defined uniquely.
    To exemplify, if some term in the parametrization (symmetric in the indices $i,k$) is 
written as $\sum_{i,k}F_{i,k}$, or if the same
   term is rewritten as $\sum_{i,k}^{i>k}F_{i,k}$, the number of addenda differs by a factor 
2 and the value of the parameter multiplying this term
   changes accordingly.\\
   \indent The masses of $\bf{8}+\bf{10}$ provide again an example to establish the rules.
   We show this here, rewriting, for convenience, the Eq.(19) of Sect.3 for the 
$\bf{8}+\bf{10}$ baryon masses:
   \begin{eqnarray}&&
   ``parametrized \ mass" = M_{0} 
+B\sum_{i}P^{s}_{i}+C\sum_{i>k}(\bsigma_{i}\cdot\bsigma_{k})+ 
D\sum_{i>k}(\bsigma_{i}\cdot\bsigma_{k})(P^{s}_{i}+P^{s}_{k}) \nonumber\\
   &&+E\sum^{i>k}_{i\neq k\neq j}(\bsigma_{i}\cdot\bsigma_{k})P^{s}_{j} + a\sum_{i>k} 
P^{s}_{i}P^{s}_{k}+b\sum_{i>k}(\bsigma_{i}\cdot \bsigma_{k}) P^{s}_{i}P^{s}_{k} \nonumber\\
   &&+ c\sum^{i>k}_{i \ne k \ne 
j}(\bsigma_{i}\cdot\bsigma_{k})P^{s}_{j}(P^{s}_{i}+P^{s}_{k})+ d P^{s}_{1}P^{s}_{2}P^{s}_{3} 
\nonumber
    \end{eqnarray}
  \indent In Eq.(19) -reproduced above- all sums are written so that
  each different addendum in a sum appears once and only once. For instance the $E$ term is 
explicitly
  \begin{equation}
E[(\bsigma_{1}\cdot\bsigma_{2})P^{s}_{3}+(\bsigma_{3}\cdot\bsigma_{1})P^{s}_{2}+(\bsigma_{3}
\cdot\bsigma_{2})P^{s}_{1}]
   \end{equation}
  \indent The hierarchy becomes apparent if one looks at the experimental values of the 
coefficients in Eq.(19); these
   were given (in MeV) in Eq.(20) using, for the (wide) decouplet resonances, both the pole 
and the conventional mass values.
   We saw that the magnitude of the coefficients, that is of the parameters,
   decreases with increasing number of indices in a term (we often say ``with increasing 
complexity").\\
  \indent Quantitatively things go as follows: Each flavor breaking $P^{s}$ factor in a sum 
(representing a term in the parametrization) introduces
  a reduction factor $\simeq 0.3$ (compare e.g. the ratio $D/C$). In addition to this factor 
-related to flavor breaking- another
  reduction factor arises from the number of the different indices in the same sum. Each 
different pair of indices or,
  as we call it, each ``gluon exchange" pair of indices, introduces a reduction factor 
$\approx 0.37$. The total reduction factor associated
   to a term is the product of the flavor and gluon exchange reduction factors.
   The numerical values of the reduction factors given above are those \cite{dimo96} obtained 
using
  for the masses of the resonances their pole values. Using the ``conventional" masses gives 
somewhat different reduction factors:
  the above $0.3$ (and $0.37$) become (in the order) $0.33$ and $0.22$.\\
  \indent As to the parameters multiplying the terms proportional to a $Trace$ (these Trace 
terms are absent in the
  mass formulas and - to 1st order flavor breaking - they are unimportant for the magnetic 
moments), their order of
  magnitude is generally much smaller than the previous ones; we will consider it in the last 
part of this Section.\\
  \indent In what follows it is useful to describe the hierarchy in field theoretical terms, 
that is to establish the correspondence
   between the General Parametrization and a Feynman diagrams description (compare the 
Appendix I for a detailed treatment).\\
   \indent Consider the field theoretical calculation of the expectation value of some field 
operator, call it $\Omega$ (expressed of course
  in terms of quark and gluon fields) in some exact state $\mid\Psi\rangle$ of the exact QCD 
Hamiltonian.
  In the GP this  calculation of $\Omega_{av}$ amounts to that of the expectation
   value of an effective operator $\tilde{\Omega}$ on the model state $\Phi$:
  \begin{equation}
  \Omega_{av}\equiv \langle\Psi\mid\Omega\mid\Psi\rangle =
  \langle\Phi\mid V^{\dag}\Omega V\mid\Phi\rangle \equiv \langle \Phi\mid 
\tilde\Omega\mid\Phi\rangle
  \end{equation}
  \indent Because $V$ can be related  to the $U$ operator of Dyson, it can be shown that 
$\Omega_{av}$
  is finally expressed in terms of Feynman diagrams (Appendix I, Eq.158 -note, in the example 
of Eq.158 $\Omega_{av}$ is in fact $\mathcal{M}_{av}$):
  \begin{equation}
  \Omega_{av}= \langle\Phi\mid T \, [\Omega(0)U(+\infty\mid -\infty)]\mid\Phi
  \rangle_{C}
  \end{equation}
  where $C$ means that only ``connected" Feynman diagrams are involved.
   Clearly, because $\mid\Phi\rangle$ is a 3-quark state, all diagrams intervening in the 
calculation of
  $\Omega_{av}$ (Eq.(50)) have three quark lines entering and three outgoing. For instance 
for the lowest $\mathbf{8+10}$ baryons
   $\Phi$ has been chosen as the product
  of a space factor $L=0$ state times a spin-flavor factor; thus $\Omega_{av}$, after 
integration on the space coordinates, becomes, as
  exemplified on parametrizing the baryon masses and magnetic moments, a combination of 
spin-flavor structures.
  The situation is illustrated graphically in Fig.2.
  \begin{figure}
    \begin {center}\includegraphics[width=10cm,height=13cm]{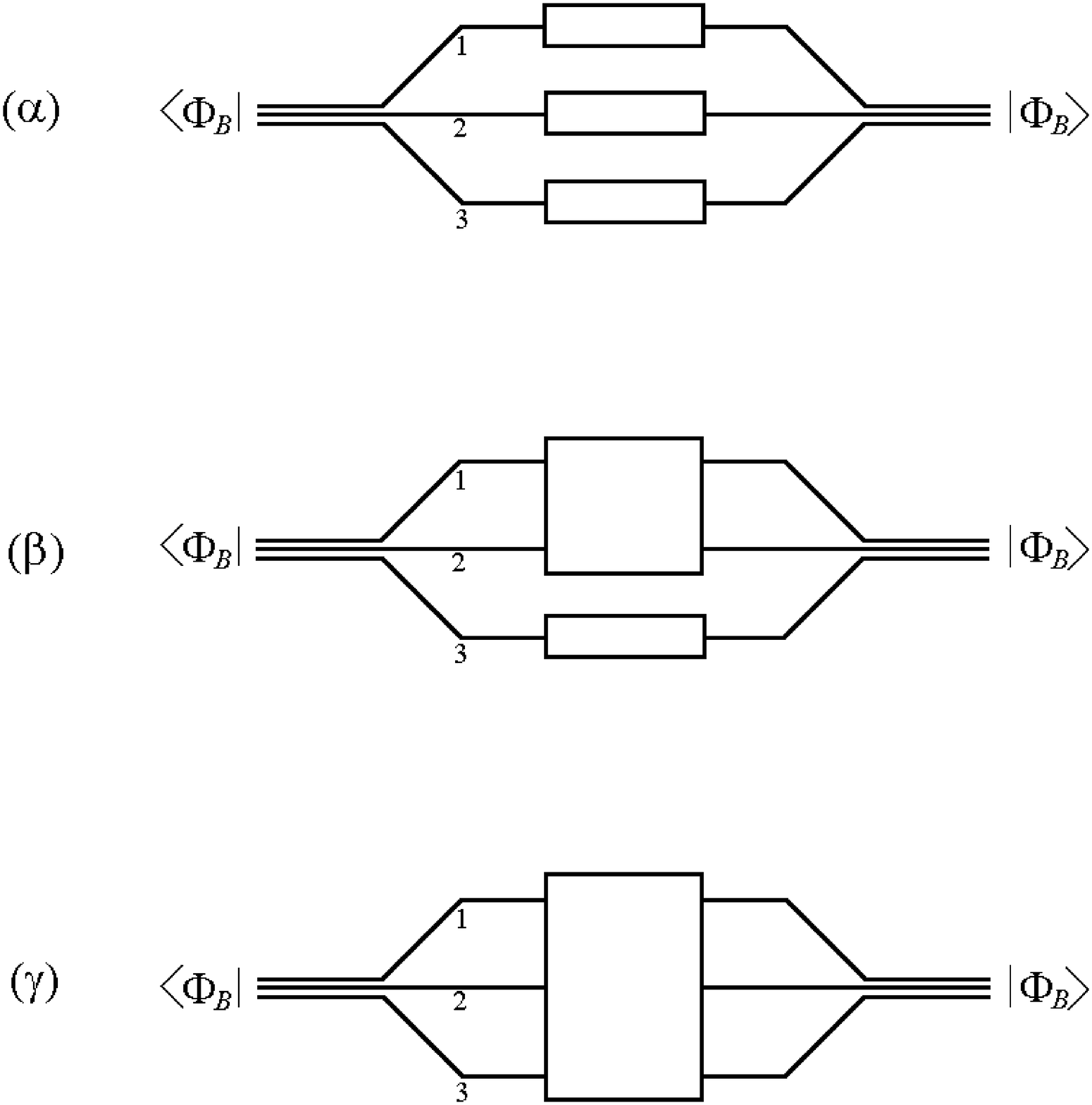}
   \end{center}
   \caption{\footnotesize{($\alpha$) The class of connected Feynman diagrams giving rise to 
zero or one-index terms in the spin-flavor space. The diagrams in this class include all 
diagrams without gluon exchange between different boxes. However in $(\alpha)$ diagrams may be present 
in wich gluon exchange is present between boxes without producing factors 
$(\bsigma_i\cdot\bsigma_k)$ or $P_i^sP_k^s$. $(\beta)$ The class of connected Feynman 
diagrams corresponding to two-index terms in the spin-flavor space. The diagrams in this 
class imply the exchange of at least one gluon, but may include diagrams with more than 
one-gluon exchange not producing spin-flavor factors with three indices. $(\gamma)$ The class 
of connected Feynman diagrams corresponding to terms with three indices in the spin-flavor 
space. These diagrams imply the exchange of at least two gluons. The boxes in 
$(\alpha),(\beta),(\gamma)$  describe the effect of the transformation $V$ on the three 
quarks in $|\Phi_B\rangle$.}}
   \end{figure}
  In Fig.2 - that refers to the baryon masses -
  the part ($\alpha$) corresponds to terms of type $\sum_{i}$, the part $\beta$  to terms of 
type
  $\sum_{i\ne k}$ , and the part $\gamma$ implies $\sum_{i\ne k \ne j}$. The coefficients 
that multiply terms
  $\sum_{i\neq k}$ imply \textit{at least} the exchange of one gluon; those multiplying terms 
$\sum_{i\neq k \neq j}$ need the exchange
  of \textit{at least} two gluons. The coefficients of the part $\alpha$ of Fig.2 include all 
diagrams without gluon exchange, but
  may receive a contribution also from diagrams exchanging gluons without producing factors 
$\bsigma_{i}\cdot\bsigma_{k}$
  or $P^{s}_{i}P^{s}_{k}$. On this basis one can list how the diagrams of the portions 
$\alpha$ or $\beta$ or $\gamma$ of Fig.2
  contribute to the the various parameters in the baryon mass formula:
  \begin{eqnarray}&&\cr
  &&M_{0}= M_{0\alpha} + M_{0\beta} + M_{0\gamma}, \quad\quad\quad C= C_{\beta} + C_{\gamma} 
\quad \quad\qquad\qquad\quad \,\, (no\quad P^{s}) \nonumber\\
  &&B= B_{\alpha} +B_{\beta} + B_{\gamma}, \qquad \qquad\quad  D= D_{\beta} + 
D_{\gamma},\qquad  E=E_{\gamma} \quad\quad \,(1\quad P^{s})\nonumber\\
  &&a =a_{\beta} + a_{\gamma},\qquad\qquad\qquad\qquad\quad  b =b_{\beta} + b_{\gamma},\qquad 
\quad  c= c_{\gamma}, \qquad   (2\quad P^{s})\nonumber \\
  &&d=d_{\gamma}\qquad\qquad\qquad\qquad 
\qquad\qquad\qquad\qquad\qquad\qquad\qquad\qquad\quad(3\quad P^{s})
  \end{eqnarray}
  \indent Note, for illustration, that $d,c$ and $E$ appear only with the suffix $\gamma$;
  this means that they imply at least the exchange of 2 gluons;
   $C,D,a,b$ are related to at least one gluon exchange; $M_{0},B$ do not show signs of gluon 
exchange (but may contain, of course,
  contributions from one or more gluon exchanges in a spin independent way). Finally, because 
$M_{0}$ is
  the spin-flavor independent part of the mass, it can be interpreted roughly as the sum of 
the masses of the
  constituent non strange quarks in a $p$ or $n$ and leads (with some assumptions) to an 
estimate
  for the value of the masses of constituent quarks.\\
  \indent However, before discussing constituent quarks, we recall that (see
   Sect.2), we analyzed in Ref.\cite{dimo96} the dependence of the GP from the quark mass 
renormalization point;
   there the GP was expressed in terms of light ``current quarks" $u,d,s$ with the 
conventional
  choice of their mass renormalization point ($q\approx 1\, GeV$); to have shown, as in 
\cite{dimo96}, that the GP can be formulated
  in this conventional QCD frame, is necessary for its consistency.\\
  \indent However, \textit{although the conventional
  choice of the quark masses is necessary for the high energy perturbative QCD, the results 
of the GP
  on the low energy properties of hadrons can be obtained both with this conventional choice
  or with different selections, say with masses of the order $\equiv 
M(\mathcal{P},\mathcal{N}) \,\approx 300\, MeV\,$ and $M_{s}\equiv M(\lambda),\approx 
500\,MeV$}- corresponding to a different mass renormalization point.
  (Such quarks may be \textit{called} ``constituents"; their usefulness in the treatment of 
the low energy hadronic properties results from
  the fig.2 \textit{plus} the hierarchy)\footnote{In 1974 a paper by Melosh appeared 
\cite{melosh} with the title:``Quarks: current and constituents"
  This paper, written at a time when current algebra was the basic description, can be 
confusing on the notion of constituent quarks. I mention it
   here only for this reason and refer to \cite{morass} for more comments.}\\
   \indent Indeed the hierarchy favors terms with few indices; therefore diagrams of type  
$\alpha$ in Fig.2 play a significant role.
   This role should be experimentally detectable, in the sense that the cases in which, say a 
proton, is in the form of three constituent quarks
   may be a measurable fraction of the total. Experiments to see this may be difficult, but, 
as discussed in \cite{petronzio} or \cite{mochalov} should
   be interesting.\\
   \indent It is clear that each
   ``box" in Fig.2, when explored on a fine grain basis (that is, analyzed in terms of the 
conventional ``light" $u,d,s$ quarks),
   contains all possible light quark-antiquark pairs -that is, quark loops- as well as 
gluons. In other words the Fig.2 describes the effect
   of the transformation $V$ on the three quarks (for a baryon) in $\vert\Phi_{B}\rangle$. 
Inside each box a quark line may zigzag as much
   as it likes - compatibly with QCD!- due to emission and reabsorption of gluons; also each 
box contains quark loops and gluon lines
    connected in all possible ways consistent the rules of QCD. \\
  \indent We now consider the determination of the parameters in the Eq.(19) giving the 
parametrized mass formula.
  We will do this in the constituent quark description, recalling that such description 
started with the ``naive" quark model (Ref.\cite{mo65}),
   where the analogy of a hadron to a nucleus composed of dressed nucleons was explicitly 
introduced;
   this description was later resurrected in Ref.\cite{der75}, where a Fermi-Breit 
approximation of QCD was introduced.\\
  \indent Here, to \textit{exemplify} how the values (Eq.(20)) of the GP baryon mass 
parameters can be determined, we will use the constituent
   treatment as in Ref.\cite{mo92}.\\
   \indent Consider first the ratio of the coefficients $D$ and $C$ in the mass 
parametrization Eq.(19); this ratio gives the magnitude
    of the reduction factor related to $1P^{s}$ flavor breaking. It is:
  \begin{equation}
  \mid D/C \mid\,\, =\frac{\displaystyle(\geq 1\,\, gluon\,\, exch.\mid 
1P^{s})}{\displaystyle(\geq 1\,\, gluon\,\, exch.\mid 0P^{s})} \cong 0.3
  \end{equation}
  \indent To obtain the \textit{reduction factor due to flavor}, we start considering the 
ratio $\vert D/C\vert$; a similar procedure can be used
  for the parameter $c$.
    The treatment of $D/C$ is given in Ref.\cite{mo92} after Eq.(6). It starts from the 
question: Why does the above ratio (53),
    which is a measure of $\Delta m/m$, where $m$ is, here, the mass of a ``constituent" 
quark in a nucleon
    and ($m +\Delta m$) the same quantity for the ``strange constituent quark", give the 
correct value of the reduction
    factor due to flavor? We might have written equally well
    the term $D\sum_{i>k}(\bsigma_{i}\cdot\bsigma_{k})(P^{s}_{i}+P^{s}_{k})$ in Eq.(19)
    as $2D\sum_{i>k}(\bsigma_{i}\cdot\bsigma_{k})P^{s}_{i}$ because only its expectation 
value on a symmetric wave function
    is relevant. Why then we used $D/C$ rather than $2D/C$? The rule adopted in writing the 
sums in Eq.(19) implies the first
    choice (and of course a uniform criterion must be adopted). Still, in view of the 
importance of this point, a direct check
    is appropriate. For this we compared our GP result with the explicit results of De 
Rujula, Georgi and Glashow
    \cite{der75}. Note that besides the comparison
    with Ref.\cite{der75}, the analysis in \cite{dimo96} confirms the above treatment.
     In Ref.\cite{mo92} we also evaluated (using the conventional values of the baryon 
masses) the ratio $\mid E/D\,\mid$ that gives the reduction
     factor due to one gluon exchange, obtaining
    $\mid E/D\,\mid = 0.22$. The same quantity calculated using the \textit{pole} values of 
the masses is:
    \begin{equation}
    \vert E/D\vert\, =\frac{\displaystyle(\geq 2\,\, gluon\,\, exch.\vert 
1P^{s})}{\displaystyle(\geq 1\,\, gluon\,\, exch.\vert 1P^{s})}\cong 0.37
    \end{equation}
    \indent We now come back to the hierarchy for the magnetic moments (Sect.4). Using the 
symbols of Eqs.(33)
    we can write:
    \begin{equation}
    \vert \tilde{g}_{2}/\tilde{g}_{1}\vert \, =\frac{\displaystyle(\geq 0\,\, gluon\,\, 
exch.\vert 1P^{s})}{\displaystyle(\geq 0\,\, gluon\,\, exch.
    \vert 1P^{s})} = \vert\, 0.934/2.793 \vert \cong 0.33
    \end{equation}
    Thus the reduction factors for flavor breaking from the analysis of the masses and
    from that of the magnetic moments are the same ($0.3$ to $0.33$).\\
    \indent \textit{The Trace terms.} Finally we consider the terms of type 
$g_{0}Tr(QP^{s})$; these terms -absent in the baryon
    mass parametrization- play a role in the GP of many quantities, e.g., in the magnetic 
moments of the octet and  the decuplet baryons.
     In Ref.\cite{mo89} these \textit{Trace}
    terms were ignored -incorrectly- for the magnetic moments of the octet baryons; but then 
(as shown in detail in Sect.4)
    they can be combined with the other terms so that, in fact, no error was made to first 
order in flavor breaking, which was
     the case considered in Ref.\cite{mo89}.\\
     \indent What is the meaning and order of magnitude of the $Trace$ terms? \\
    \indent A) \textit{Meaning} - Looking at the GP in terms of Feynman diagrams a $Trace$ 
term corresponds to a closed quark loop. There may be,
    as already stated, a large number of quark loops inside each box in fig.2; but it is not 
to such loops -in which only gluons attach to the
    loop- that we refer here. We refer to
    quark loops where an external photon or some other external particle (say a Z meson for 
weak interactions) attaches to a vertex
    in the loop quark circuit. For the magnetic moments one has to do with a photon (the 
external magnetic field). These Trace terms, that imply
    the presence of an external field, are not present in Fig.2 that refers to the baryon 
masses and should be completed
    in the presence of external fields (for instance, just having a photon attached to one of 
the boxes). A case of loop with
    a photon at a vertex will be considered in Sect.12 (compare there the Fig.3). These quark 
loops are represented by Trace terms.\\
     \indent \textit{If a photon vertex is on the loop}
      we meet, in circling the loop, the photon vertex plus, at least, three gluons vertices. 
The reason for the above ``three"
    is that one gluon is forbidden by color (``forbidden" means that the loop gives zero) and 
two gluons (+ a photon) is forbidden by the Furry theorem,
     so that three gluons are the minimum number that can be present.\\
    \indent B) \textit{Order of magnitude} - A general estimate is not easy, but the 
circumstance that such a loop is connected to the rest
    of the diagram by at least three gluons suggests, due to the hierarchy, that its 
contribution
    should be comparatively small. One may expect a reduction factor of the order 
$(0.33)^{3}\cdot (1/3)\cdot 3\approx \,3.6\cdot10^{-2}$,
     where the $(1/3)$ appearing in the product is the value of  $Tr(QP^{s})$ and the 3 stays 
for $N_{c}$. (This rough estimate disregards the
     permutations between the gluons from the loop). An order of magnitude (consistent with 
the above estimate) of the loop reduction factor
      derived from specific cases was given near to the end of Sect.3.

    \vskip 30pt
     \noindent{\bf 8. The parametrization of the masses of the lowest Ps and V meson 
nonets.}\\

    We apply now the GP to the lowest nonets of Pseudoscalar and Vector mesons (compare 
Ref.\cite{mo90}).
    As we will see, the GP leads to formulas that look very similar to those of the NRQM, 
although now the description is
    fully relativistic. We will proceed as for baryons, beginning with the model states
    for mesons. Because the model Hamiltonian for the mesons, $\mathcal{H}$(mesons) (written 
simply $\mathcal{H}$ below in this Section), is assumed
    to conserve the orbital angular momentum and to be spin and flavor independent, the
    model states of the Ps and V nonets are degenerate states of $\mathcal{H}$ ($M^{0}_{0}$ 
is the same for all $i$ in Eq.(56)):
    \begin{equation}
    \mathcal{H} \mid\Phi_{i}\rangle \mid no\,gluons\rangle = M^{0}_{0}\mid \Phi_{i}\rangle 
\mid no\, gluons\rangle \quad\quad
    \quad (i=\pi,\eta,\eta',K;\, \rho,\omega,\Phi,K^{*})
    \end{equation}
    \indent In Eq.(56) $\mid \Phi_{i}\rangle$ is a state of a meson at rest with $L=0$ and 
$M_{0}^{0}$ is the common value of the masses of the $18$
    model states of the $0^{-}$ and $1^{-}$ lowest mesons.
    From now on we omit the factor $\mid no\,gluons\rangle$ in (56).
     Because the model Hamiltonian $\mathcal{H}$ is spin and flavor independent,
    the wave function $\Phi_{i}$ of a meson has the form:
    \begin{equation}
    \Phi_{i} = S_{c}W_{i}\varphi(r)
    \end{equation}
    where $S_{c}$ is the color singlet factor for a meson (that we will not write in what 
follows, unless necessary), $W_{i}$ is
    the $q \overline{q}$ spin-flavor wave function of the $i$-th meson and $\varphi(r)$ is a 
function of the relative distance $r$
    of the quark and antiquark.\\
    \indent Operating as we did for baryons, we define a unitary transformation $V$ acting on 
the above meson model states and transforming
     them into the exact states (to be indicated $\mid \psi_{i} \rangle$). Recall, once more, 
that (as for baryons, Sect.2)
      the model states are chosen as simple as possible; all the ``complications" are hidden 
in the operator $V$:
     \begin{equation}
                 \mid \psi_{i}\rangle = V \mid \Phi_{i}\rangle
     \end{equation}
     Again our aim will be to parametrize the results. As done for baryons, $V$ is 
constructed in terms of the exact QCD Hamiltonian $H_{QCD}$;
      the problem of quark mass renormalization  is similar to that considered in Sect.2 for 
baryons.
      Because the QCD hamiltonian is the same, the flavor breaking parameter is decently 
small and can be treated perturbatively (as for baryons).\\
     \indent In the above notation the mass of the $i$-th meson is:
     \begin{equation}
     M_{i} = \langle \psi_{i}\mid H\mid \psi_{i}\rangle = \langle \Phi_{i}\mid 
V^{\dagger}HV\mid \Phi_{i}\rangle
     \end{equation}
     \indent Because the $\Phi_{i}$'s are, by construction, two body states (one quark-one 
antiquark), only the projection
     $\tilde{H}$ of the operator $V^{\dagger}HV$ in the subspace of these two body states 
intervenes in the calculation of (59):
     \begin{equation}
      \tilde{H}\equiv \sum\mid 1q,1\bar{q}\rangle \langle 1q,1\bar{q}\mid V^{\dagger}H V\mid 
1q',1\bar{q}'\rangle \langle 1q',1\bar{q}'\mid
     \end{equation}
     \indent The above statement, that the only part of $V^{\dagger}HV$ intervening in the 
calculation of $M_{i}$ (Eq.(59)) is the two body part
     $\tilde{H}$, is trivial but essential; in fact calculating the expectation value of the 
field operator $V^{\dagger}HV$ in the state
     $\mid\Phi_{i}\rangle$ \textit{becomes equivalent to calculate the expectation value of a 
certain quantum mechanical two-body operator}
     $\tilde{H}$ on the wave function $\Phi_{i}$ in ordinary non relativistic two-body 
quantum mechanics:
     \begin{equation}
     M_{i}= \langle \Phi_{i}\mid \tilde{H}\mid \Phi_{i}\rangle
     \end{equation}
     The most general parametrization of the meson masses amounts then to write in ordinary 
non relativistic two body quantum mechanics
     the most general operator $\tilde{H}$ of the relative coordinates and momentum 
$\mathbf{r},\mathbf{p}$, of the spins
     $\bsigma_{1},\bsigma_{2}$ (1=quark, 2=antiquark) and of the flavor operators $f$ (see 
below) invariant with respect to translations
     and rotations. It follows from Eq.(61) that this most general operator has necessarily 
the form:
     \begin{equation}
     \tilde{H} = \sum_{\nu} R_{\nu}(\b\r, \mathbf{p})G_{\nu}(\bsigma,f)
     \end{equation}
     where the $G_{\nu}$'s (that of course \textit{have nothing to do with the same 
(boldface) symbols used treating the baryon magnetic moments
     in Sect.4}) is a set of independent operators (specified by the index $\nu$) constructed 
in terms of $\bsigma_{1},\bsigma_{2}$ and the flavor
     operators $f$ of the quark and the antiquark; $\mathbf{R}_{\nu}(\b\r,\mathbf{p})$ are 
operators (of which it is not necessary to know
     the expression) constructed in terms of the relative coordinate and momentum $\b\r$ and 
$\mathbf{p}$ of the quark and the antiquark pair
      in the model state.\\
     \indent Calculating the expectation values of $\tilde{H}$ (62) on the various mesons 
$i$, their masses $M_{i}$ can be written:
     \begin{equation}
     M_{i}= \sum_{\nu}g_{\nu}\langle W_{i}\mid G_{\nu}(\bsigma,f)\mid W_{i}\rangle
     \end{equation}
     where $g_{\nu}$ stays for:
     \begin{equation}
     g_{\nu}= \langle \varphi(r)\mid R_{\nu}(\b\r, \mathbf{p})\mid \varphi(r)\rangle
     \end{equation}
     $\varphi(r)$ being the space dependent factor of $\Phi_{i}$ in Eq.(57).\\
     \indent Because the model Hamiltonian $\mathcal{H}$ was chosen to be independent from 
the flavor and from $J$,
      the space wave function $\varphi(r)$ in Eq.(64) is independent of the meson index $i$; 
the same is true for the coefficients
     $g_{\nu}$ in Eqs.(63, 64). Therefore Eq.(63) leads to the mass operator $M$ (65)(of 
which, to get the masses, one must take the
     expectation value on the spin-flavor factor $W_{i}$ of the wave function):
     \begin{equation}
     M= \sum_{\nu}g_{\nu}G_{\nu}(\bsigma,f)
     \end{equation}
     Note that this is a general result; part of its interest (as for the analogous 
expression for baryons, Eq.(17))
    is its simplicity (and the fact that it appears in a notation directly related to the 
NRQM).\\
    \indent Listing the possible $G_{\nu}$ in Eq.(65) means to find their possible spin and 
flavor dependence. Ignoring, for the moment,
     the electromagnetic interactions, the only $\lambda$ matrix that intervenes is 
$\lambda_{8}$, that appears in the flavor breaking
     part of the Hamiltonian. Instead of $\lambda_{8}$ we consider the projection operator 
$P^{s}= \frac{1}{3}(1-\lambda_{8})$ that
     gives zero when applied to $u$ or $d$ quarks (or antiquarks) and is $1$ when applied the 
$s$ quark (or antiquark). For a system
     of one quark and one antiquark, and if the isospin $I$ of the state is $\neq 0$ the list 
of flavor operators is:
    \begin{equation}
                              1,\qquad \qquad P_{1}^{s},\qquad \qquad P_{2}^{s}
    \end{equation}
    In (66) and in what follows the index $1$ will refer to the quark and the index $2$ to 
the antiquark.\\
    \indent For the mesons with $I=0$ the list above (66) is however incomplete. Indeed the 
flavor operators in Eq.(66) connect only a
    $q\overline{q}$ to a $q\overline{q}$ of the same type, e.g. a $u\overline{u}$ with a 
$u\overline{u}$, or $d\overline{d}$ with
    a $d\overline{d}$, or a $u\overline{s}$ with a $u\overline{s}$ and so on. The sum in 
Eq.(65), however, include, for mesons with isospin
    $I=0$, also matrix elements of $\tilde{H}$ that connect a $u\overline{u}$ state with a 
$d\overline{d}$; or a $s\overline{s}$ with a
    ($u\overline{u}$+$d\overline{d}$) state. A gluon exchange in QCD may produce these 
processes. Thus, for $I=0$, there are other
    flavor operators possible in addition to those listed in (66). To write them we first 
introduce the $I=0$ flavor states:
    \begin{equation}
     \mid z\rangle =\mid u\overline{u}+d\overline{d}+ s\overline{s}\rangle, \qquad  \mid 
w\rangle=\mid s\overline{s}\rangle
    \end{equation}
     Note that $\mid z\rangle$ given above is not normalized to 1 (differently from the other 
states); to normalize to 1,
      we must be multiply it by $1/\sqrt{3}$. Here we leave it in the above form, keeping in 
mind this point.\\
    Taking into account the expressions (67) we must add to the list (66) the operators:
    \begin{equation}
    (a)\quad \mid z\rangle \langle z\mid \quad; \quad (b)\quad \mid z\rangle \langle w\mid + 
\mid w\rangle \langle z\mid \quad;\quad (c)\quad
    \mid w\rangle \langle w\mid \equiv P_{1}^{s}P_{2}^{s}
    \end{equation}
    As to the $\bsigma$ dependence of the $G_{\nu}(\bsigma,f)$ only the following expressions 
(69) are possible:
    \begin{equation}
                                   1,\qquad \qquad \qquad  \bsigma_{1}\cdot\bsigma_{2}
    \end{equation}
    \indent We can now discuss the parametrization of the meson masses. We begin with the 
(lowest mass) mesons with $I\neq0$,
    ($\pi,\,K; \rho,\, K^{*}$). The flavor breaking parts of $G_{\nu}(\bsigma,f)$ can contain 
only $P_{1}^{s}$ and $P_{2}^{s}$.
    The charge conjugation invariance of the Hamiltonian implies that the $G_{\nu}$'s must be 
invariant for the exchange
    of 1 and 2; thus the possible $G_{\nu}$'s are $1,\,\, \bsigma_{1}\cdot\bsigma_{2}, \,
    \,P_{1}^{s}+P_{2}^{s}, \, \,\bsigma_{1}\cdot\bsigma_{2}(P_{1}^{s}+P_{2}^{s})$. Therefore 
the \textit{most general expression of the mass} of
    a meson with $I\neq0$ \textit{correct to all orders in flavor breaking} is:
    \begin{equation}
    M_{I\neq 0}= A+B\bsigma_{1}\cdot\bsigma_{2} +C(P_{1}^{s}+P_{2}^{s}) 
+D\bsigma_{1}\cdot\bsigma_{2}(P_{1}^{s}+P_{2}^{s})
    \end{equation}
    where $A,B,C,D$ are four real coefficients. Because $\bsigma_{1}\cdot\bsigma_{2}= -3$ for 
$J=0$ and
    $\bsigma_{1}\cdot\bsigma_{2}= +1$ for $J=1$, the masses (indicated by the meson symbols) 
are:
    \begin{eqnarray}&
    \pi= A - 3B\,\, ( =\,138 ),\quad K= A- 3B +C -3D\,\, ( =\,495 )\nonumber\\
    &\rho= A + B\,\,\,\, ( =\,770 ),\qquad K^{*}=A+B+C+D\,\, ( =\,894)
    \end{eqnarray}
    where the values given in MeV are approximate values averaged over the charges.\\
    The Eqs.(71) imply (in MeV):
    \begin{equation}
    \qquad\qquad A=612,\qquad\qquad B=158,\qquad\qquad C=182,\qquad\qquad D= -58
    \end{equation}
    \indent We now consider the mesons with $I=0$. Their most general mass formula is 
obtained adding to (70) with $A,B,C,D$
    just determined, another part obtained
    multiplying the flavor operators (68) with the spin operators (69). We have:
    \begin{eqnarray}&
    M_{I=0} = A+B\bsigma_{1}\cdot\bsigma_{2} +C(P_{1}^{s}+P_{2}^{s}) 
+D\bsigma_{1}\cdot\bsigma_{2}(P_{1}^{s}+P_{2}^{s})\nonumber\\
     &+ (E+F\bsigma_{1}\cdot\bsigma_{2})\mid z\rangle\langle z\mid + 
(H+G\bsigma_{1}\cdot\bsigma_{2})(\mid z\rangle\langle w\mid
     +\mid w\rangle\langle z\mid)\nonumber\\
     &+(N+T\bsigma_{1}\cdot\bsigma_{2})P^{s}_{1}P^{s}_{2}
    \end{eqnarray}
    The Eq.(73) contains $A,B,C,D$ \footnote{We use here the same symbols of Ref.\cite{mo90}; 
in Ref.\cite{dimo96}, Eq.(28) $C$ and $B$
    were -unfortunately- interchanged: $C$ there is our $(B)$ here
    and viceversa.} (determined [Eq.(72)] from $\pi,K,\rho,K^{*}$) and six additional 
parameters $E,F,H,G,N,T$.
     These might be determined from the masses of $\eta,\eta^{'},\omega,\phi$ + two mixing 
angles (the vector $\theta_{V}$, and pseudoscalar
    $\theta_{P}$), if these angles were known. Alternatively one can limit to the first-order 
flavor-breaking approximation,
     that is disregard the last term in Eq.(73)(this means that the $N,T$ terms, of 2nd-order 
in flavor breaking -proportional
     to $P_{1}^{s}P_{2}^{s}$- are disregarded; that is we set: $N=0,T=0$). In this 1st order 
flavor breaking approximation the masses
     of $\eta,\eta',\omega,\phi$ fix $E,F,G,H$; then one can determine, to 1st order in 
flavor-breaking, the mixing angles.\\
     \indent Here we will first \cite{mo90} consider the Pseudoscalar $P$ mesons 
$\eta,\eta'$; for them in Eq.(73) $\bsigma_{1}\cdot\bsigma_{2}=-3$.
      Introducing the abbreviations:
     \begin{equation}
     b=A-3B\, (=138 MeV),\quad d=C-3D\,(=357 MeV),\quad f=E-3F,\quad g=H-3G
     \end{equation}
     where $b$ and $d$ are known from (72), the Eq.(73) for the $P$ mesons with $I=0$ takes 
the form:
     \begin{equation}
     M_{I=0}(P)= b +d(P_{1}^{s}+P_{2}^{s}) +f\mid z\rangle\langle z\mid +g(\mid 
z\rangle\langle w\mid +\mid w\rangle\langle z\mid)
      +O(\Delta m/m)^{2}
     \end{equation}
     To determine the masses of $\pi^{0},\eta,\eta'$ correct to 1st-order in flavor-breaking 
we must diagonalize the Equation \footnote{In an
     arXiv paper \cite{duma01} L.Durand, while stating that the matrix (76) below was the 
most general form, added that it did not contain
     the physical identification of the various contributions. At a question of G.M. on the 
meaning of this, he answered that the last statement certainly needed a clarification- if that arXiv
paper was published, which was uncertain.}
    \begin{eqnarray}
    \left (\begin{array} {ccc}
    b+f-M& \quad\quad\, f \qquad\qquad\quad\qquad   f+g&\\
    f& \quad\quad b+f-M \quad\qquad\qquad         f+g&\\
      f+g&\quad\qquad\, f+g\quad\quad\qquad    b+2d+f+2g-M&
    \end{array} \right)
    =0
    \end{eqnarray}
    One solution of Eq.(76) is the $\pi^{0}$ mass; we get, of course for it: $\pi^{0}= \,b\, 
=A-3B$; the masses of
    $\eta$ and $\eta'$ are the two roots of:
    \begin{equation}
    (b+2f -M)(b+2d+f+2g -M) - 2(f+g)^{2}=0
    \end{equation}
    The solutions of Eq.(77) are $M=b+k\pm(k^{2}+2g^{2} -4fd)^{1/2}$ where 
$2k\equiv(3f+2d+2g)$. Equating the two solutions to
    the masses of $\eta(547)$ and $\eta'(958)$ we obtain $\eta'+\eta=2b+3f+2d+3g$ and
    $(\eta'-\eta)/2 =[2g^{2}-4fd+\frac{1}{4}(3f+2d+2g)^{2}]^{1/2}$ where $b$ and $d$ have the 
value
    given in Eq.(74). Solving for $g$ and $f$ we obtain two possible solutions $(g;f)_{1}$ 
and $(g;f)_{2}$.
    Expressed in MeV, they are:
    \begin{equation}
    (g;f)_{1}=(-133,+261),\qquad \,\qquad (g;f)_{2}=(-343,+400)
    \end{equation}
    The ratio ($g/f\sqrt{3}$) between the coefficient $g\sqrt{3}$ of the (normalized) 
flavor-breaking term
    and the coefficient $3f$ of the (normalized) unitary singlet term is substantially 
smaller for the solution $N.1$
    than for the solution $N.2$. Thus we must choose the solution $N.1$ (in order to be 
consistent with the assumption of neglecting terms
    of order higher than the first in ($\Delta m/m$)). On choosing $(g;f)_{1}$ it is 
straightforward to check that the
    diagonalization of (77) implies [in addition to $\pi^{0}= 
(1/\sqrt{2})(u\overline{u}-d\overline{d})$]:
    \begin{equation}
    \eta=0.603(u\overline{u}+d\overline{d}) - 0.522\, s\overline{s},\qquad 
\eta'=0.367(u\overline{u}+d\overline{d}) + 0.854\, s\overline{s}
    \end{equation}
    With the usual definition $\eta=\eta_{1}\,\sin\theta_{P}+\eta_{8}\,\cos\theta_{P}$ and 
$\eta'=\eta_{1}\,\cos\theta_{P}-\eta_{8}\,\sin\theta_{P}$ where
    $\eta_{1}= (1/\sqrt{3})(u\overline{u}+d\overline{d}+s\overline{s})$ and 
$\eta_{8}=(1/\sqrt{6})(u\overline{u}+d\overline{d}-2s\overline{s})$,
    we obtain from Eqs.(79):
    \begin{equation}
                     \sin\theta_{P}\simeq -0.39      \qquad\qquad
                     (\mathrm{ that \,\, is,}\, \,\theta_{P}\simeq\,-23^{\circ})
    \end{equation}
    This value can be compared with $\theta_{P}\simeq -20^{\circ}$ and $\theta_{P}\simeq \pm 
24^{\circ}$ obtained respectively
    in Refs.\cite{gika87},\cite{karlNC}. \\
    \indent We refer here to the part B of Sect.V of \cite{mo90} for a discussion (in our 
opinion now totally obsolete)
     on the values of the above angle
     obtained using, instead of the Hamiltonian, its square or other powers. The discussion 
is obsolete for the reason already stated
     (Sect.2,footnote 4), that the square of the QCD Hamiltonian is, almost certainly, a non 
renormalizable operator.
    It is obvious, from the previous derivation of $\theta_{P}$,
    that the value given above ($-23^{\circ}$) is that obtained from QCD neglecting terms of 
second order in flavour breaking.\\
    \indent Other work comparable to our results described above is that in Ref.\cite{der75} 
and in Ref.\cite{zs66},
    the latter closely related to the NRQM.
    As to Ref.\cite{der75},
    if we identify our $f$ with the $\beta$ in the mass-matrix of \cite{der75} and include 
the 1st-order flavor breaking correction $g$
    (that should have been included in the work of \cite{der75}-compare the end of Sect.V in 
Ref.\cite{mo90}, our mass matrix and that of
     Ref.\cite{der75} coincide.
    Also our results coincide with those of \cite{zs66}. Here, aside from other minor points, 
there seems to be a numerical mistake in the
    Eqs.(5) and (5a) of \cite{zs66} for the $\eta$ and $\eta'$, but, after its correction, 
the result coincides (again compare for the details
    the footnote $^{8}$ of \cite{mo90}). \\
    \indent More interesting, is the following remark. We have seen that the pion mass is:
    \begin{equation}
    \pi= A-3B
    \end{equation}
    with $A=612, B=158$. These values -that refer to $I\neq0$ mesons- are \textit{exact QCD 
values} (there is no higher order correction omitted).
    The smallness of the pion mass,
    $\pi=A-3B$ might be just an accident depending critically on the values of the parameters
    $A$ and $B$ in Eq.(72), with $B$ multiplied by $(-3)$ in Eq.(71). Assume, for instance, 
that $B$ had a value $100\,MeV$
    instead of $158\,MeV$ (Eq.(72)). Then, if $A$ maintains its value $612\, MeV$,
    the pion would have a mass of $312\, MeV$, no longer so small (the $\rho$ would then have 
mass $712\, MeV$).\\
    \indent In QCD a reduction of $B$ by the above amount $\approx 40\%$ could arise from an 
even smaller percentage change
    of $\alpha_{s}$, the quark-gluon coupling. Thus we do not attribute a profound meaning
    to the smallness of the pion mass. \textit{We differ substantially on this from the 
standard point of view in chiral QCD},
     where the pion is seen as a quasi-Goldstone boson,
    getting its mass from explicit breaking of chiral symmetry due to the small $u,d$ masses.
    This remark does not question at all, obviously, the phenomenological treatments and 
results related to PCAC (partial conservation of
    axial-vector current), insofar as they simply take note of the empirical value of the 
pion mass.\\
    \indent We now come back to the Vector mesons $V$ with $I=0$; their discussion is similar 
to that of the $Ps$ mesons;
    one has to put in Eq.(70) $\bsigma_{1}\cdot\bsigma_{2}=+1$
    (instead of $-3$) and therefore one replaces in the previous treatment of the $Ps$ mesons 
$A-3B$ with $A+B$; $C-3D$ with $C+D$;
    $E-3F$ with $E+F$; $H-3G$ with $H+G$. As is well known, the parametrization with 
$A,B,C,D$ alone is almost sufficient for the $I=0$
    Vector mesons; this means that in the Equation that replaces (74) $E+F\, \approx 0$ and 
$H+G \, \approx 0$.
     (From this follows that, for instance, the value of the
    rate $\Phi\rightarrow \pi^{0}\gamma$ is s very small; we will come back to this in the 
next Section.)\\
    \indent \textit{A few comments on the relation of the above results with those of the 
NRQM}. To conclude this section we come back to the
     question raised at the start: Why does the NRQM work so well?
     The following remarks, that conclude Ref.\cite{mo90}, provide a summary of the answer; 
of course, for more details, the whole
    Ref.\cite{mo90} should be consulted.\\
    \indent(1) Almost all the features of the mass formulas for the lowest meson nonets 
currently used in a NRQM description are general consequences
     of the GP; they are similar to those of the NRQM, but they do do not depend on it; the 
only exception is the $(m_{1}m_{2})^{-1}$ multiplying
     factor of the $\bsigma_{1}\cdot\bsigma_{2}$ term
    in the De Rujula et al. expression \cite{der75} for the meson masses, a result that 
depends specifically on the one gluon exchange potential.
    Stated differently, the meson mass formulas used in the NRQM are more general than one 
might have thought. In this respect the situation for
     the meson masses is quite different, for instance, from that found for the baryon 
magnetic moments. We recall that the GP expression
      for the magnetic moments, correct to first order in flavor breaking, had seven 
parameters whereas the fairly successful
      parametrization of the NRQM had only two.\\
    \indent(2) The procedure developed above to derive the meson mass formulas clarifies the 
old question of the mixing angle for the $I=0$
    meson nonets. The mixing angle (say $\theta_{P}$) can be determined from the GP (that is 
in a model independent way) knowing the masses
     of the $P$ mesons provided terms of order higher than the first in the flavor breaking 
expansion parameter are negligible
     and are neglected. According to this procedure the linear angle is much more natural 
than the quadratic or square-root ones;
      it is the angle resulting from a QCD calculation on expanding the exact result in 
series of $(\Delta\,m/m)$ and neglecting terms
      of order higher than the first.

     \vskip 30pt
    \noindent {\bf 9. The radiative $\bV\rightarrow \bP+ \bgamma $ meson decays.}\\

     An early (1965) test of the NRQM consisted in comparing with the data the calculated 
\cite{bmo65} radiative decays of Vector mesons;
     the evaluation of these ($M1$) $\,\gamma $ transitions was possible in 1965 because
     the magnetic moments of the $\mathcal{P}$ and $\mathcal{N}$ quarks had been deduced 
\cite{mo65} from
     those of the proton and neutron. We list below many $M1$ transitions of interest, though 
only a few were
     measured in those years and the magnetic moment of the $\lambda$\ quark was still 
unknown:\\
    \begin{math}
    (1)\,\,\omega\rightarrow \pi^{0}+\gamma, \qquad \qquad(2)\,\,\omega\rightarrow 
\eta+\gamma,\qquad \quad\,\,\,
    (3)\,\,\,\rho \rightarrow \pi+\gamma,\,\,\qquad
    \,(4)\,\rho^{0}\rightarrow\eta+\gamma,\\ \,\, (5)\,\,K^{*+}\rightarrow K^{+}+\gamma, 
\qquad (6)\,K^{*0}\,\rightarrow K^{0}+\gamma,\qquad
    (7)\,\,\rho^{0}\rightarrow\eta'+\gamma, \qquad\, (8)\,\omega\rightarrow 
\eta'+\gamma,\\\qquad \qquad (9)\,\phi\rightarrow \eta+\gamma,\qquad
    \qquad(10)\,\phi\rightarrow \eta'+\gamma,\,\qquad\quad\,(11)\,\phi\rightarrow 
\pi^{0}+\gamma\\
    \end{math}\\
    \indent The $\omega\rightarrow\pi^{0}\gamma$ transition had been measured reasonably 
well. The NRQM calculation
    reproduced it and gave the orders of magnitude of the others not too far from reality.
    Ref.\cite{bmo65} (compare also \cite{moVien68, moerice}) contains a list of the 
transitions calculated at that time.\\
    \indent Because these transitions need in fact a full QCD calculation, one can apply the 
GP also
    to this problem. The formalism is developed in Ref.\cite{moVPgam90}(General 
parametrization of
    the $V\rightarrow P\gamma$ meson decays), that contains a full description of the 
procedure.
    Here we only summarize some of the main results. It will appear that the hierarchy
    is useful also here; and, once more, this explains why the NRQM works.\\
    \indent To clarify this statement, before giving the details, consider
    the ratio\\ $\Gamma(\omega\rightarrow\pi^{0}\gamma)/\Gamma(\rho\rightarrow 
\pi^{0}\gamma)$; the GP predicts this to be $9$ (to all orders
     in flavor breaking) plus the contributions from processes
    where the initial $q\overline{q}$ state $\omega$ transforms into 3 gluons and then gives 
rise to the final state.
    The hierarchy, of course, implies that such contribution should be a minor one and, in 
fact, it is not visible
     (at the level $\pm 15\%$, the experimental error). Other processes and, in particular,
    $\phi\rightarrow\pi^{0}\gamma$, lead to the same conclusion. In general, if the 3-gluon
    diagrams are negligible, the important terms in the GP calculation of the 
$V\rightarrow\pi\gamma$ decays (that being an exact QCD calculation,
    includes automatically the configuration mixing and all the complexities of the Fock 
$q\bar{q}$, gluon expansion of the hadron states) reproduce
    in practice the result of the NRQM.\\
    \indent We now give a short summary of the procedure; that is we will give the main steps 
of
    the GP for a transition of type
    $A_{i}\rightarrow B_{j}+\gamma$ where $A_{i}$ is a vector meson (e.g. 
$\rho,\omega,\phi,K^{*0},\bar{K}^{*0},K^{*\pm}$), and
    $B_{j}$ is a pseudoscalar meson (e.g $\pi,\eta,\eta',K^{0},\bar{K}^{0},{K}^{\pm}$).\\
    \indent The matrix element for the transition $A_{i}\rightarrow B_{j}+\gamma$ in the rest 
frame of $A_{i}$ is:
    \begin{equation}
    M_{ji}= \frac{1}{\sqrt{2k\cdot2E_{j}(P)}}\int dt\,\, \exp\,(-ikt)\big\langle 
B_{j}(\mathbf{P})\big\vert \int d^{3}\b\r\,\,
    \exp\,(i\mathbf{k}\cdot\b\r)\mathbf{j}(\b\r,t)\big\vert A_{i}(0)\big\rangle 
\cdot\mathbf{\epsilon}
    \end{equation}
    where $\mathbf{j}(\b\r,t)$ is the quark e.m. current and $\mathbf{\epsilon,k}$ and $k$ 
are the photon polarization, momentum
    and energy; $\vert B_{j}(\mathbf{P})\rangle$ are respectively the true states of the 
pseudoscalar meson with momentum
    $\mathbf{P}$ and of the vector meson at rest (the index $i$ will always refer to the $V$ 
mesons and $j$ to the $P$ mesons);
     $E_{j}(P)$ is the energy of the $P$ meson (until further notice  $V$ is assumed to be
    heavier than $P$); $[2E_{j}(P)]^{-1/2}$ is required by Lorentz invariance, if, as we do, 
we normalize both $\vert B_{j}(\mathbf{P})\rangle$
    and $\vert A_{i}(0)\rangle$ to one meson per unit volume in the rest system of $A_{i}$.\\
    \indent On expressing the exact states $\vert B_{j}(\mathbf{P})\rangle$ and $\vert 
A_{i}(0)\rangle$ (each of which,
     being an exact state, is
    a superposition of an infinite number of Fock states with $q,\bar{q}$ and gluons) as 
$\vert B_{j}(\mathbf{P})\rangle=
    V\vert \Phi_{B_{j}}(\mathbf{P})\rangle$ and $\vert A_{i}(0)\rangle$ = $V\vert 
\Phi_{A_{i}}(0)\rangle$ where $V$ is the
    unitary transformation already introduced repeatedly (transforming in this case the 
($1q,1\bar{q}$) model states $\Phi$ into the exact states),
    the Eq.(82) becomes:
    \begin{equation}
    M_{ji}= \frac{1}{\sqrt{2k\cdot2E_{j}(P)}}\int dt\,\, \exp\,(-ikt)\big\langle 
\Phi_{B_{j}}(\mathbf{P})\big\vert V^{\dagger} \int d^{3}\b\r\,\,
    \exp\,(i\mathbf{k}\cdot\b\r)\mathbf{j}(\b\r,t) V \vert\Phi_{A_{i}}(0)\big\rangle 
\cdot\mathbf{\epsilon}
    \end{equation}
    In Eq.(83) it is:
    \begin{equation}
    \mathbf{j}(\b\r,t)= 
\exp\,[i(Ht-\bG\cdot\b\r)]\,\,\mathbf{j}(0)\,\,\exp\,[-i(Ht-\bG\cdot\b\r)]
    \end{equation}
    where $\bG$ is the momentum operator.\\
    \indent Recall that $\bG$ commutes with $V$, whereas, of course, $H$ does not; this 
corresponds to the fact that the momentum
    (but not the energy) of the model state must be equal to the momentum of the exact state. 
For the exact Hamiltonian $H$ one has:
    \begin{equation}
    H\vert B_{j}(\mathbf{P})\rangle = [P^{2}+M^{2}(B_{j})]^{1/2}\vert 
B_{j}(\mathbf{P})\rangle, \quad H\,\vert A_{i}(0)\rangle=
     M(A_{i})\vert A_{i}(0)\rangle
    \end{equation}
    We repeat that \textit{the index $i$ will always refer to a Vector meson (in this case 
$A_{i}$ (at rest) -mass $M_{i}$-
    and the index $j$ to a $P$ meson ($B_{j}$)}, in this case with energy 
$\sqrt{\mathbf{P_{j}^{2}+M_{j}^{2}}}$\,).
     Inserting Eqs.(83)(84) in Eq.(82) we get:
    \begin{equation}
    M_{ji}= 
\frac{1}{\sqrt{2k\cdot2E_{j}(P)}}(2\pi^{4})\delta^{(3)}(\mathbf{P+k})\delta(M_{i}-k-E_{j}(P))
\langle \Phi_{B_{j}}(\mathbf{P})
    \vert\,V^\dagger \mathbf{j(0)}V\,\vert \Phi_{A_{i}}(0)\rangle \cdot \bepsilon
    \end{equation}.
    \indent The model states of the $P_{j},V_{i}$ mesons are written in the usual way, that 
is the simplest one compatible with the good
    quantum numbers. They are (ignoring the color factor):
    \begin{eqnarray}&&
    \vert\Phi_{A_{i}}(0)\rangle = \vert\chi(A_{i})\varphi(r)\rangle = 
\sum_{\bp}\sum_{\rho_{1}\rho_{2}}\varphi(p)\chi_{\rho_{1}\rho_{2}}
    (A_{i})a^{\dagger}_{\bp,\rho_{1}}b^{\dagger}_{-\bp,\rho_{2}}\vert 0\rangle \\
    &&\vert\Phi_{B_{j}}(\bP)\rangle=\vert\chi(B_{j})\varphi(r)\, exp(i\bP\cdot\bR)\rangle=
    \sum_{\bp}\sum_{\rho_{1}\rho_{2}}\varphi(p)
    \chi_{\rho_{1}\rho_{2}}(B_{j})a^{\dagger}_{\bp+(\bP/2),\rho_{1}}b^{\dagger}_{-\bp+\bP/2,
\rho_{2}}\vert 0\rangle \nonumber
    \end{eqnarray}
    $\vert 0\rangle$ in the above formulas (87) is the bare vacuum of quarks, antiquarks and 
gluons; $a^{\dagger}_{\bp,\rho}$ and
    $b^{\dagger}_{\bp,\rho}$ are creation operators of a quark and, respectively, an 
antiquark (the question of the renormalization
     of their masses has been discussed
    previously); $\bp$ is their momentum and the index $\rho$ characterizes the spin-flavor 
status; $\bP$ is the momentum of the Ps meson.
    For a transition $A_{i}\rightarrow B_{j}+\gamma$ it is (writing $P=\mid \bP \mid$):
    \begin{equation}
    P\equiv P_{ij} = (M_{i}^{2}-M_{j}^{2})/(2M_{j})
    \end{equation}
    In the Eqs.(87) $\rho_{1}$ and $\rho_{2}$ are indices referring to the spin-flavor state 
of the quark (1) and antiquark (2) ($1$ will
    always refer to the quark $q$ and $2$ to the antiquark $\overline{q}$); 
$\chi_{\rho_{1},\rho_{2}}$ are the spin-flavor functions
    (of course with spin $1$ for the Vector mesons and $0$ for the $Ps$ mesons; in spite of the 
fact that the GP is fully relativistic,
    the model functions are constructed with Pauli spinors, as already discussed). We recall 
(Sect.2) that the unitary transformation $V$
    [here this $V$ is not the symbol of the vector meson!] operating on the model states, has 
the property of leading from Pauli to Dirac spinors;
     $\varphi(r)$ and its Fourier transform $\varphi(p)$, equal for all the states (that is 
independent from the indices $i$ and $j$),
    is the (rotation invariant ($L=0$)) space or momentum part of the model wave function. 
For more details compare the Ref.\cite{moVPgam90},
    Sect.III.\footnote{In Ref.\cite{moVPgam90} the notation ($\mathcal{N,P},\lambda$) and the 
expression ``constituent quarks" were used for objects
    that might have been called simply ``quarks" (as we do here, after the analysis of 
\cite{dimo96}). We remarked this already;
    we note it again here for the readers of Ref.\cite{moVPgam90}.}\\
    \indent The most general vertex for a $V\rightarrow P\,\gamma$ decay is:
    \begin{equation}
    G_{ij}\partial_{\alpha}A_{\beta}\partial_{\mu}V_{\nu}P\epsilon_{\alpha\beta\mu\nu}
    \end{equation}
    where $A_{\alpha}$, $V_{\nu}$, and $P$ are the electromagnetic, vector and pseudoscalar 
fields, $\epsilon_{\alpha\beta\mu\nu}$ is the
     Levi-Civita symbol and $G_{ij}$ is a real constant with the dimensions
     of a magnetic moment depending on the $i,j$ pair;
      $G_{i,j}$ = $G_{i,j}(p_{1}^{2},p_{2}^{2},p_{3}^{2})$
     is a Lorentz invariant that can depend only on invariants constructed with the four 
momenta
     $p_{1},p_{2},p_{3}$ of the three external ``legs" of the $V\leftrightarrow P\,\gamma$ 
diagram; because it is $p_{1}+p_{2}=p_{3}$ only
     two such invariants exist, the masses of the $V$ and $P$ mesons, so that:
     \begin{equation}
     G_{ij}\equiv G_{ij}(M_{i}^{2},M_{j}^{2})
     \end{equation}
    Once one has  $G_{i,j}$ expressed as the parametrized expression for the decay 
$i\rightarrow j$ under consideration, the
    rate of the decay $V_{i}\rightarrow P_{j}+\gamma$ is given by:
    \begin{equation}
    \Gamma(A_{i}\rightarrow B_{j}\,\gamma) = G_{ij}^{2}k^{3}/(12\pi)\qquad\qquad\quad 
(V\rightarrow P\gamma)
    \end{equation}
    If $P$ (index $j$) is heavier than $A$ (index $i$) the same formula holds (except for a 
factor 3) (See the remarks on this in the
    Appendix of Ref.\cite{moVPgam90}, after Eq.A5):
    \begin{equation}
    \Gamma(B_{j}\rightarrow A_{i}\,\gamma) = G_{ij}^{2}k^{3}/(4\pi)\qquad\qquad\quad 
(P\rightarrow V\gamma)
    \end{equation}
    \indent Now we should ``summarize" the contents of Ref.\cite{moVPgam90}. As a matter of 
fact this would be too long here
     and we can just give the final result for the parametrization of $G_{ij}$ appearing in 
Eq.(92):
    \begin{equation}
    G_{ij}(M_{i}^{2},M_{j}^{2}) = 
2(\overline{M}_{V}/M_{i})^{1/2}\sum_{\nu=I}^{7}\mu_{\nu}f_{\nu}(P)\Gamma_{\nu}(B_{j}A_{i})
    \end{equation}
    On the r.h.s. the notation is as follows:\\
    \indent 1) $\overline{M}_{V}$ is an average mass of the Vector mesons (irrelevant if we 
consider the ratio between the decay rates of
    two $V$ mesons of the same family),\\
    \indent 2) ${M}_{i}$ is the mass of the decaying $V$ meson,\\
    \indent 3) The $\mu_{\nu}$'s have the dimensions
    of a magnetic moment of the decaying $V$ meson; except for $\mu_{\nu_{1}}$ all the 
$\mu_{\nu}$'s multiply $\Gamma_{\nu}$'s
    that -as we will see- are either reduced by flavor breaking, or reduced by gluon 
exchange. This will play a role when we will
    consider the ratio between two different decay modes (see below),\\
    \indent 4)The seven $\Gamma_{\nu}(B_{j}A_{i})$ will be listed and discussed
    in a moment (compare the Eqs.(59,60) of Ref.\cite{moVPgam90}, where the notation 
$\Pi^{\mathcal{P}}$ etc. was used for what we now call ${P^{u}}$
    etc.). The values of the $\Gamma_{\nu}$'s depend (as the notation indicates) on the 
transition considered (recall that $1$ always refers to the
    quark and $2$ to the antiquark; note also that with the definition 
$Q=(2/3)P^{u}-(1/3)P^{d}-(1/3)P^{s}$ the charge of an antiquark is $-Q$). These
    values of the $\Gamma_{\nu}$'s appear in the Table at the end of this Section for each 
transition $V\rightarrow P\gamma$ or $P\rightarrow V\gamma$.
    \begin{eqnarray}
    \Gamma_{1} & = & (Q_{1}+Q_{2}) \nonumber\\
    \Gamma_{2} & = & (Q_{1}+Q_{2})(P_{1}^{s}+P_{2}^{s}) \nonumber\\
    \Gamma_{3} & = & (Q_{1}-Q_{2})(P_{1}^{s}-P_{2}^{s})\equiv QS \nonumber\\
    \Gamma_{4} & = & (Q_{1}+Q_{2})P_{1}^{s}\cdot P_{2}^{s} \nonumber\\
    \Gamma_{5} & = & [(Q_{1}+Q_{2})\vert z'\rangle \langle z'\vert + \vert z'\rangle z'\vert 
(Q_{1}+Q_{2})] \nonumber\\
    \Gamma_{6} & = & [\vert z'\rangle \langle w\vert(Q_{1}+Q_{2})+(Q_{1}+Q_{2})\vert 
w\rangle\langle z'\vert] \nonumber\\
    \Gamma_{7} & = & [\vert w\rangle \langle z'\vert(Q_{1}+Q_{2})+(Q_{1}+Q_{2})\vert 
z'\rangle\langle w\vert]
    \end{eqnarray}
    In the last form of $\Gamma_{3}$, $Q$ and $S$ stay for the total charge and strangeness. 
The symbols $\vert z'\rangle$ and
    $\vert w \rangle$ stay respectively for:
    \begin{equation}
    \vert z'\rangle = 
(1/\sqrt{3})\vert{u_{1}\overline{u}_{2}+d_{1}\overline{d}_{2}+s_{1}\overline{s}_{2}}\rangle,
    \qquad \vert w\rangle = \vert{s_{1}\overline{s}_{2}}\rangle
    \end{equation}
    and represent flavor structures corresponding to transitions taking place through an 
intermediate gluon (indicated in what follows
    by $g$)\footnote{The accent on $z'$ is to remind that $\langle z'\vert z'\rangle =1$, while 
the $\vert z\rangle$ in
    Sect.8 was normalized to 3.} that, for instance, can occur in processes like:
    \begin{equation}
    \phi\rightarrow 3g\rightarrow \pi^{0}+\gamma, \qquad \phi\rightarrow 2g+\gamma\rightarrow 
\eta+\gamma, \qquad
    \omega\rightarrow 3g\rightarrow \pi^{0}+\gamma, \qquad \omega\rightarrow 
2g+\gamma\rightarrow \eta+\gamma
    \end{equation}
    \indent It should be remarked that the $\Gamma$'s in Eq.(94) are all the flavor 
expressions contained in the exact transition operator
     $V^{\dagger}\mathbf{j}(0)V$; $\Gamma_{5},\Gamma_{6},\Gamma_{7}$ (the gluon exchange 
terms)
     add to the flavor terms $(Q_{i};\,Q_{i}P^{s}_{k};\,Q_{i}\cdot P_{1}^{s}\cdot P_{2}^{s})$ 
contained in
     $\Gamma_{1},\Gamma_{2},\Gamma_{3},\Gamma_{4}$. Forgetting gluon exchange, these would be 
the
     only ones present.\\
     \indent \textit{Here we will limit the discussion to some processes that can depend in a 
clear way from the \textbf{contribution of gluon
     exchange} -compare for this the Table I}.\\
     \indent Let us start with the ratio ($\omega^{0}\rightarrow \pi 
\gamma$)/($\rho^{0}\rightarrow \pi \gamma$)
     discussed at the beginning of this Section as one of the first applications of the NRQM. 
One has from the Eqs.(94) that only $\Gamma_{1}$
     and $\Gamma_{5}$ intervene in $\omega\rightarrow \pi \gamma$ and only $\Gamma_{1}$ in 
$\rho\rightarrow \pi \gamma$:
     \begin{equation}
     \frac{\Gamma(\omega^{0}\rightarrow \pi\gamma)}{\Gamma(\rho^{0}\rightarrow \pi\gamma)}= 
\bigg(3 + \frac{2\mu_{5}f_{5}(P)}
     {\mu_{1}f_{1}(P)}\bigg)^{2}\times 1.06
     \end{equation}
     In the formula (97) above, the factor 1.06 is due to the different momenta in the two 
cases (it comes from the ratio of the third powers
     of the momenta); when the data will improve, this factor should be recalculated because 
of the large width of the the $\rho$.\\
     \indent Any -at present non appreciable- deviation from 9.5 of the r.h.s.of (97) (of course
     after the errors are duly 
taken into account) is a measure of the
      contribution to $\,\omega\rightarrow \pi \gamma$ via gluons
     represented by the term $\mu_{5}f_{5}(P)$. Note that Eq.(97) is an exact consequence of 
any relativistic field theory
     that satisfies the assumptions stated in Sect.1. (In particular it is correct to all 
orders in flavor breaking). Of course QCD belongs
     to such theories and, in a sense, the Eq.(97) provides a confirmation of the smallness 
of these 3-gluons contributions in QCD.\\
     \indent A similar example comes from the ratio of the $\eta'\rho\gamma$ and 
$\eta'\omega\gamma$ decays; these depend only
     on $\mathbf{\Gamma_{1}}$ and on $\mathbf{\Gamma_{5}}\,$, $\mathbf{\Gamma_{6}}$, 
$\mathbf{\Gamma_{7}}\,$, all implying the intervention of $3$
      gluons, the last two processes with first order flavor breaking. Also here the 
experimental error is large, but the indications are for
     a negligible contribution from the gluonic diagrams. Finally a strong confirmation of 
the smallness of the gluon diagrams contribution
     comes from smallness of the $\phi\rightarrow \pi^{0}\gamma$ decay. The order of 
magnitude of this decay can be reproduced by the small deviation
     ($\approx 1^{0}$) of the Vector meson mixing angle $\theta_{V}$ in the $\phi$ meson from 
its ideal value ($\theta_{V}=35.3^{0}$).
     The theoretical uncertainty, particularly from the form factor, is significant; still, 
because a sensible estimate of the order
     of magnitude of the rate $\phi\rightarrow \pi^{0}\gamma$ is obtained without invoking 
the gluon
     annihilation diagrams appearing in the ninth row of table I, there is no evidence, 
inside the errors,
     for the relevance of such diagrams. Even assuming that the deviation of $\theta_{V}$ 
from its ideal value is $2^{0}$, the gluon contributions,
     if present, cannot certainly be larger (indeed should be smaller) than the tiny value 
calculated with this $2^{0}$ deviation.\\

     \textbf{TABLE I.} The values of $\Gamma_{\nu}(B_{i},A_{j})$ in Eq.(94). The 
abbreviations are indicated at the bottom of the table;
      the flavor wave function of each meson is assumed to be normalized to one; $\theta_{V}$ 
is taken to have its ideal value,
      except in the calculation of $\Gamma_{1}$ for $\phi\rightarrow \pi^{0}\gamma$.\\ \\
     {\footnotesize
      \begin{tabular}{lccccccc}

      \hline\hline
      &$\Gamma_1$&$\Gamma_2$&$\Gamma_3$&$\Gamma_4$&$\Gamma_5$&$\Gamma_6$&$\Gamma_7$\\
      \hline
      $\rho\pi\gamma$ &1/3 &0&0&0&0&0&0\\
      $\omega\pi\gamma$ &1&0&0&0&2/3&0&0\\
      $\rho\eta\gamma$ &$K$ &0&0&0&$-s\sqrt{2/3}$&0&$-H\sqrt{2/3}$\\
      $\omega\eta\gamma$ &$K/3$ &0&0&0&$L\sqrt{2/9}$&$H\sqrt{8/27}$&$-H\sqrt{2/27}$\\
      $\rho\eta'\gamma$ &$H$ &0&0&0&$c\sqrt{2/3}$&0&$K\sqrt{2/3}$\\
      $\omega\eta'\gamma$ &$H/3$ &0&0&0&$N\sqrt{2/9}$&$-K\sqrt{8/27}$&$K\sqrt{2/27}$\\
      $\phi\eta\gamma$ &$2H/3$ &$4H/3$&0&$2H/3$&$N\sqrt{2/9}$&$R$&$W$\\
      $\phi\eta'\gamma$ &$-2K/3$ &$-4K/3$&0&$-2K/3$&$-L\sqrt{2/9}$&$Z$&$T$\\
      $\phi\pi^0\gamma$ &$\sin(\theta^*_V-\theta_V)$ &0&0&0&$\sqrt{2/9}$&0&$\sqrt{2/3}$\\
      $K^{*0}K^0\gamma$&-2/3&-2/3&0&0&0&0&0\\
      $K^{*+}K^+\gamma$&1/3&1/3&1&0&0&0&0\\
\hline
\multicolumn{5}{l}{$s=\sin\theta_P$ ; $c=\cos\theta_P$}\\
\multicolumn{2}{l}{$K=1/\sqrt{3}(c-s\sqrt{2})$ ;}&
\multicolumn{2}{l}{$H=1/\sqrt{3}(s+c\sqrt{2})$ ;}& 
\multicolumn{2}{l}{$N=1/\sqrt{3}(c+s\sqrt{2})$ ;}&
\multicolumn{2}{l}{$L=1/\sqrt{3}(c\sqrt{2}-s)$ }\\
\multicolumn{2}{l}{$R=\frac{2}{9}(4s+c\sqrt{2})$ ;}&
\multicolumn{2}{l}{$T=\sqrt{2}/9(5s-c\sqrt{2})$ ;} &
\multicolumn{2}{l}{$W=\sqrt{2}/9(5c+s\sqrt{2})$ ;}&
\multicolumn{2}{l}{$Z=\frac{2}{9}(-4c+s\sqrt{2})$} \\
\hline\hline
\end{tabular}\bf
}\\ \\
Before concluding this section we note that an impressive amount of work, both experimental 
and theoretical, has been done in the last decade
by the groups of Achasov et al. and Benayoun et al. and by others; unfortunately it is 
impossible to discuss in detail all this; even limiting to the
$\gamma$ decays of Vector mesons, we had to be necessarily most synthetic in the presentation 
of our work \cite{moVPgam90}). However we must
cite some papers to refer at least to part of the work mentioned above. For the contributions of 
Benayoun et al. we refer to: \cite{benadb99, benapr99,
bena01, bena00}. For Achasov et al. \cite{Acha00, Acha06}. Compare also \cite{akh01} and 
other contributions by the same group.

\vskip 30pt
     \noindent \textbf{10. The baryons electromagnetic mass differences and the 
Coleman-Glashow equation.} \\

     Another test of the hierarchy \cite{dimoplb00} is provided by the Coleman-Glashow 
formula \cite{cg61} for the baryon octet mass differences.
     (Below we indicate again \textit{the masses} with the baryon symbols). The 
Coleman-Glashow formula is:
    \begin{equation}
      p - n = \Sigma^{+} - \Sigma^{-} +\Xi^{-} - \Xi^{0}
    \end{equation}
    The present data (after a comparatively recent measurement of the $\Xi^{0}$ mass 
\cite{NA48}) give:
    \begin{equation}
    l.h.s= -1.29\, MeV\qquad\qquad\qquad r.h.s.= -1.58\pm 0.25\, MeV
    \end{equation}
    Because the mass difference $(\Sigma^{-} - \Sigma^{+})$ in (98) is $\approx 8\, MeV$, the 
agreement is amazing [before
    the measurement in Ref.\cite{NA48} it was already good ($1.29$ to be compared with 
$1.67\pm 0.6$)]. Note that the Coleman-Glashow formula was
    derived (Ref.\cite{cg61}) assuming unbroken flavor; but the $SU(3)$ violation in the 
baryon octet masses is significant
    $[(M_{\Xi}-M_{p})/(M_{\Xi}+M_{p})\approx 17\%$]. Here we will examine how, due to the 
hierarchy, this violation has a rather small effect.\\
    \indent Note again that it was once more the hierarchy, to produce the improved Gell 
Mann-Okubo
     mass formula \cite{mopl92} of Sect.1 (Eq.1), with an excellent agreement with the 
data.\\
    \indent In this Section we shall consider also some relations due to Gal and Scheck  
(though there the experimental errors are much larger).\\
    \indent Consider now the Coleman-Glashow (CG) equation. As mentioned, in their derivation 
Coleman and Glashow neglected entirely
    the flavor breaking of the strong interactions. But it was shown in Ref.\cite{moem92} 
that the CG formula holds also taking
    into account \textit{all} the flavor breaking terms, except those with three quark 
indices (negligible because of the hierarchy).\\
    \indent Here we summarize the derivation of the CG equation taking into account flavor 
breaking, without performing all the calculations,
     (see \cite{dimoplb00}).\\
     \indent It must be underlined that the $m_{u}-m_{d}$ terms and
     the so called Trace terms were discussed in Sect.4 of Ref.\cite{dimoplb00}; they do not 
affect the CG
    equation. (Incidentally we remark that the $m_{u}-m_{d}$ terms for the generalized
    Gell Mann-Okubo mass formula (Eq.(1)) were duly taken into account in 
Ref.\cite{mopl92}).\\
    \indent To derive the CG equation using the GP, call $\Omega$ the exact QCD operator -to 
2nd order in the charge-
    expressed in terms of the quark fields; $\Omega$ will represent the e.m. contribution of 
interest to the baryon mass,
     that is the e.m. current-current interaction of the quarks in a baryon (proportional to 
their charge-charge interactions);
     with respect to the original Coleman-Glashow paper, the new thing is that we take into 
account the flavor breaking contributions.\\
    \indent Call, as usual, $\vert \Psi_{B}\rangle$ and $\vert \Phi_{B}\rangle$ the exact and 
model states of the baryon $B$, writing, once more,
    in the usual notation: $\vert \Phi_{B}\rangle = \vert X_{L=0}\cdot W_{B}\cdot 
S_{c}\rangle$ (and omitting from now on the
    color factor $S_{c}$). In Ref.\cite{moem92} (Eqs.13-16) the e.m. interactions were 
written explicitly and discussed.\\
    \indent They lead to the terms listed below (because the $\Omega^{-}$ has no role in the 
CG equation, the last line is useless
    here, but we included it for completeness):\\
    $Q_{i}^{2},\qquad Q_{i}Q_{k}$\hspace{9cm} (no flavor breaking)\\
    $Q_{i}^{2}P_{i}^{s},\qquad Q_{i}^{2}P_{k}^{s},\qquad Q_{i}Q_{k}P_{i}^{s},\qquad 
Q_{i}Q_{k}P_{j}^{s}$,\hspace{2.5cm} (1st order flavor breaking)\\
    $Q_{i}^{2}P_{i}^{s}P_{k}^{s},\qquad Q_{i}^{2}P_{k}^{s}P_{j}^{s},\qquad 
Q_{i}Q_{k}P_{i}^{s}P_{k}^{s}
    \qquad Q_{i}Q_{k}P_{j}^{s}P_{k}^{s}$,\hspace{0.5cm} (2nd order flavor breaking)
    $Q_{i}^{2}P_{1}^{s}P_{2}^{s}P_{3}^{s},\qquad 
Q_{i}Q_{k}P_{1}^{s}P_{2}^{s}P_{3}^{s}$,\hspace{5cm} (3rd order flavor breaking)\\ \\
    \indent As to the spin dependence, we proved in \cite{mo89} that only the following 
scalars exist:
    \begin{equation}
    1; \qquad (\bsigma_{i}\cdot\bsigma_{k}) \qquad\qquad\qquad (i,k=1,2,3)
    \end{equation}
    We recall that the scalar $(\bsigma_{1}\times \bsigma_{2})\cdot \bsigma_{3}$ has 
vanishing expectation value on a spin flavor state having a real
    wave function $W_{B}(1,2,3)$. [The  $W_{B}$'s are the usual spin-unitary spin functions 
of the baryons $B$.]\\
    \indent After this list of charge, flavor and spin functions, we display the results of 
the $V$ transformation:
    \begin{equation}
    \langle \Psi_{B}\vert \Omega \vert\Psi_{B}\rangle = \langle \Phi_{B}\vert V^{\dagger} 
\Omega V\vert\Phi_{B}\rangle
    \equiv\langle W_{B}\vert\widetilde{\Omega}\vert W_{B}\rangle
    \end{equation}
    where :
    \begin{equation}
    \widetilde{\Omega}=\sum_{\nu}t_{\nu}\Gamma_{\nu}(s,f) \qquad \mathrm{with:} \qquad 
t_{\nu}\equiv \langle X_{L=0}\vert G_{\nu}(\b\r)\vert X_{L=0}\rangle
    \end{equation}
     The $\Gamma_{\nu}$'s depend only on the spin and flavor variables of the quarks in 
$\Phi_{B}$; the
     $t_{\nu}$'s are a set of parameters coming from the integration of the space factors 
$G_{\nu}(\b\r)$ of $\widetilde{\Omega}=
     \sum_{\nu}G_{\nu}(\b\r)\Gamma_{\nu}(s,f)$ on the space part $X_{L=0}$ of the baryon 
model factor.
     The Eqs.(101),(102) reproduce the usual procedure of the GP. As mentioned,
     the $m_{u}-m_{d}$ and Trace terms do not affect the CG relation; thus below we 
transcribe simply
     (in the Eqs.(103)) the $\Gamma_{\nu}$ at zero order (call them $\delta_{0}B$) and at 
first order ($\delta_{1}B$) in flavor breaking;
     Note that the three quark terms with no flavor breaking \textit{are already included 
-that is,  they are taken
    into account-} in the original CG (no flavor breaking) formula.\\
    \indent In this case (no flavor breaking) define:
    \begin{eqnarray}
    \Gamma_{1}=\sum[Q_{i}^{2}],\qquad\Gamma_{2}=\sum[Q_{i}^{2}(\bsigma_{i}\cdot\bsigma_{k})],
    \qquad\Gamma_{3}=\sum[Q_{i}^{2}(\bsigma_{k}
    \cdot\bsigma_{j})],\qquad \Gamma_{4}=\sum[Q_{i}Q_{k}],\nonumber\\  
\qquad\Gamma_{5}=\sum[Q_{i}Q_{k}(\bsigma_{i}\cdot\bsigma_{k})],
    \qquad\qquad\Gamma_{6}=\sum[Q_{i}Q_{k}(\bsigma_{i}+\bsigma_{k})\cdot \bsigma_{j}] \qquad 
\qquad
    \end{eqnarray}
    In Eq.(103) the sum symbols referring to terms with 1,2,3 indices  are defined 
respectively as:
    \begin{equation}
    \sum[i]= \sum^{3}_{i=1},\qquad \sum[i,k]= (1/2)\sum^{3}_{i,k=1\,(i\neq 
k)},\qquad\sum[i,k,j]= (1/6)\sum^{3}_{i,k,j=1\,(i\neq k\neq j)}
    \end{equation}
    We display also the $\Gamma$'s at first order in flavor breaking. (The Ref.(\cite{moem92} 
contains the list of $\Gamma$'s at 2nd order
    flavor breaking). Of course the 1st order $\Gamma$'s are present only in 
$\Lambda,\Sigma,\Sigma^{*},\Xi,\Xi^{*},\Omega$.\\
    \begin{eqnarray*}&&
    \Gamma_{7}=\sum[Q_{i}^{2}P_{i}^{s}]; \quad\quad \,\,\, 
\Gamma_{8}=\sum[Q_{i}^{2}P_{i}^{s}(\bsigma_{i}\cdot \bsigma_{k})];\quad\quad
    \Gamma_{9}=\sum[Q_{i}^{2}P_{i}^{s}(\bsigma_{j}\cdot \bsigma_{k})];\quad\quad\quad\\&& 
\Gamma_{10}=\sum[Q_{i}^{2}P_{k}^{s}];\quad
    \quad\Gamma_{11}=\sum[Q_{i}^{2}P_{k}^{s}(\bsigma_{i}\cdot\bsigma_{k})];\,\quad\quad 
\Gamma_{12}=\sum[Q_{i}^{2}P_{k}^{s}(\bsigma_{i}+\bsigma_{k})
    \cdot \bsigma_{i}]\quad\\&&\Gamma_{13}=\sum[Q_{i}Q_{k}P_{i}^{s}]; 
\quad\Gamma_{14}=\sum[Q_{i}Q_{k}P_{i}^{s}(\bsigma_{i}\cdot\bsigma_{k})];\quad
    \Gamma_{15}=\sum[Q_{i}Q_{k}P_{i}^{s}(\bsigma_{i}+\bsigma_{k})\cdot\bsigma_{i}]\quad\\&&
    \Gamma_{16}=\sum[Q_{i}Q_{k}P_{j}^{s}];\quad
    \Gamma_{17}=\sum[Q_{i}Q_{k}P_{j}^{s}(\bsigma_{i}\cdot\bsigma_{k})];\quad\Gamma_{18}=\sum[
Q_{i}Q_{k}P_{i}^{s}(\bsigma_{i}+\bsigma_{k})
    \cdot\bsigma_{j}]\quad
    \end{eqnarray*}
    \indent In the above list of 1st order flavor breaking terms, the three quark terms 
$\Gamma_{9},\Gamma_{12},\Gamma_{15}$ do not contribute
    to the left and right hand side of the CG formula; the terms 
$\Gamma_{16},\Gamma_{17},\Gamma_{18}$, do no contribute to $n$ and $p$ while
     the correction for $\Sigma^{+}$, $\Sigma^{-}$, $\Xi^{-}$, $\Xi^{0-}$ are in each case of 
the order ``something"/3 where
     the magnitude of ``something" is estimated by the hierarchy
     as $1/9$; this is due, in all cases, to the product of a reduction factor $\approx 1/3$ 
-presence of a $P^{s}$- and a factor
     $\approx 1/3$, due to one more gluon exchange. Thus for each of the above three terms 
$\Gamma_{16},\Gamma_{17},\Gamma_{18}$ we have a reduction
     of the order $(1/3)^{3}\simeq 4.10^{-2}$ with respect to the dominant no-flavor breaking 
contribution. (Note: The $\Gamma_{18}$ here is $1/2$
     that listed in Ref.\cite{moem92}; this was wrong by a factor 2 -but produced no error 
because, in Ref.\cite{moem92} it was not used).\\
     \indent Because experimentally $\Sigma^{-}-\Sigma^{+}\simeq 8\,MeV$ and 
$\Xi^{-}-\Xi^{0}\simeq 6.4\,MeV$ the reduction given above
     $(1/3)^{3}\simeq 4.10^{-2}$ implies an expected difference between the left and right 
hand sides of the
     C.G. formula (due to 1st-order flavor breaking terms), of $\approx 0.2\div 0.3 \, MeV$ that 
does not disagree with the data.\\
    \indent  The $\Gamma$'s necessary to calculate the 2nd order flavor breaking correction 
are given in \cite{moem92}, Eq.(19).
    The order of magnitude of the terms of interest are estimated using the hierarchy. \\
     \indent Though we do not list here the 2nd order flavor breaking terms, we mention that 
the estimate of their contribution to the difference
    between the left and right hand sides of the CG formula is $\approx 0.02-0.1\, MeV$.\\ \\
    \indent We finally note the Gal-Scheck relations, derived long ago using the NRQM; again 
the hierarchy
    plays an important role (for more details see the Ref.\cite{moem92}).\\
    \indent The Gal and Scheck relations considered here are those for the baryon masses; 
they were
    derived from the NRQM assuming that 3-quark terms were negligible. Three of these 
relations deal with
    the masses of wide resonances (the $\Delta$'s); thus they are not easily verifiable. Two, 
the only ones to be
    displayed below, imply resonances not so wide ($\Sigma^{*\pm}$ and $\Xi^{*-,0}$) and can 
be checked more easily.
    They are:
    \begin{eqnarray}&&
    (\Sigma^{*+}- \Sigma^{*-})+ (\Xi^{*-}-\Xi^{*0})= p - n  \qquad\qquad   (-1.2\pm 0.9= 
-1,29) \\
    &&(1/2)(\Sigma^{*+} + \Sigma^{*-}) - \Sigma^{*0}=(1/2)(\Sigma^{+} + \Sigma^{-}) - 
\Sigma^{0}\qquad (1.3\pm 1.2 = 0.85\pm 0.12)\nonumber
    \end{eqnarray}
    The above relations hold to all orders in flavor breaking.\\
    \indent Finally (independently of the Gal Scheck relations), to stimulate more precise 
measurements,
    we write a relation between the baryon electromagnetic masses \cite{dimoplb00}, which is 
the analogous for the decuplet of
    the Coleman Glashow equation for the octet. It can be easily verified using the
    Eqs.(27)(28) of Ref.\cite{moem92}; it holds to all orders in flavor breaking:
    \begin{equation}
    \delta\Delta^{+} - \delta\Delta^{0} = 
\delta\Sigma^{*+}-\delta\Sigma^{*-}+\delta\Xi^{*-}-\delta\Xi^{*0}
    \end{equation}
    The above Eq.(106) (as well as the one below) might have been written -more simply- 
suppressing all the $\delta$ symbols,
    since they are independent of the $u,d$ quark mass differences. We kept however the 
notation used in Ref.\cite{dimoplb00}.\\
    \indent The Eq. (106), plus the Equation: $\delta\Delta^{++} -\delta\Delta^{-}= 
3(\delta\Delta^{+}-\delta\Delta^{0})$
    (also true to all orders in flavor breaking) might be useful for determining the mass 
differences between the $\Delta$'s.
    \vskip 30pt
    \textbf{11.Two relations: a) Between the charge radii of $\bp,\bn$,$\mathbf{\Delta^{+}}$; 
b)Between the radii
     of $\bpi^{+},\mathbf{K}^{+}$,$\mathbf{K}^{0}$.}\\

    \textit{1.-On the charge radii of $\,\mathbf{p,n,\Delta^{+}}$.}\\
    Buchmann, Hernandez and Faessler \cite{buch97} derived, using an elaborate
    quark model including two body gluon and pion exchange, the following relation between 
the electric charge radii of proton, neutron and $\Delta$:
    \begin{equation}
    r^{2}(p) - r^{2}(n) = r^{2}(\Delta^{+})
    \end{equation}
    We will show that the relation (107) can be reproduced using the GP if one neglects terms 
with three indices and a closed loop contribution
    (a Trace term); these are indeed absent in the model of Buchmann et al. and are expected 
to be small in the GP due to the hierarchy.\\
    \indent The quantities $r^{2}(p),\, r^{2}(n),\, r^{2}(\Delta^{+})$ calculated in the rest 
frame of the baryon considered are scalars  under
    space rotations. Indeed the charge square radius of the baryon $B$ is defined as:
    \begin{equation}
    r^{2}(B) = \langle W_{B}\vert \Big[parametrized\,\,r^{2}\Big]\vert W_{B}\rangle
    \end{equation}
     where the $W_{B}$ are the standard spin-flavour functions, used previously 
(Sects.2,3).\\
     \indent The most general $\,parametrized\,\, r^{2}$ for $p,n,\Delta^{+}$ is linear 
\cite{dimoplbuch} in the quark charges $Q$, and
     it can only contain rotation invariant spin expressions of the type 
$(\bsigma_{i}\cdot\bsigma_{j})$. As to the projection operator
     $P^{s}$ this can only be contained, for $p,n,\Delta^{+}$, in the term $Tr[QP^{s}]$ 
arising from closed internal loops. Thus the most general form
     of [$parametrized\,\,r^{2}$] is:
     \begin{equation}
     [parametrized\,\,r^{2}]= A\sum_{i}Q_{i} +B\sum_{i\neq 
k}Q_{i}(\bsigma_{i}\cdot\bsigma_{k}) +C\sum_{i\neq j\neq k}Q_{i}(\bsigma_{j}\cdot
     \bsigma_{k}) +D\,Tr[QP^{s}]
     \end{equation}
     where $A,B,C,D$ are four parameters. (As noted repeatedly, the scalar 
$(\bsigma_{i}\times\bsigma_{j}\cdot\bsigma_{k})$
     cannot be present in Eq.(109)). Note that in principle there should be two different 
$D$'s in Eq.(108) multiplying $Tr[QP^{s}]$
     for $n,p$ and for $\Delta$; but because the coefficient $D$ is negligible in all cases, 
we wrote \cite{dimoplbuch} the Eq.(109)
     introducing just one $D$ (there is of course no difficulty in writing the correct 
expression).\\
     \indent With a few steps one finally obtains, calculating
     the expectation values of the appropriate expressions on $p,n,\Delta$:
     \begin{equation}
     r^{2}(p) = A -3C -D/3; \qquad r^{2}(n) = -2B +2C -D/3; \qquad r^{2}(\Delta^{+}) = A + 2B 
+C -D/3;
     \end{equation}
     Note the following: If we only take into account in Eq.(110) the $A$ (additive) and $B$ 
(two index) terms we get:
     \begin{equation}
     r^{2}(p) = A ; \qquad r^{2}(n) = -2B ; \qquad r^{2}(\Delta^{+}) = A + 2B;
     \end{equation}
     This means that if the NRQM result $\Big(r^{2}(p) = A; \,r^{2}(n) = 0;\, 
r^{2}(\Delta^{+}) = A \Big)$ is corrected by the two index terms
     we get the Buchmann et al. result $r^{2}(p) - r^{2}(n) = r^{2}(\Delta^{+})$. If now we 
keep also the 3-index terms having
     $C$ as coefficients as well as the $D$ terms $r^{2}(p) - r^{2}(n) = r^{2}(\Delta^{+})$ 
is replaced by:
     \begin{equation}
     r^{2}(p) - r^{2}(n) = r^{2}(\Delta^{+}) -6C + D/3
     \end{equation}
     The coefficient D characterizes the contributions of internal quark closed loops with 
the probe photon line (the photon exchanged
     in order to measure the radii) ending on the loop. The reduction factor is around 30 
(see the end of Sect.7)
      so that the $D$ term is negligible.\\
     \indent As to the value of $C$, we have $C/B\approx 1/3$. Because, from the experimental 
knowledge of the radii of $n,p$ we get,
      $\vert B/A\vert \cong 0.08$, and, from the hierarchy we expect $6\vert C/A \vert \cong 
0.16$ we obtain:
     \begin{equation}
     r^{2}(p) - r^{2}(n) \cong r^{2}(\Delta^{+})(1\pm 0.16)
     \end{equation}
     We conclude that the main result found in the model of \cite{buch97} discussed above is 
correct in QCD except for terms possibly
     of order $10\%$ to $20\%$. \\
     \indent As to the derivation from the same model of the quadrupole moment of $\Delta$ 
and of the
     $\Delta\rightarrow p\gamma$  quadrupole transition, they do not appear to follow from 
QCD, as far as we can see.
     (see the remarks in Sect.1 and 4 of Ref.\cite{dimoplbuch}).\\
     \indent Following the results described above, Buchmann (with Henley, in some papers) 
used the GP to treat several problems -
     mainly on the e.m. properties
     of hadrons. A partial list of contributions is: Refs.\cite{buch00, buchnp00, 
buch01,buch02}; we do not share however the point of view of Buchmann
     on the quasi-equivalence between the GP and the large $N_{c}$ method underlying some of 
these papers (compare Sect.13).\\
     \textit{2.-On the charge radii of $\,\mathbf{\bpi^{+},\,K^{+}, \,K^{0}}$.}\\
      For $\pi^{+},\,K^{+}\,K^0$ the calculation \cite{dimoepl01}
     is even simpler than the above
     one because we are dealing with zero-spin mesons. As to the experimental values of the 
radii listed above their errors are still
     comparatively large, except (recently) for the $K^{0}$.\\
     \indent The square radius $r^{2}(M)$ of a meson $M$ with e.m. form factor $F(q^{2})$ is:
     \begin{equation}
     r^{2}(M)= -6\frac{dF(q^{2})}{dq^{2}}\Big \vert_{q^{2}=0};\qquad F(q^{2})=\langle 
M(\mathbf{q}/2)\vert \rho(0)\vert M(-\mathbf{q}/2)\rangle
     \end{equation}
     Here $\vert M(\mathbf{p})\rangle$ is the exact eigenstate of the QCD Hamiltonian for the 
meson $M$ with total momentum
     $\mathbf{p},\,\rho(0)\,=\,i \overline{\psi}(0)Q\gamma_{4}\psi(0)$, with $\psi(x)$ the 
quark field and $Q=(1/2)(\lambda_{3}+(1/3)\lambda_{8})$
     is the charge operator. {We recall, incidentally, that the $q^{2}$ dependence of the 
e.m. form factors was given for $p,n$ in Ref.\cite{dimoplesc}.
     The exact $r^{2}(M)$ derived from QCD for a meson with an $L=0$ auxiliary state (as the 
lowest pseudoscalar mesons) is, in the GP:
     \begin{equation}
     r^{2}(M)= \langle W_{M} \vert``parametrized\,\, r^{2}"\vert W_{M}\rangle
     \end{equation}
     where $W_{M}$ are the standard spin-flavor functions of the $\pi$ or $K$ mesons. Due to 
the linearity of $r^{2}$ in $\rho(0)$ the most general
     $``parametrized\,\, r^{2}"$ for $\pi^{+},K^{+},K^{0}$ is a scalar linear in the quark 
charges $Q_{i}\,(i=1,2)$ (1=quark, 2= antiquark).
     As to the spins, the $``parametrized\,\, r^{2}"$ can contain only 
$\bsigma_{1}\cdot\bsigma_{2}$ which, applied to the spin singlet factor in
     $W_{M}$ is just $-3$. Therefore the most general $``parametrized\,\, r^{2}"$ (a scalar 
under rotations)is:
     \begin{equation}
     ``parametrized\,\, r^{2}" = A\sum_{i}Q_{i} +B\sum_{i} Q_{i}P^{s}_{i}+ C\sum_{i\neq 
k}Q_{i}P^{s}_{k} +D Tr[QP^{s}]
     \end{equation}
     In Eq.(116) $A,B,C,D$ are four real parameters; the sums (on $i,k=1,2$) are exemplified 
by the following case for $K^{+}$:
     $\sum_{i\neq k}Q_{i}P^{s}_{k}$=(for $K^{+}$) $\langle W_{K^{+}}\vert 
Q_{1}P^{s}_{2}+Q_{2}P_{1}^{s}\vert W_{K^{+}}\rangle = 2/3$.
     Neglecting the Trace term in Eq.(116) for the reasons already stated (depressed by the 
Furry theorem at least 30 times with respect to the
     dominant term $A$),
      the equation (116) contains three parameters $A,B,C$ and must fit three quantities.  
$\vert B/A \vert$ = the flavor reduction factor =
     0.3 to 0.33. As to $\vert C/B \vert$, this is governed by the one gluon exchange 
reduction factor factor (from 0.22 to 0.37)
     for which we also take $1/3$ although this value is less ``universal" than the previous 
(flavor) one. In conclusion we set:
     \begin{equation}
     \vert C \vert \approx (1/3)|B| \cong (1/9)|A|
     \end{equation}
     We thus obtain:\\
     \begin{eqnarray}&
     r^{2}(\pi^{+})= A\qquad\qquad\qquad(=0.44\pm 0.02)&\\
     r^{2}(K^{+})= & A+(1/3)B +(2/3)C\qquad (=0.34\pm 0.05)& \nonumber \\
     r^{2}(K^{0})=&(1/3)B - (1/3)C\qquad \qquad \qquad \qquad& \nonumber
     \end{eqnarray}
     \indent In general the Eqs.(118) lead to the relation:
     \begin{equation}
      r^{2}(\pi^{+})-r^{2}(K^{+})= -r^{2}(K^{0})- C \qquad\qquad (0.10\pm 0.05 = 0.077\pm 
0.007 \pm 0.05)
     \end{equation}
     Experimentally $r^{2}(\pi^{+})-r^{2}(K^{+})= 0.10\pm 0.05$; on the left hand side 
$r^{2}(K^{0})=-0.077\pm 0.007$ and $C$ (calculated from the
     hierarchy, Eq.(117)) is $\pm 0.05$; more precise measurements for 
$r^{2}(\pi^{+}),r^{2}(K^{+})$ would be useful and one would like to have
     a confirmation of the very precise value of $r^{2}(K^{0})$.
     The interest in obtaining precise values is due also to the following circumstance. From 
the Eqs.(118) one has:
     \begin{equation}
     \vert B/A \vert = \vert r^{2}(K^{+})-r^{2}(\pi^{+})+2r^{2}(K^{0})\vert/r^{2}(\pi^{+})= 
\vert -0.10\mp 0.05 -0.154\pm 0.014\vert/(0.44\pm 0.01)
     \end{equation}
     \indent It is important to note that the hierarchy leads to $\vert B/A \vert\cong 1/3$; 
this prediction should be rather solid because
     based only on flavor breaking (it is also independent of $C$). However with the values 
that appear in Eq.(120) this expectation is fulfilled
     only using the $\pm 0.05$ error at two standard deviations or/and if the error in 
$r^{2}(K^{0})$ has been underestimated in Ref.\cite{abou06}.
     \vskip 30pt
      \noindent \textbf{12. Parametrization of the $\rho\gamma, \,\omega\gamma$ and 
$\phi\gamma$ couplings:
     Why $f_{\rho\gamma}$:$f_{\omega\gamma}$ differs from 3:1 only by $\sim 10\%$ in spite of 
flavor breaking?}\\

     This Section is based on Ref.\cite{dimo94}. Some numbers appearing in the original 
treatment are modified
      but the conclusion for the $\rho\gamma, \,\omega\gamma$ couplings summarized by the 
title above, is practically the same
      as that of \cite{dimo94}. [However the conclusion of \cite{dimo94} on a negative 
$\phi-\omega$ mixing angle
      was incorrect: the angle is positive, $\approx +1^{0}$]. \footnote{We apologize for the 
mistake on this point.}\\
     \indent To determine the $\rho\gamma, \,\omega\gamma$ and $\phi\gamma$ couplings, we 
analyze below, using the GP,
     the $V-\gamma$ couplings $f_{V\gamma}$ in the decays $\rho^{0},\omega,\phi\rightarrow 
e^{+}e^{-}$. The above $V-\gamma$ couplings
     are, in fact, presented sometimes as an illustration of how well the NRQM + $SU_{3}$ 
predicts the ratios of the
     $f_{V\gamma}$ for $V=\rho^{0},\omega$ and, in part $\phi$. But $\vert f_{\rho 
\gamma}/f_{\omega \gamma}\vert\approx 3$
     needs an explanation, because the value $3$ is the perfect $SU_{3}$ prediction with no 
flavor breaking, whereas
     flavor is broken. Yet we will see, using the GP, that
     the statement is almost correct (see the title of this section) although this is not 
obvious. (From now on in this section the vector mesons
     will be indicated by $v$
     to avoid confusion with capital $V$, the unitary transformation of the GP, transforming 
the model states into the exact QCD states).
     The coupling $f_{v\gamma}$ of a neutral vector meson $v$ to the photon
      is proportional to the matrix element $\langle v\vert j^{\alpha}(x)\vert 0\rangle$, 
where $j^{\alpha}(x)$ is the quark current:
     \begin{equation}
     j^{\alpha}(x)=\frac{ie}{2}[\overline{\psi}(x)(\lambda_{3}+ 
\frac{1}{3}\lambda_{8})\gamma_{\alpha}\psi(x)]
     \label{12.1}
     \end{equation}
     in the usual notation, with the color index omitted and $\psi(x)= u(x),\,d(x),\,s(x)$.\\
     \indent Introducing the charge $Q=\frac{2}{3}P^{u}-\frac{1}{3}P^{d}-\frac{1}{3}P^{s}$ 
the Eq.(\ref{12.1}) becomes
     \begin{equation}
     j^{\alpha}(x)=\frac{ie}{2}\overline\psi(x)Q\gamma_{\alpha}\psi(x)
     \label{12.2}
     \end{equation}
     \indent The decays of $\rho,\omega,\phi$ into $e^{+}e^{-}$ are expressed in terms of the 
quantities $f_{v\gamma}$ defined
     \textit{in the rest system of the decaying v meson}, by
     \begin{equation}
     f_{v\gamma}\delta_{ik}= \langle v^{i}\vert j^{k}(0)\vert\,0\rangle,\qquad \qquad 
i,k=1,2,3
     \label{12.3}
     \end{equation}
     where $\vert v^{i}\rangle$ is the exact state of the QCD Hamiltonian representing the 
vector meson $v$ with polarization $i$
     and $j^{k}$ is the $k$-th component of the quark current (\ref{12.2}); similarly $\vert 
0\rangle$ is the exact vacuum of QCD,
     that is the exact vacuum of quarks, antiquarks and gluons.
      Of course the General Parametrization connects the exact states to the model states by 
the unitary transformation
     $V$, introduced in Sect.2; the structure of $V$ depends on the problem considered, and 
we have to construct $V$ for our problem here.
     First we must clarify the meaning of ``exact state" introduced above. The states  $\vert 
v\rangle$
      and $\vert  0\rangle$  on the r.h.s. of Eq.(\ref{12.3}), when written in terms of Fock 
states in the space of quarks $q$, antiquarks
       $\overline{q}$ and gluons $G$, are a super-position of an infinite number of Fock 
states (compare Sect.2). Schematically,
       \begin{eqnarray}
      \vert v\rangle \equiv \vert v\rangle_{exact}= \vert q\overline{q}\rangle +\vert 
q\overline{q}q\overline{q}\rangle +\vert q\overline{q},G\rangle
      +.... \nonumber \\
      \vert \, 0\rangle \equiv \vert\, 0\rangle_{exact}= \vert\, 0_{bare}\rangle + \vert 
q\overline{q},G\rangle +....
      \label{12.4}
      \end{eqnarray}
      where $\vert\,0_{bare}\rangle$ [perhaps the name $\vert \,0_{model}\rangle$ instead of  
$\vert \,0_{bare}\rangle$ would have been better,
       but we preferred not to change on this the notation adopted in Ref.(\cite{dimo94}), is 
a zero energy state of the model Hamiltonian with no
      quarks, antiquarks or gluons.
       As we exemplified abundantly, the GP starts with the introduction of a model 
Hamiltonian and of a unitary transformation $V$
      that transforms the simple model states of the model Hamiltonian into the complicated 
exact states of the exact QCD Hamiltonian.
      A large freedom exists in the selection of the model Hamiltonian in a given problem, as 
seen for baryons and mesons in
      the Refs.\cite{mo89, mo90, dimo96} and in the past sections of this survey. In the 
present case the model Hamiltonian is chosen as a
       function of $1q-1\overline{q}$; one of its eigenstates is a state of zero energy 
without quarks, antiquarks and
      gluons that in the Eqs.(\ref{12.4}) above we called $\vert \,0_{bare}\rangle$ and in 
the following we will call $\vert \,0_{b}\rangle$.\\
      \indent The model states of mesons for the lowest nonet
      are pure $\vert q\overline{q}\rangle$ states with $L=0$, all degenerate in mass. The 
wave function $F(v^{i})$ of the $\vert q\overline{q}
      \rangle$ Fock state $\vert 1q,1\overline{q},\,model \rangle$ is factorizable:
      \begin{equation}
      F(v^{i})= \varphi_{L=0}(r)\cdot\chi^{i}\cdot C
      \label{12.5}
      \end{equation}
      a product of a space factor $\varphi_{L=0}(r)$, a spin-flavor factor $\chi^{i}$ and a 
color factor $C$.
       (In what follows, to simplify the notation, we will omit writing the factor $C$ and 
the color structure of the e.m. current;
      they lead to a multiplicative factor $\sqrt{3}$ in the matrix element $\langle 
v^{i}\vert j^{k}(0)\vert\,0\rangle$ that will be reinserted
      in Eq.(\ref{12.12}) below). The exact state, obtained from it via the unitary 
transformation $V$, is:
      \begin{equation}
      \vert v \rangle= V\vert 1q-1\overline{q},\,model\rangle
      \label{12.6}
      \end{equation}
      The transformation $V$ has in this case also the task of transforming the model vacuum 
state $\vert\,0_{b}\, \rangle$ into the exact
       vacuum $\vert \, 0 \rangle$
      \begin{equation}
      \vert \,0 \rangle = V\vert\, 0_{b}\rangle
      \label{12.7}
      \end{equation}
      Thus, \textit{in the rest system of the v}, it is:
      \begin{equation}
      V = V_{0}\vert\,0_{b}\rangle \langle 0_{b}\,\vert + V_{1}\sum \vert 
1q,1\overline{q}\rangle \langle 1q,1\overline{q}\vert +....
      \label{12.8}
      \end{equation}
      In the Eq.(\ref{12.8}) above $\vert\,0_{b}\rangle \langle 0_{b}\,\vert$ is the 
projector on the $\vert \,0_{b}\rangle$ vacuum Fock state and
      $\sum \vert 1q,1\overline{q}\rangle \langle 1q,1\overline{q}\vert$ is the projector on 
the states of the $1q,1\overline{q}$
      no-gluon Fock sector. The dots refer to the projectors of $V$ on the other Fock states, 
not relevant here.\\
      \indent The matrix element  $\langle v^{i}\vert j^{k}(0)\vert\,0\rangle$ in 
Eq.(\ref{12.3}) is
      \begin{equation}
       \langle v^{i}\vert j^{k}(0)\vert\,0\rangle= \langle \,model \,1q,1\overline{q}\vert 
\,V^{\dagger}j^{k}(0)V\vert\,0_{b}\rangle=
       \langle F(v^{i})\vert V^{\dagger}j^{k}(0)V\vert\,0_{b}\rangle
       \label{12.9}
       \end{equation}
       On the r.h.s. of (\ref{12.9}) we inserted the state $\vert\,F(v^{i})\rangle$  
corresponding to the wave function $F(v^{i})$ (\ref{12.5})
       of the model state of the $v$ meson:
       \begin{equation}
       \vert\,F(v^{i})\rangle= 
\sum_{\mathbf{p}}\sum_{\rho_{1},\rho_{2}}g(p)\chi^{i}_{\rho_{1},\rho_{2}}
       a^{+}_{\mathbf{p},\rho_{1}} b^{+}_{\mathbf{-p},\rho_{2}}\vert\,0_{b}\rangle
       \label{12.10}
       \end{equation}
       where   $\bp=(1/2)(\bp_{1}-\bp_{2})$ is the relative   momentum of the
       quark 1 and antiquark 2 in the model state. In Eq.(\ref{12.10}) $a^{+}_{\bp,\rho}, 
\,b^{+}_{\bp,\rho}$ are creation operators of a quark
       and, respectively, antiquark with momentum $\bp$ and in the spin-flavor state $\rho$; 
$\rho_{1}$ and $\rho_{2}$ refer to the spin-flavor
       state of the quark 1 and antiquark 2 [$\rho=s,f; \,s=spin,\,f=flavor$]; \textit{here 
and in the following,  1 always refers to the quark
       and 2 to the antiquark.} As to $\varphi_{L=0}(r)$ (the same for all nonet states) and 
its Fourier transform $g(p)$,
       they are the space or momentum factors of the $L=0$ model wave functions.\\
       \indent Inserting (\ref{12.10}) in the last member of (\ref{12.9}) we obtain:
       \begin{equation}
       \langle v^{i}\vert 
j^{k}(0)\vert\,0\rangle=\sum_{\bp}\sum_{\rho_{1},\rho_{2}}g^{*}(p)\chi^{i*}_{\rho_{1},
\rho_{2}}\langle 0_{b}\vert
       a_{\bp,\rho_{1}}b_{-\bp,\rho_{2}}(V^\dagger j_{k}(0)V)\vert\,0_{b}\rangle
       \label{12.11}
       \end{equation}
       The only part of $(V^\dagger j_{k}(0)V)$ contributing to the r.h.s. of (\ref{12.11}) 
has, clearly, the form $G^{k}_{\rho_{1},\rho_{2}}(\bp)
       a^{+}_{\bp,\rho_{1}}b^{+}_{-\bp,\rho_{2}}$, where $G^{k}_{\rho_{1},\rho_{2}}(\bp)$ is 
some function of $\bp$. Using this expression
       of $(V^\dagger j_{k}(0)V)$ we obtain:
       \begin{equation}
       \langle v^{i}\vert j^{k}(0)\vert\,0\rangle= 
\sqrt{3}\sum_{\bp}\sum_{\rho_{1},\rho_{2}}g^{*}(p)\chi^{i*}_{\rho_{1},\rho_{2}}
        G^{k}_{\rho_{1},\rho_{2}}(\bp)
       \label{12.12}
       \end{equation}
        where we reintroduced the factor $\sqrt{3}$ mentioned above (after 
Eq.(\ref{12.5})).\\
        \indent At this stage our aim is to display the most general flavor dependence of the 
right hand side of Eq.(\ref{12.12}). To achieve this
        we must first eliminate the $p$'s summing over $\bp$. Calling:
        \begin{equation}
        \xi^{k}_{\rho_{1},\rho_{2}} = \sum_{\bp}g^{*}(p)G^{k}_{\rho_{1},\rho_{2}}(\bp)
        \label{12.13}
        \end{equation}
        we have
        \begin{equation}
        \langle v^{i}\vert j^{k}(0)\vert\,0\rangle=\sqrt{3}\sum_{\rho_{1},\rho_{2}}
        \chi^{i*}_{\rho_{1},\rho_{2}}\xi^{k}_{\rho_{1},\rho_{2}}
        \label{12.14}
        \end{equation}
        The indices $\rho\equiv s,f$, as already stated, specify both spin $s$ and flavor 
$f$. To obtain the flavor dependence it is convenient
        to use the fact that both $\chi$ and $\xi$ in Eq.(\ref{12.14}) can be written as 
products of a spin and a flavor factor. We can then perform
        the sum over the spins obtaining:
        \begin{equation}
        \sqrt{3}\sum_{s_{1},s_{2}}\chi^{i*}_{\rho_{1},\rho_{2}}\xi^{k}_{\rho_{1},\rho_{2}}= K 
\delta_{ik}\mathcal{F}_{f_{1},f_{2}}
        \label{12.15}
        \end{equation}
        where $K$ is a constant and we have put
        \begin{equation}
        \mathcal{F}_{f_{1},f_{2}}=\langle v_{f_{1},f_{2}}\vert \mathcal{F}\vert\,0\rangle
        \label{12.16}
        \end{equation}
        Here $v_{f_{1},f_{2}}$ is the flavor factor of the model wave function of the meson 
$v$ and $\mathcal{F}$ is the most general
        flavor structure corresponding to the operator $(V^\dagger j(0)V)$ (j(0) here is any 
component of the current). We write
        \begin{equation}
        \mathcal{F}= \sum_{f^{'}_{1},f^{'}_{2}}\vert \,{f^{'}_{1},f^{'}_{2}}\rangle \langle 
f^{'}_{1},f^{'}_{2}\vert\,(V^\dagger
        j(0)V)_{flavor}\vert \,0\rangle \langle 0\vert
        \label{12.17}
        \end{equation}
        where the projector $\sum_{f^{'}_{1},f^{'}_{2}}\vert \,{f^{'}_{1},f^{'}_{2}}\rangle 
\langle f^{'}_{1},f^{'}_{2}\vert$
        refers only to the flavor space
        of $1q,1\overline{q}$ and $(V^{\dagger} j(0)V)_{flavor}$ is the most general operator 
that can result in QCD when $(V^{\dagger}j(0)V)$
        is calculated and integrated over all variables except flavor. In (\ref{12.17}) it is 
of course irrelevant to write $\vert\,0\rangle$ or
        $\vert\,0_{b}\rangle$ because the bare and exact vacuum have the same flavor 
properties.\\
        \indent \textit{The most general expression of} $\mathcal{F}$ in the calculation of 
the
        $v\gamma$ couplings is easily found from the Eqs. given above. It is (as usual, 
$Q=(2/3)P^{u}-(1/3)P^{d}-(1/3)P^{s}$):
        \begin{equation}
        \mathcal{F}= A(Q_{1}+Q_{2})(\vert\,u\overline{u}\rangle + 
\vert\,d\overline{d}\rangle)\langle 0\vert + B(Q_{1}+Q_{2})\vert s\overline{s}
        \rangle \langle 0\vert\, +C Tr(2QP^{s})\cdot 
\vert\,u\overline{u}+d\overline{d}+s\overline{s}\rangle\langle 0 \vert  +h.c.
        \label{12.18}
        \end{equation}
       Here $A,B,C$ are three coefficients; a fourth parameter in front of $s\overline{s}$ in 
the last term of the expression is unnecessary
       because it can be absorbed in the coefficient $B$. Note that there are no terms 
proportional to $(Q_{1}-Q_{2})$ because this quantity
       -not $(Q_{1}+Q_{2})$- is the total charge of the meson, and therefore vanishes for 
neutral states. Also we introduced a factor 2
       in the $Trace$ term (not present in \cite{dimo94}) to normalize all terms properly. 
Clearly $Tr(2QP^{s})\equiv-2/3$, but we left,
        so far, this operator in this form to keep track of its origin due to a closed loop  
of type shown in Fig.3.

\begin{figure}
    \begin {center}\includegraphics[width=8cm,height=4cm]{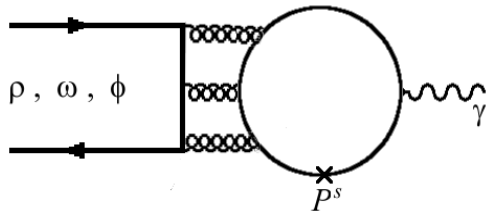}
   \end{center}
   \caption{\footnotesize{ The closed loop three gluons contribution to the $\rho^0\gamma ,\ 
\omega\gamma ,\ \phi\gamma$ couplings. The loop may be circled by a quark  of flavor $u$, $d$ 
or $s$. The net result is nonvanishing because of flavor breaking, as indicated by the 
operator $P^s$, and  corresponds to the term $CTr(2 QP^s)$ in Eq.(\ref{12.18}).}}
   \end{figure}
        \indent On circling the closed loop we must meet, for a $v$ meson, in addition to the 
photon vertex, at least three gluon-quark
        vertices. Three is due to color and to spin 1 (or to Furry theorem).
        We stress (this is in fact the basic point of the present discussion) that 
(\ref{12.18}) is the most general parametrization
         that can emerge from an exact complete QCD calculation. Nothing more complicated 
than this can be present.\\
       \indent At this point one natural question is: So what? There are three parameters and 
three mesons ($\rho,\omega,\phi$) and, without further
       information, nothing useful can be predicted. However we expect that, due to the 
hierarchy (Sect.6), the $Trace$ term
       in (\ref{12.18}) contributes much less than the others; therefore, it may be a fair 
approximation to neglect it in a first approximation.
       We will see that the data on the ($f_{\rho\gamma}$/$f_{\omega\gamma}$) ratio confirm 
this point and allow to explain why, in spite of
       flavor breaking, the above ratio is near to $3$.\footnote{In Ref.\cite{dimo94} we 
estimated the ratio $C/A\approx 10^{-2}$; as we will see
       it is instead $\approx 5\cdot 10^{-2}$, but this difference does not change the 
conclusion just stated.} Note that
       for $C=0$ (and in the absence of $\rho-\omega$ mixing) we would obtain:
       \begin{equation}
       f_{\omega\gamma}= KA \frac{2}{3\sqrt{2}}; \qquad f_{\rho\gamma}= KA 
\frac{6}{3\sqrt{2}}
      \label{12.19}
       \end{equation}
       which produces the old $NRQM$ or $SU_{3}$ result:
       \begin{equation}
        f_{\rho\gamma}:f_{\omega\gamma} = 3:1
        \label{12.20}
       \end{equation}
       What is important, \textit{however}, is that
       \textit{we have shown that this is an almost exact QCD result obtained neglecting only 
the (small) contribution due to
       the Trace term in Eq.(\ref{12.18}).}\\
       \indent We now discuss the experimental value of  $f_{\rho\gamma}/f_{\omega\gamma}$. 
We have:
       \begin{equation}
       (f_{\rho\gamma}/f_{\omega\gamma})^2=(M_\rho/M_\omega)^2\Gamma(\rho\rightarrow e^+e^-)
       /\Gamma(\omega\rightarrow e^+ e^-)
       \label{12.21}
       \end{equation}
       Here the $(M_\rho/M_\omega)^2$ is due to the square of the photon propagator in the 
decay $v\rightarrow e^+ e^-$, proportional to $M_v^{-4}$,
       and to the phase space factor proportional to $M_v^2$. Note that in constructing the 
$v$ meson states with the general parametrization method
       in the rest frame of the decaying meson, one starts from a set of normalized model 
states which are mass-degenerate for all mesons of the nonet. The differences in mass of the 
exact states are incorporated and automatically produced by the flavor breaking part of the 
$V$ operator which, operating on the model states, transforms them into the exact states. 
Therefore we must not insert explicitly the normalization factor $(2M_v)^{-1/2}$ of the state 
of the decaying $v$ meson. For this reason the r.h.s. of (\ref{12.21}) contains 
$(M_\rho/M_\omega)^2$ instead of $(M_\rho/M_\omega)$ that often appears in this formula.
       We adopt the values $\Gamma(\rho\rightarrow e^+e^-)=6.77\pm 0.32\ KeV$ and 
$\Gamma(\omega\rightarrow e^+ e^-)=0.60\pm 0.02\ KeV$.
       Thus, according to Eq.(\ref{12.21}), one gets:
       \begin{equation}
       |f_{\rho\gamma}/f_{\omega\gamma}|_{exp}=3.35\pm 0.07
       \label{12.22}
       \end{equation}
       The deviation from the ``perfect''(3:1) result  Eq.(\ref{12.20}) can be due to the 
$Trace$ term and to the $\rho-\omega$ mixing (for this
       compare e.g. Ref. \cite{bena01},\cite{oconnel}).
       Taking into account only the $Trace$ term 
      , the Eq.(\ref{12.19})
        is replaced by:
       \begin{equation}
       f_{\omega\gamma}= K(A-2C)\frac{2}{3\sqrt{2}} ; \quad \qquad f_{\rho\gamma}= KA 
        \frac{6}{3\sqrt{2}}
       \label{12.23}
       \end{equation}
       that is:
       \begin{equation}
        f_{\rho\gamma}:f_{\omega\gamma} = \frac{3}{1-2C/A}
        \label{12.24}
       \end{equation}
         From the experimental value (\ref{12.22})-and having omitted, as stated, the 
$\rho-\omega$ mixing- one gets: $C/A \approx 0.05$.  \\
       \indent Considering now the $\phi$ decay one has:
        \begin{equation}
        f_{\phi\gamma}:f_{\rho\gamma} = \frac{\sqrt{2}}{3}\frac{B}{A}
        \label{12.25}
       \end{equation}
       and, from the experimental value $\Gamma(\phi\rightarrow e^+e^-)=1.32\pm 0.05\ KeV$, 
one obtains:
       \begin{equation}
       |B/A|=1.24\pm 0.09,
       \label{12.26}
       \end{equation}
        a value consistent with that due to the flavor breaking. Note that a positive flavor 
breaking correction (i.e. $B/A>1$)
        in this case is expected from a ``quarkonium'' model, since the quarkonium wave 
function at the origin $\varphi_Q(0)$ is
       larger for a heavier  $q\bar{q}$ pair increasing the annihilation probability 
amplitude.\\
       \indent To summarize: The general parametrization (with neglect of the $Trace$ term) 
predicts $f_{\rho\gamma}:f_{\omega\gamma}=3$ in spite of
       flavor breaking and, at the same time, accounts for the deviation of 
$|f_{\phi\gamma}:f_{\rho\gamma}|$ from $\sqrt{2}/3$.
       It is misleading to say, as done sometimes, that the experimental values of
       the ratios $|f_{\rho\gamma}:f_{\omega\gamma}:f_{\phi\gamma}|$ are $3:1:\sqrt{2}$, as 
if flavor breaking were absent.
       It is true that $|f_{\rho\gamma}:f_{\omega\gamma}|$ is not far from 3:1, but, as 
expected, $|f_{\phi\gamma}:f_{\rho\gamma}|$ differs
       appreciably from $\sqrt{2}/3$. That is, flavor breaking in the expected amount is 
necessary to account for the last ratio.\\
       \indent A last remark is in order. The above analysis refers to   $\omega$ and $\phi$ 
particles with the  vector mixing angle at
       the ideal value $\theta_v=35.3^o$. Allowing for a small deviation  $\delta\theta_v$ 
one has:
       \begin{equation}
f_{\omega\gamma}(35.3^o+\delta\theta_v)/f_{\omega\gamma}(35.3^o)=(\cos\delta\theta_v+\sqrt{2}
(B/A)\sin\delta
\theta_v)
       \approx(1+1.75\delta\theta_v)
       \label{12.27}
       \end{equation}
       In Ref.\cite{dimo94} we did use erroneously the Eq.(\ref{12.27}) and got a negative 
mixing angle
       $\delta\theta_v$. We now see that Eq.(\ref{12.27}) is compatible with a small {\it 
positive} $\delta\theta_v\approx 1^o$.
        \vskip 10pt
     \noindent \textbf{13. The GP and chiral theories: A few remarks.}\\

     In a comparatively recent paper \cite{du02} Durand, Ha and Jaczko re-derived, using 
heavy-baryon chiral perturbation
     theory, the generalized Gell Mann-Okubo mass formula (Eq.(1),Sect.1) obtained
     in 1992 by one of us (Ref.\cite{mopl92}) using the GP. We found their result 
interesting, especially because it shows that some kind of hierarchy
      plays  a role in chiral QCD.\\
     \indent In an Erratum, Ref.\cite{duE02}, Durand and his collaborators 
acknowledged the coincidence of their result
      with \cite{mopl92}. Before illustrating an interesting aspect that we find in the 
derivation of Durand et al., we wish to
      reproduce a few words of clarification to \cite{duE02} already contained in 
Ref.\cite{dimo03}(footnote[3]).\\
      \indent 1) The $T$ in the Eq.(1) -of the present survey- is called in \cite{duE02} the 
``parameter" $T$. However, as we saw,
       $T$ is not a parameter; it is a well defined quantity: $T=\Xi^{*-} 
-(1/2)(\Omega+\Sigma^{*-})$ [the symbols are the baryon masses].\\
      \indent 2) Also the statement in \cite{duE02} from ``so is not to be used" to ``Our 
approaches differ in that respect"  is not too clear
       because our $T$ is identical to their $\hat{\alpha}_{MM'}$ -except that the Eq.(1) 
(Sect.1) includes the e.m. corrections, essential,
       in this case, to arrive to the level of precision noted in Sect.1.\\
      \indent Of course the fact that some of our results can be obtained from chiral 
theories is not unexpected, because
      the only properties of QCD that the GP exploits are the obvious ones indicated in the 
points a) to d) of Sect.1 of this survey.
      Thus any relativistic theory (chiral or non chiral) compatible with the general
      quark-gluon description of QCD and satisfying the above points a)-d) could, if used 
properly, produce our results.\\
      \indent But the main reason of our interest in the derivation of Durand et al. is the 
following:
       The re-derivation of the generalized Gell Mann-Okubo mass formula
       by Durand et al. using chiral QCD implies that also their chiral description predicts 
that
      certain terms of second order in flavor breaking are very small. (Precisely those terms 
corresponding to our coefficients $c,d$
      in Eq.(19) of Sect.3.) The smallness of $c$ due to the hierarchy (and established 
directly - see the values of the parameters in Eq.(20))
      is all that was used in the General Parametrization to derive the new mass formula: 
compare Sect.3.\\
      \indent In fact the result of Durand et al. seems
      to be a case where a chiral procedure, even if after a very long calculation, leads to 
a prediction in low energy QCD depending only from the
      hierarchy of the parameters. As we stated repeatedly, the hierarchy appears naturally 
in the General QCD Parametrization and, we feel,
      \textit{should have appeared earlier and more generally and ``spontaneously" in all 
theories that (as the chiral ones claim) intend to be
      a good approximation to QCD or at least to provide its main results}. Our interest in 
the result of Durand et al. centered basically on this
      aspect.\\
      \indent The above point (Why in the usual treatments of chiral QCD the hierarchy 
remains so hidden?) was the main one that we intended
       to raise in this Section on the relation between the GP and chiral theories (and 
this question, after all, remains open).
       We have mentioned other problems related to chiral theories both
       in this survey -Sects.6 and 8- and in Ref.\cite{dimo96}, but we will not come back to 
them here.\\
     \vskip 15 pt
        \noindent {\bf 14. Comparing the GP with the $1/N_{c}$ method; some comments.}\\

         We inserted this section only for completeness; its contents is a summary of two 
         arXiv reports \cite{moar00,dimoar00} to which we refer
         for all details. Here we limit to a few comments.\\
         \indent The basis of the ``Large $N_{c}$ method" was a paper of `t Hooft 
         \cite{hooft74} on  an hypothetical QCD with an
         increasingly large $N_{c}$ (number of colors).
         It should be added that `t Hooft's aim was to try to understand quark confinement; he 
         did not use
          the results of his paper for the dynamical QCD problems, treated later by others.\\
          \indent In the large $N_{c}$ method, $N_{c}$ is considered a parameter; one assumes 
          that in the limit $N_{c}\rightarrow \infty$,
          the QCD strong coupling
          constant $g$  decreases proportionally to $1/\sqrt{N_{c}}$. The question if
          this expansion is meaningful at $N_{c}=3$ does not certainly have an obvious answer 
          (see \cite{ssar00}); no one knows the behaviour
          of QCD for $N_{c}\rightarrow\infty$.
           However, this expansion in $1/N_{c}$  (and also in flavour breaking) was widely 
          used in the past years
         (see e.g. \cite{jenarns}). In Ref.\cite{bulpr62} this
         expansion and also the GP were discussed. This fact (plus the popularity of the 
         large $N_{c}$ method) have been the reason for inserting here
         a few remarks on the method.\\
      \indent Let us compare the GP parametrized baryon mass with the same quantity obtained 
        in the $1/N_{c}$ method. Also there \cite{jl95},
       the parametrization of the baryon masses is expressed in terms of 8 parameters (from 
       $c_{(0)}^{1,0}$
       to $c_{(3)}^{64,0}$), but these parameters multiply collective rather than individual 
        quark variables. Again, setting to zero the
       smaller coefficients, one finds a relation between octet and decuplet baryon masses 
        (Eq.(4.6) in \cite{jl95}). This coincides with the result
        obtained several years before in \cite{mopl92} and reproduced here as Eq.(1), the 
generalized Gell Mann Okubo mass formula. But this result
         was ignored both in \cite{jl95} (and in \cite{jenarns}, \cite{bulpr62}).\\
       \indent The re-derivation of the generalized Gell Mann-Okubo mass formula implies, of 
course, that the $1/N_{c}$ method is characterized by some
       hierarchy, at least for the masses, similar to that of the GP; but note the following: 
Whereas in the large $N_{c}$ description
       \textit{the reduction factor in the hierarchy is precisely $1/N_{c}=1/3$},
       in the GP $1/3$ is only an \textit{order of magnitude} for the reduction factor. The 
above feature (``precisely 1/3") is clearly
       very restrictive, perhaps too much, in the large $N_{c}$ procedure. (One can, of 
course, always find some
       way out from such problems, but, in so doing, one loses  the basic feature of the 
theory). In this situation, the statement (see
       Ref.\cite{bulpr62}) that the GP gives a reasonbly good reproduction of the QCD 
results, but imposes ``mild physical constraints" is,
        to say the least, unclear [for the meaning of the ``mild" above see in \cite{bulpr62} 
the remarks between Eqs.(3.5) and (3.6)]. The same
      lack of clarity applies to the expression ``very general quark model" that Lebed used 
for the GP in a previous paper \cite{lebed94}.\\
      \indent A question arises, of course: Does the large $N_{c}$ method lead always to the 
same results of the GP method? The answer is no: below,
      we will illustrate why the $1/N_{c}$ method can be incorrect in (at least) some 
cases.\\
\indent We comment first on the Coleman-Glashow (CG) relation that we treated (sec.10)  by the GP
      in Ref.\cite{dimoplb00}. There we showed that neither the $u-d$ mass difference, nor 
the Trace terms, modify the conclusion,
      reached in \cite{moem92}, that only a few, small, three index terms violate the CG 
relation.
      This explains the ``miraculous'' precision of the CG relation, originally derived by CG 
in exact
      $SU(3)_{f}$; a precision confirmed by a new measurement of the $\Xi^{0}$ mass 
\cite{NA48}.\\
     \indent After the appearance of \cite{dimoplb00} ``On the miracle of the Coleman-Glashow
     and other baryon mass formulas", a preprint by Jenkins and Lebed \cite{jele} implied  
that according to the
     large $N_{c}$ description
     it is ``natural" (not ``miraculous") that the CG relation is so beautifully verified. It 
is implied in \cite{jele} that the
     terms neglected are ``naturally" expected to be small.\\
     \indent This confidence has no basis. For the CG relation the terms in the GP are many 
\cite{moem92,dimoplb00}. It is totally unjustified to
     estimate their global contribution only through the order in $1/N_{c}$ of a typical 
term, as done in the $1/N_{c}$ method. Thus
      the predictions of the $1/N_{c}$ expansion do not have a real QCD foundation.\\
      \indent Another simple case where the results of the large $N_{c}$ method clearly 
differs from the GP analysis (in the wrong sense)
      is that of the magnetic moments of
     $p,n$ and $\Delta$ 's. The $1/N_{c}$ results do not account for the facts, contrary to 
the statements in  Refs.\cite{plb335, prd51, prd53}.
      Due to the omission of effects of order $1/N_{c}^{2}$, the $1/N_{c}$ expansion cannot 
explain
     for instance, the $\mu(p)/\mu(n)$ ratio and the $\Delta\to p\gamma$ transition (Sect.5 
in this survey).\\

     \vskip 20pt
     \noindent \textbf{15. Appendix I - A field theoretical derivation of the GP.}\\

     We now relate the General QCD Parametrization to a conventional Feynman diagrams 
description. Essentially we will
     show how the unitary transformation $V$ connecting the exact state, say of a baryon, 
$\vert \Psi_{B}\rangle$, to its model state
     $\vert \Phi_{B}\rangle$ can be constructed in a field theoretical frame. We did the same 
in the Appendix
     to Ref.\cite{mo89}, except that there we assumed to identify $m$ (the renormalized mass 
of a quark)
     with the (not well defined) mass of a constituent quark. As shown in Ref.\cite{dimo96}, 
this is unnecessary; here we will not keep
     this limitation. In principle the renormalization point for the quark masses can be 
selected arbitrarily; although it
     is fixed at some value, we do not need at this stage to specify it explicitly. For the 
rest we proceed as in the Appendix of Ref.\cite{mo89}.
     Here we will deal only with the relation between the $V$ transformation and the 
conventional field theoretical description.
     We will illustrate the construction of $V$ only for baryonic 3-quark states, but clearly 
the same arguments hold in all cases.
      Call $\vert qqq\rangle$ a state of three quarks and no gluon and
     let $\eta = \sum\vert qqq\rangle\langle qqq\vert\,$ be the projection operator into any 
state (of 3 quarks and no gluons):
     \begin{equation}
     \eta\vert\, qqq\rangle = \vert\, qqq\rangle; \qquad\qquad \eta\,\vert\neq qqq\rangle=0
     \end{equation}
     We rewrite $H$ identically as:
     \begin{equation}
     H= \eta H\eta +(1-\eta)H(1-\eta) +\eta H(1-\eta) +(1-\eta)H\eta
     \end{equation}
     Introduce now the model Hamiltonian $\eta\mathcal{H}\eta$ which is a typical 
non-relativistic quark model Hamiltonian acting only in the Fock
     space of the states of three quarks and no gluons. We decompose $H$ as:
     \begin{equation}
     H=K_{0}+K_{1}
     \end{equation}
     with:
     \begin{equation}
     K_{0} = (1-\eta)H(1-\eta) +\eta \mathcal{H}\eta \,;\qquad
     K_{1}=\eta H(1-\eta) +(1-\eta)H\eta +\eta H\eta -\eta \mathcal{H}\eta
     \end{equation}
     having added to $K_{0}$ and subtracted from $K_{1}$ the model Hamiltonian 
$\eta\mathcal{H}\eta$. Referring to the baryons
     we assume that $\eta\mathcal{H}\eta$ has degenerate eigenvalues $M^{0}_{0}$ for all the 
octet and decuplet baryon states:
     \begin{equation}
     \eta \mathcal{H}\eta \, \Phi_{B}= M^{0}_{0}\Phi_{B} 
\qquad(B=N,\Lambda,\Sigma,\Xi,\Delta,\Sigma^{*},\Xi^{*},\Omega)
     \end{equation}
     where $\vert\Phi_{B}\rangle \vert 0\,gluons\rangle$ are the $L=0$ model states.
     Because in the three-quark sector $K_{0}$ and $\mathcal{H}$ coincide, $\vert 
\Phi_{B}\rangle$ are the degenerate eigenstates of $K_{0}$:
     \begin{equation}
     K_{0}\vert\Phi_{B}\rangle=M^{0}_{0}\vert\Phi_{B}\rangle
     \end{equation}
     In the part $\eta\mathcal{H}\eta$ of $K_{0}$ the masses of the $u,d,s$ quarks are 
assumed to be equal [as implied by Eq.(152)]; the flavor
     breaking mass term (Eq.(9) of Sect.2) appears in the term $(1-\eta)H(1-\eta)$ of 
$K_{0}$ and in the term $\eta H\eta$ of $K_{1}$.
     The term $(1-\eta)H(1-\eta)$ of $K_{0}$ includes in particular the Hamiltonian of the 
non interacting gluons; $K_{1}$ contains the
     interaction terms $\eta H(1-\eta)$ and ${1-\eta}H\eta$ of the quark-gluon Hamiltonian.\\
     \indent We now treat $K_{0}$ as the unperturbed Hamiltonian and $K_{1}$ as the 
perturbation; imagine inserting $K_{1}$ adiabatically
      and construct the true states $\vert\Psi\rangle$ with the procedure of Gell Mann and 
Low
     (Ref.\cite{gellml})(this procedure is not compulsory, but it shows that at least one 
method of
     construction exists). Writing $K_{1}(t)=exp\,(+iK_{0}t)K_{1}\,exp\,(-iK_{0}t)$ the 
adiabatic $U(t,t_{0})$ satisfies:
     \begin{equation}
     i\dot{U_{\alpha}}(t,t_{0})= \exp\,(-\alpha\,|t|)\cdot K_{1}(t)U_{\alpha}(t,t_{0})\qquad 
(\mathrm{with}\,\, \alpha>0,\, U_{\alpha}(t_{0},t_{0})=1)
     \end{equation}
     and the $\vert \Psi\rangle$'s for the lowest bound states corresponding to the lowest 
$\vert \Phi_{B} \rangle$'s are:
     \begin{equation}
     \vert\Psi_{B}\rangle = \lim_{\alpha \rightarrow 0}\exp\,(-w_{B}/\alpha)\cdot 
U_{\alpha}(0,-\infty)\vert\Phi_{B}\rangle
     \end{equation}
     where $w_{B}$ is purely imaginary ($w_{B}+w^{*}_{B}=0$) [so that the factor 
$\exp(-w_{B}/\alpha)$ in front of Eq.(155) is a pure
     phase factor that eliminates the singularity coming  from the $\lim_{\alpha \rightarrow 
0}U_{\alpha}(0,-\infty)$; $w_{B}$ in Eq.(155)
     is related to the $S=U(+\infty,-\infty)$ matrix element of the $\Phi_{B}\rightarrow 
\Phi_{B}$ transition by
     $\lim_{\alpha\leftrightarrow 0}\langle \Phi_{B}\vert\,S\,\vert \Phi_{B} \rangle = 
\exp(2w_{B}/\alpha)$]. The basic operator $V$ introduced
     in the text can be therefore written explicitly as:
     \begin{equation}
     V=\lim_{\alpha \rightarrow 0}\exp\,(-w_{B}/\alpha)\cdot U_{\alpha}(0,-\infty)
     \end{equation}
     Thus, \textit{for instance}, the formula for the magnetic moments (omitting Trace terms) 
has the form used in the text, namely:
     \begin{equation}
     \bM=\langle \Phi \vert V^{\dagger}\mathbf{\mathcal{M}}V\vert \Phi \rangle \quad \qquad
     \end{equation}
     It can be seen easily that the above formula (157) is the same as that used frequently 
for practical calculations:
     \begin{equation}
     \bM= \frac{\langle\Phi\vert T(\mathbf{\mathcal{M}}(0)S)\vert 
\Phi\rangle}{\langle\Phi\vert S \vert\Phi\rangle}\equiv\\
      \langle\Phi\vert T(\mathbf{\mathcal{M}}(0)S)\vert \Phi\rangle_{C}
     \end{equation}
     where the index $C$ means ``connected" .However the formula (156) for $V$ is not that 
written more frequently:
     \begin{equation}
     \vert\Psi\rangle=\lim_{\alpha\rightarrow 
0}\frac{U_{\alpha}(0,-\infty)\vert\Phi\rangle}{\langle\Phi\vert 
U_{\alpha}(0,-\infty)\vert\Phi\rangle}
     \end{equation}
     \indent Although the final formulas for the physical quantities are always the same, in 
the Eq.(159) the denominator is not a pure
     phase factor, as it is the factor multiplying $U_{\alpha}(0,-\infty)$ in Eq.(156).
      For the GP this might create the problem considered at the end of Sect.8 of 
\cite{mo89}, that is, the need for an
     additional parameter in the parametrization of the magnetic moments. But with the $V$ in 
Eq.(156) this problem does not arise, as discussed
     in some detail also at the end of the Appendix of Ref.\cite{mo89}.\\

      \vskip 20pt
     \noindent \textbf{16. Appendix II - A summary of the main steps in the derivation of the 
spin-flavor dependence of the GP terms.}\\

     A)\textit{The spin algebra}\\
     \indent How can one obtain the spin-flavor structures that appear, e.g., in the 
parametrized expression
     of the baryon masses (Eq.19) or in
     that (Eqs.28,29) of the baryon octet magnetic moments? \footnote{The reader should 
consult \cite{mo89} for more details on
     the points treated in this Appendix; in fact this summary cannot replace such a 
consultation, but we felt it necessary to list
     here some important points, to give an idea of the methods used.}\\
     \indent In this Appendix we will illustrate, as an example, the derivation of the 
spin-flavor terms appearing in the magnetic
     moments, at 1st order in flavor breaking (recall that, as we stated in Sect.2, the model 
state $\vert \Phi_{B}\rangle$ has been selected with
     orbital angular momentum $L=0$; this is the reason why, after the $V$ transformation, 
only the spins of the quarks appear in the construction
     of the expressions for the magnetic moment of the baryons). Indeed, all three body space 
axial vectors vanish
     on evaluating their expectation value on $\vert \Phi_{B;L=0}\rangle$; therefore in 
evaluating the magnetic moments
     one has to do with the expectation value of something of the form 
$\sum_{\nu}\sum_{\bp,\bp'}R_{\nu}(\bp,\bp')\bG_{\nu}(\bsigma,f)$,
     where $\bp,\bp'$ are the two independent momenta of three bodies in the rest system. 
Because $R_{\nu}(\bp,\bp')$ is a scalar under space
     rotations, the whole axial vector contributing to the magnetic moment of  $B$ is due to 
the spins.\\
     \indent Setting $\langle X_{L=0}\vert R_{\nu}(\bp,\bp')\vert X_{L=0}\rangle\equiv 
g_{\nu}$, one obtains for the magnetic moments the
      expression $\sum_{\nu}g_{\nu}\bG_{\nu}(\bsigma,f)$. (Compare Sect.IV of \cite{mo89}).\\
     \indent We will start examining the spin dependence of the most general axial vector 
operator formed
      with the spins $\bsigma_{1},\bsigma_{2},\bsigma_{3}$ of three spin
     $1/2$ particles (the quarks) and also its dependence on the flavor operators $f$. This 
is given in Eq.(160), to which one
      has to add all terms, possibly with different coefficients, obtained performing any 
permutation on 1,2,3.
     \begin{equation}
     \bsigma_{1}[a(f)+b(f)(\bsigma_{2}\cdot\bsigma_{3})] +c(f)(\bsigma_{1}\times\bsigma_{2})+ 
d(f)(\bsigma_{1}\times\bsigma_{3})
     \end{equation}
     In (160) $a(f),b(f),c(f),d(f)$ are Hermitian operators constructed with the flavor 
variables and having real matrix elements between real
     functions (we call such operators ``real").\\
     \indent One can show [for the proof compare Ref.\cite{mo89}, page 3001] that when 
calculating the expectation value of the
     operator (160) on a spin-flavor state with a real wave function
     and a given value of the total angular momentum $J$, (a) the cross products terms in 
Eq.(160) give no contribution, and (b) the term
     $\bsigma_{1}(\bsigma_{2}\cdot \bsigma_{3})$ can be rewritten purely in terms of 
$\bsigma_{1}$ and of the $c$ number $J$. It follows that
      the most general Hermitian axial-vector operator $\bG(\bsigma,f)$ constructed in terms 
of the $\bsigma_{i}$ of the three quarks
      (and of the flavor operators $f$) is, when used for evaluating an expectation value as 
specified above, a combination of
      $\bsigma_{1}\Gamma^{J}_{1}(f)$, $\bsigma_{2}\Gamma^{J}_{2}(f)$, 
$\bsigma_{3}\Gamma^{J}_{3}(f)$, where $\Gamma^{J}_{i}(f)$, are three
      operators depending, for a given $J$, only on $f$. With the same type of proof one also 
finds that the scalar:
      \begin{equation}
      (\bsigma_{1}\times\bsigma_{2})\cdot \bsigma_{3}\cdot F(f)
      \end{equation}
      where $F(f)$ is any Hermitian real flavor-dependent operator has a vanishing 
expectation value on any real spin-flavor state of three particles.
      (compare the Eq.(26) in \cite{mo89} and the proof after it).\\
      \indent We now list a set of relations involving $\bsigma_{i}$'s that are useful to 
develop the GP.\\
      Consider the term $\bsigma_{1}(\bsigma_{2}\cdot\bsigma_{3})$ that appears in Eq.(160) 
(multiplied with a real flavor operator $b(f)$)
      and take its $z$ component. It is: $\sigma_{1z}({\bsigma_{2}\cdot 
\bsigma_{3})}=(1/2)\sigma_{1z}[(\bsigma_{2}+\bsigma_{3})^{2}-6]$.\\
      \indent Writing: $2\bJ=\bsigma_{1}+\bsigma_{2}+\bsigma_{3}$ we obtain:
      \begin{equation}
      \sigma_{1z}(\bsigma_{2}\cdot 
\bsigma_{3})=\frac{1}{4}\sigma_{1z}\Big[(2\bJ-\bsigma_{1})^{2}-6\Big]+\frac{1}{4}
      \Big[(2\bJ-\bsigma_{1})^{2}-6\Big]\sigma_{1z}
      \end{equation}
      and, with some algebra:
      \begin{equation}
      \sigma_{1z}(\bsigma_{2}\cdot \bsigma_{3})=\frac{1}{4}\Big[(4\vert \bJ^{2}\vert 
-7)\sigma_{1z} + \sigma_{1z}(4\vert \bJ^{2}\vert -7)\Big]- 2J_{z}
      \end{equation}
      In calculating the expectation value of $\sigma_{1z}(\bsigma_{2}\cdot \bsigma_{3})$ on 
a state with a given $J$ we can write
      $\vert\bJ\vert^{2}=J(J+1)$, a $c$ number.\\
      \indent In conclusion the most general axial-vector operator \textit{(as far as its 
expectation value on a real spin-flavor state
       with a given J is concerned)} is:
       \begin{equation}
       \bG(\sigma,f)=\bsigma_{1}\Gamma^{J}_{1}(f)\quad \,or\, 
=\bsigma_{2}\Gamma^{J}_{2}(f)\quad \,or\,=\bsigma_{3}\Gamma^{J}_{3}(f)\quad \,
       \end{equation}
      where the $\Gamma^J_{i}(f)$ are real flavor operators. We did prove, essentially, that 
the most general axial vector formed with
      three spin $\frac{1}{2}$ particles (under the italicized condition above) is a 
combination of $\bsigma_{1},\bsigma_{2},\bsigma_{3}$
      and nothing else. It might seem strange that the only axial vectors are those listed in 
(164), since we can for instance, multiply
       $\bsigma_{1}$ by $(\bsigma_{1}\cdot\bsigma_{2})$ or by any other scalar product of the 
spin matrices and remain with an axial vector.
      The answer appears from the Eqs.(165) and (166) below where we have limited to the $z$ 
components (the $x$ and $y$ behave similarly).
      \begin{eqnarray}
      \sigma_{1z}(\bsigma_{1}\cdot\bsigma_{2})= \sigma_{2z} +i(\bsigma_{1}\times 
\bsigma_{2})_{z},&\\ \nonumber\\
      \sigma_{1z}\Big[\bsigma_{1}\cdot(\bsigma_{2}\times \bsigma_{3}) 
\Big]=(\bsigma_{2}\times \bsigma_{3})_{z} +i\Big[\bsigma_{1}\times\nonumber
      (\bsigma_{2}\times \bsigma_{3})\Big]_{z}&=\\
      =(\bsigma_{2}\times \bsigma_{3})_{z} + 
i(\bsigma_{1}\cdot\bsigma_{2})\sigma_{3z}-i(\bsigma_{1}\cdot\bsigma_{3})\sigma_{2z}
      \end{eqnarray}
      because the expectation value of $(\bsigma_{i}\times\bsigma_{k})$ is zero as stated 
above, Eqs.(165) and (166)[in conjunction with (163)]
      exemplify how the multiplication of $\bsigma_{i}$ by scalar products of $\bsigma$'s 
does not create new axial vectors in addition to (164)
      with a non zero expectation value.\\
     B)\textit{The flavor algebra}\\
     \indent To write all $\bG_{\nu}(\bsigma,f)$'s that appear, to all orders in flavor 
breaking, in the construction of the magnetic moments
     (and were listed in Eq.(29) of Sect.4 to first order in flavor breaking), we still need 
the expression of the most general
     $\Gamma^{J}(f)$'s appearing in Eq.(164). The $\Gamma^{J}(f)$ have been discussed in 
Sect.VI of Ref.\cite{mo89}. That presentation
      is simple and remains valid for the octet magnetic moments to first order flavor 
breaking. Here we will indicate some changes
      in notation of this survey with respect to \cite{mo89} and also some corrections to the 
list of $\Gamma^{J}(f)$'s;
      these changes do not affect the 1st order flavor breaking baryon octet magnetic moments 
given in Ref.\cite{mo89} and here in Sect.4.\\
      \indent The changes in notation are the following: 1) The projection operators on the
       $\mathcal{P}, \mathcal{N},\lambda$ quarks 
($P^{\mathcal{P}},P^{\mathcal{N}},P^{\lambda}$) have been now
     rewritten respectively as $P^{u},P^{d},P^{s}$. 2)The projection operator $P^{q}$ in the 
e.m. current is now rewritten as
     $[(2/3)P^{u}-(1/3)P^{d}-(1/3)P^{s}]$. 3) We usually write $Q_{i}$ instead of 
$P^{q}_{i}$.\\
     \indent As to the corrections (mentioned above), they consist in the following additions 
to the $\Gamma(f)$'s with respect to those
     listed in Ref.\cite{mo89}: In addition to the $\Gamma(f)$'s listed in the Eq.(36) of 
that paper [and, for the $\Omega^{-}$, in the line below
     that of Eq.(36)], a new class of $\Gamma(f)$ (1st order -or more- flavor breaking) must 
be considered of the form
      $Tr(QP^{s})$, $P^{s}_{i}Tr(QP^{s})$, $P^{s}_{i}P^{s}_{k}Tr(QP^{s})$, and, for the 
$\Omega^{-}$, $P^{s}_{1}P^{s}_{2}P^{s}_{3}Tr(QP^{s})$.
      Note that the treatment of the magnetic moments to 1st order flavor breaking given in 
\cite{mo89} remains correct, because the
      1st order Trace term, as shown in Sect.4, Eq.(30), can be expressed in terms of 
$\bG_{1}$....$\bG_{7}$.\footnote{In the list of Eq.(39)
      of Ref.\cite{mo89} the term $\bG_{7b}$=$\sum_{i\neq 
k}P^{q}_{i}P^{\lambda}_{i}P^{\lambda}_{k} \bsigma_{i}$ should also have appeared;
      it is classified however among the 2nd order flavor breaking terms that were not 
considered there.}\\
      \indent A last remark on the meaning of ``order in flavor breaking". When we refer to 
first order flavor breaking, this means to keep
      only flavor breaking terms linear in the $P^{s}$. But, because it is 
$(P^{s}_{i})^{n}=P^{s}_{i}$ for any (integer positive) $n$,
       the GP includes in fact all orders of the flavor breaking contributions 
\textit{additive} in the quarks; flavor-breaking effects
       of 2nd order or more coming from terms carrying the indices of two different quarks 
are not included.\\
       C)\textit{The coefficients in the baryon mass parametrization}\\
       \indent This subject is treated in Sect.XII of Ref.\cite{mo89} and in the Appendix B 
of Ref.\cite{dimo96}. Both should be consulted (and,
        in particular, the first order flavor breaking mass formulas (82)(83)of 
Ref.\cite{mo89} can be useful), but here
       we limit to list the formulas (given in \cite{dimo96}) leading to the values of the 
coefficients $M_{0},B,C,D,E,(a+b),c,d$
       displayed in Sect.3 of this survey.\\
       \indent Because in Sect.3 the treatment of the general parametrization (and therefore 
of its parameters)
        refers to the strong interactions only [that is the masses
       in Eq.(19),Sect.3 are the eigenvalues of $H_{QCD}$, without the e.m. interaction], it 
is necessary, especially for the smaller
       coefficients, to extract from the experimental mass values  their strong part.
       Stated otherwise, to determine the coefficients of the parametrization (19), one must 
use mass values independent
        of the e.m. and isospin breaking ($m_{u}\neq m_{d}$), at least to first order.
        We did this already when writing the generalized Gell Mann-Okubo mass formula 
(Eq.(1)).\\
        \indent For the large coefficients $M_{0},B,C$,
        this precision is usually not necessary and we will just use an average mass on the 
different baryons of a multiplet, e.g. take
        $\overline{N}\equiv(n+p)/2$ for the mass of a nucleon or 
$(\overline{\Delta}\equiv\Delta^{++} +\Delta^{+}+\Delta^{0}+\Delta^{-})/4$
        for the average mass of a $\Delta$. Thus we will determine $M_{0},B,C$ as:
        \begin{equation}
\       M_{0}=(\overline{N}+ \overline{\Delta})/2, \qquad B=\Lambda - \overline{N}+3E, \qquad  
C=(\overline{\Delta}-\overline{N})/6
        \end{equation}
        where the parameter $E$ in (167) (the same appearing in Eq.(168)) will be expressed 
in terms of the masses in the second equation
        below; $D,E,(a+b),c,d$ are determined from the following equations (168) that are 
Coulomb and isospin independent to first order:
        \begin{eqnarray}
        D=&(1/6)[(\Sigma^{*-}-\Delta^{-})+(\Sigma^{+}-p)] \\
        E=&(1/6)(\Sigma^{*-}-\Delta^{-})+(1/12)(\Sigma^{0}-3\Lambda+2n)\nonumber\\
        c=&(1/3)\big[(\Xi^{*-}+\Xi^{-})-(\Sigma^{*-}+\Sigma^{-})\big]-2E \nonumber\\
        a+b=&\Xi^{-}-\Sigma^{-}+(1/2)(\Sigma^{0}-3\Lambda+2n) +2c \nonumber\\
        d=&\Omega^{-}-\Delta^{++} -3(\Xi^{*0}-\Sigma^{*+}) \nonumber
        \end{eqnarray}
        where in the above formulas $\Delta^{-}$ stays for:
        \begin{equation}
        \Delta^{-}=\Delta^{++} +3(n-p)
        \end{equation}


\end{document}